UNIVERSITY OF GENOVA
PHD PROGRAM IN BIOENGINEERING AND ROBOTICS

# Development of a Novel Platform for *in vitro* Electrophysiological Recording

Giovanni Melle

| | |
|---|---|
| Francesco de Angelis | Supervisor |
| Giorgio Cannata | Head of the PhD program |
| Thesis Jury: | |
| Andrea Barbuti | University External examiner |
| Vittoria Raffa | University External examiner |

Department of Informatics, Bioengineering, Robotics and Systems Engineering

# Table of Contents

# Abstract

The accurate monitoring of cell electrical activity is of fundamental importance for pharmaceutical research and pre-clinical trials that impose to check the cardiotoxicity of all new drugs.
Traditional methods for preclinical evaluation of drug cardiotoxicity exploit animal models, which tend to be expensive, low throughput, and exhibit species-specific differences in cardiac physiology [1]. Alternative approaches use heterologous expression of cardiac ion channels in non-cardiac cells transfected with genetic material.
However, the use of these constructs and the inhibition of specific ionic currents alone is not predictive of cardiotoxicity.
Drug toxicity evaluation based on the human ether-à-go-go-related gene (hERG) channel, for example, leads to a high rate of false-positive cardiotoxic compounds, increasing drug attrition at the preclinical stage. Consequently, from 2013, the Comprehensive in Vitro Proarrhythmia Assay (CiPA) initiative focused on experimental methods that identify cardiotoxic drugs and to improve upon prior models that have largely used alterations in the hERG potassium ion channel.
The most predictive models for drug cardiotoxicity must recapitulate the complex spatial distribution of the physiologically distinct myocytes of the intact adult human heart. However, intact human heart preparations are inherently too costly, difficult to maintain, and, hence, too low throughput to be implemented early in the drug development pipeline. For these reasons the optimization of methodologies to differentiate human induced Pluripotent Stem Cells (hiPSCs) into cardiomyocytes (CMs) enabled human CMs to be mass-produced in vitro for cardiovascular disease modeling and drug screening [2]. These hiPSC-CMs functionally express most of the ion channels and sarcomeric proteins found in adult human CMs and can spontaneously contract.
Recent results from the CiPA initiative have confirmed that, if utilized appropriately, the hiPSC-CM platform can serve as a reliable alternative to existing hERG assays for evaluating arrhythmogenic compounds and can sensitively detect the action potential repolarization effects associated with ion channel–blocking drugs [3].
Data on drug-induced toxicity in hiPSC-CMs have already been successfully collected by using several functional readouts, such as field potential traces using multi-electrode array (MEA) technology [4], action potentials via voltage-sensitive dyes (VSD) [5] and cellular impedance [6]. Despite still under discussion, scientists reached a consensus on the value of using electrophysiological data from hiPSC-CM for predicting cardiotoxicity and how it’s possible to further optimize hiPSC-CM-based in vitro assays for acute and chronic cardiotoxicity assessment. In line with CiPA, therefore, the use of hiPSC coupled with MEA technology has been selected as promising readout for these kind of experiments.
These platforms are used as an experimental model for studying the cardiac Action Potentials (APs) dynamics and for understanding some fundamental principles about the APs propagation and synchronization in healthy heart tissue. MEA technology utilizes recordings from an array of electrodes embedded in the culture surface of a well. When cardiomyocytes are grown on these surfaces, spontaneous action potentials from a cluster of cardiomyocytes, the so called functional syncytium, can be detected as fluctuations in the extracellular field potential (FP). MEA measures the change in FP as the action potential propagates through the cell monolayer relative to the recording electrode, neverthless FP in the MEA do not allows to recapitualte properly the action potential features.
It is clear, therefore, that a MEA technology itself is not enough to implement cardiotoxicity assays on hIPSCs-CMs. Under this issue, researchers spread in the world started to think about solutions to achieve a platform able to works both at the same time as a standard MEA and as a patch clamp, allowing the recording of extracellular signals as usual, with the opportunity to switch to intracellular-like signals from the cytosol.
This strong interest stimulated the development of methods for intracellular recording of action potentials. Currently, the most promising results are represented by multi-electrode arrays (MEA) decorated with 3D nanostructures that were introduced in pioneering papers [7,8], culminating with the recent work from the group of H. Park [9] and of F. De Angelis [10–16]. In these articles, they show intracellular recordings on electrodes refined with 3D nanopillars after electroporation and laser optoporation from different kind of cells. However, the requirement of 3D nanostructures set strong limitations to the practical spreading of these techniques. Thus, despite pioneering results have been obtained exploiting laser optoporation, these technologies neither been applied to practical cases nor reached the commercial phase.
This PhD thesis introduces the concept of meta-electrodes coupled with laser optoporation for high quality intracellular signals from hiPSCs-CM. These signals can be recorded on high-density commercial CMOS-MEAs from 3Brain characterized by thousands of electrode covered by a thin film of porous Platinum without any rework of the devices, 3D nanostructures or circuitry for electroporation[7]. Subsequently, I attempted to translate these unique features of low invasiveness and reliability to other commercial MEA platforms, in order to develop a new tool for cardiac electrophysiological accurate recordings.
The whole thesis is organized in three main sections: a first single chapters that will go deeper in the scientific and technological background, including an explanation of the cell biology of hiPSCs-CM followed by a full overview of MEA technology and devices. Then, I will move on state-of-the-art approaches of intracellular recording, discussing many works from the scientific literature.
A second chapter will describe the main objectives of the whole work, and a last chapter with the main results of the activity. A final chapter will resume and recapitulate the conclusion of the work.

# Scientific and Technological Background

# Dissociated Cardiomyocytes Coupled to Micro-Electrode Array Devices

In this chapter, I will outline some basics about human-derived cardiomyocytes and their specific biological features. I will also discuss the main electrophysiological approaches used to record and monitor their electrical activity. Subsequently, I will discuss about both intracellular and extracellular recording techniques that offer different advantages and disadvantages, focusing more about features and peculiarities of these approaches. Multi electrode array (MEA) technology will be revised more in details, with a final focus about the applications of MEA in cardiotoxicity and pharmacology.

## Human Pluripotent Stem Cells-Derived Cardiomyocytes

One of the most remarkable discoveries of the last few years is that human adult somatic cells can be reprogrammed, by the transient expression of just a few key transcription factors, into human-induced Pluripotent Stem Cells (hiPSCs) [17,18]**.** Various methods have been then described to induce differentiation of iPSCs in various cell lineage.

The growth of undifferentiated cells as aggregates in suspension, which causes them to form structures called embryoid bodies, is the most widely used approach. Within the embryoid body, derivatives of the three primary germ layers (ectoderm, endoderm and mesoderm) develop spontaneously [19].

Cardiomyogenic mesodermal progenitors are normally formed in the embryo during gastrulation as cells of the epiblast pass through the primitive streak. Their development is induced by the regulation between chemical inductive signals, like the Fibroblast Growth Factor (FGF), and repressive signals, mostly Wnt and beta-catenin. After that, the early beating heart tube is formed, composed by synchronously beating small myocytes [20].

After huge efforts, recent protocols have been developed to initiate differentiation of hiPSCs into cardiomyocytes using small-molecule regulators of the Wnt signaling pathway. This is usually an efficient process, and one can readily obtain >90% purity hiPSC-CMs after differentiation and metabolic selection using low-glucose media enriched with lactate. In addition, in the human system the use of anti VCAM1 antibody in the early differentiation steps allowed the sorting of the CM population, comprising ventricular as well as pacemaker cells [21].

hiPSC-CM vary in maturity, thus, we will define them as either early phase, defined as contractile cells, with some proliferative capacity and with embryonic like electrophysiology (i.e., depolarized maximum diastolic potential (MDP) and small action potential amplitude), or late phase, defined by loss of proliferative capacity and more adult-like electrophysiology [22]. The former, in particular, show primitive characteristics in morphology, cytoskeletal proteins, and the inability to develop a well-organized excitation– contraction coupling machinery [23].

The differentiated hiPSC-CMs will spontaneously contract and express most sarcomeric proteins and ion channels normally found in human cardiac tissue. For example, hiPSC-CMs express cardiac-specific sarcomeric markers such as cardiac troponin T (TNNT2) and alpha-actinin (ACTN2) [24]. Once purified, hiPSC-CMs can be replated onto Matrigel or a comparable matrix in high-throughput plates for downstream imaging and contractility assays.

In general, differentiation of pluripotent stem cells to cardiomyocytes results in mixtures of ventricular-like, atrial-like and pacemaker-like cells defined by patch clamp electrophysiology of action potentials (APs) (Dell'Era *et al.,* 2015). Ventricular- like action potentials are defined by a relatively fast upstroke velocity and a plateau phase that results in longer repolarization compared with the more triangular shaped atrial-like APs. Relative slow upstroke velocities and much smaller amplitudes characterize pacemaker-like cells. Interestingly, specific differentiation protocols seem to affect the ratios of the different cardiac cell types formed in culture.

The establishment of methodologies to convert hiPSCs to CMs has enabled human cardiomyocytes to be mass-produced *in vitro* for cardiovascular disease modeling and drug screening. In addition, hiPSC-CMs have been able to recapitulate, at the cellular level, phenotypes for a variety of cardiovascular diseases [25,26].

# Biological Features of Cardiomyocytes

The electrophysiological behavior of the heart is characterized by synchronized propagation of excitatory stimuli that result in rapid depolarization and slow repolarization of various excitable cell types. These rapid depolarization events are known as Action Potentials (APs).

The AP represents the time-dependent changes of the potential across the cellular membrane in CMs that occur during the contraction of heart tissue. This process requires a well-defined orchestration of numerous ion channels. Since the human heart comprises different cardiomyocyte subtypes, the APs patterns and shapes are significantly different, depending on the regions of heart (e.g., atrium, sinus node and ventricle) [27]. Despite this electrical heterogeneity, each subtype specific pattern consists of five different phases, reflecting the activity of certain ion channels as show in Figure 1.

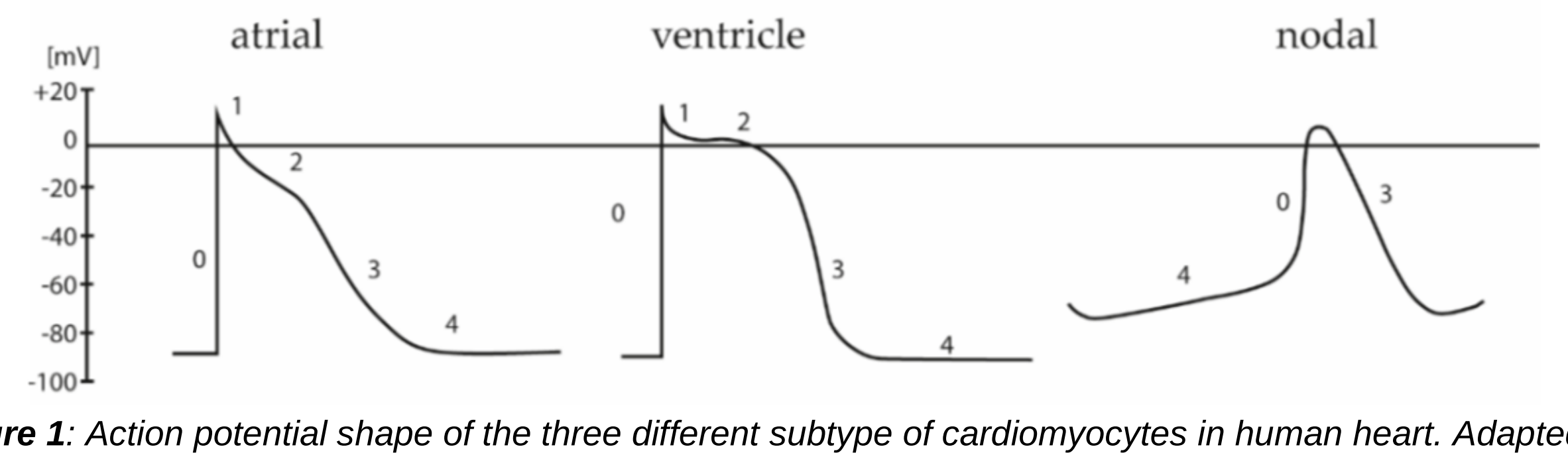

***Figure 1**: Action potential shape of the three different subtype of cardiomyocytes in human heart. Adapted from [28].*

After a depolarization stimulus, opening of voltage-gated Na+ channels promotes sodium influx into the cytoplasm, causing in a rapid depolarization of the membrane potential up to +20 to +40 mV (Phase 0). After that, a time and voltage-dependent opening and closing of various ion transporters permeable to Na+, Ca2+ and K+, leading to a transient hyperpolarization (phase 1) of the membrane potential (−10 to −30mV). The following phase 2 is primarily driven by depolarization dependent L-type Ca2+ channels. Due to an equilibrium between inward currents of calcium and efflux of potassium phase 2 shows a plateau period, which is particularly prominent in ventricular muscle cells [29]. During the plateau phase, Ca2+ channel conductance decreases while the outward current of K+ inclines. This in turn promotes further repolarization (phase 3) leading to a resting potential of ~−85 mV (phase 4).
As stated above, AP patterns are unique for each cardiac subtype resulting from different ion channel composition within the cellular membrane. Nodal cells, found in sinoatrial node, are capable to generate their own AP without an additional depolarizing stimulus. Compared to the resting potential of atrial and ventricular cells, nodal cells have a highly unstable potential, begins at ~−60 mV (MDP) and gradually increases towards a threshold. This "pacemaker potential" is generated by the opening, at negative potential of about 60mV, of the funny channels that are members of the family of potassium channel that are however non-selective and at physiological potentials conduct a slow sodium inward currents. Once a threshold is reached, opening of voltage-gated Ca2+ channels induce a strong upstroke. This is in contrast to atrial and ventricular cells where a Na+ influx mainly contributes to the rapid depolarization in phase 0 [27].
Analysis and classification of AP patterns of iPSC-CMs is challenging as they demonstrate a large amount of variability, which supports the notion that CMs derived from iPSCs are a mixture of different cardiac subtypes of distinct maturation level [30]. Although multiple differentiation protocols have been established, researchers failed to generate fully mature cardiomyocytes *in vitro* possessing identical electrophysiological properties as their native adult counterparts. For example, Ronaldson-Bouchard *et al*. have shown structural and metabolic maturity and adult-state like gene-expression of three iPSC cell lines after cultivation as cardiac tissues for four weeks; it remains to be seen whether this approach can be generally transferred to any iPSC cell line [31]. In addition, it is difficult to compare APs among different studies because of various experimental conditions used. Nevertheless, a common feature of hiPSC-CMs compared to native CMs is their ability to generate APs without the need of an external, depolarizing signal, indicating the presence of funny channel and related currents that drive spontaneous activity.
Nowadays, several different techniques exist to study the electrophysiological properties of cardiac cells; each of these techniques has its own advantages and limitations, which are described in detail in the following section.

# Cell-Electrode Interface: Electrical Modelling

Electrophysiologists frequently described the interface between cells and electrode as equivalent circuits, with a combination of resistors and capacitances.
The cell membrane is a double lipid layer that separates charges in the extracellular space from charges in the cytoplasm. A pure lipid membrane is an excellent insulator, but a real membrane consists of a mosaic of lipids and structural proteins that span the whole surface and work also as ion channels, allowing for charge trafficking. Therefore, these proteins reduce the total resistance of the membrane. If one wants to apply a given voltage across the membrane, for example by injecting current with an electrode, the current required is determined by the Ohm's law ($V = R_{membrane} * I$).

The higher the membrane resistance, the lower the injected current. Because the membrane is an insulator that separates opposing charges, it not only has a resistance, but also a capacitance. Therefore, to change the voltage across it, one need to charge the capacitance. Since both membrane resistance and capacitance occur over the membrane, they are electrically parallel.
The structural connection between a cell and a surface-integrated electrode is resumed in Figure 2, which shows a generic circuit model of a bio-hybrid interface, composed of a cell, an electrode and a cleft in between. The cell membrane is divided into the non-junctional membrane that interacts with the extracellular medium (with Rnj as non-junctional resistance and Cnj as non-junctional capacitance) and a junctional membrane that faces the electrode (with Rj as junctional resistance and Cj as junctional capacitance). The gap between cell and electrode generates a seal resistance (Rs). The voltage formed over Rs directly modulates the charge dispersion across the metal electrode. The electrode uses an access resistance (Ra) to measure intracellular potential.
Changing the quantitative relationships between the seal resistance, junctional membrane properties and the electrode impedance will therefore change the mode (extracellular to intracellular), quality (time derivative of the AP to Ohmic recordings) and amplitude (from microvolts to tens of mV) of the recorded potentials.

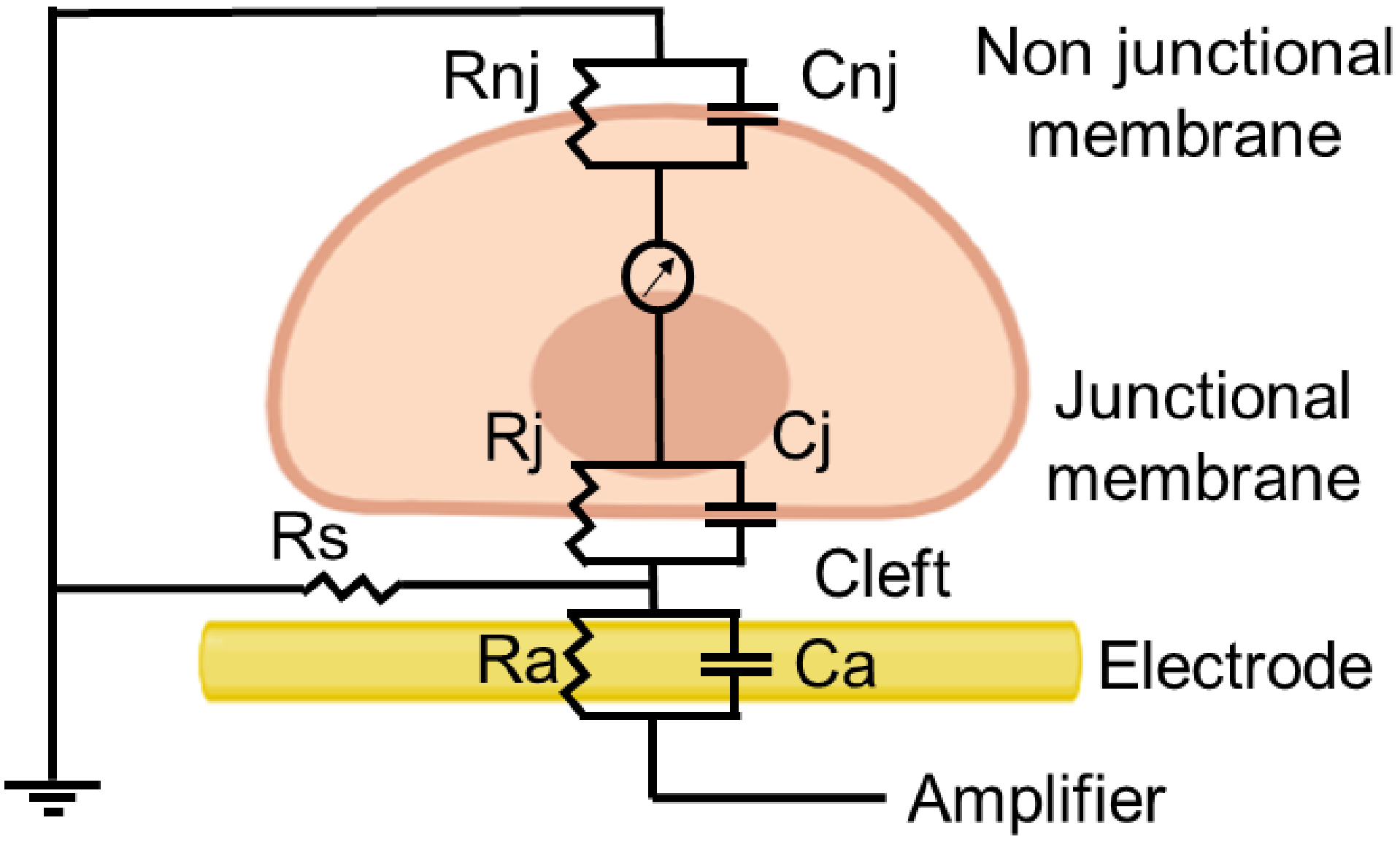

***Figure 2**: Schematic model depicting the relationships between a cell and a substrate-integrated electrode and the analogue passive electrical circuit.*

# Platforms for Functional Analysis of hiPSCs-CMs

Many techniques have been adopted for measuring the electrophysiological activity of in vitro electroactive cell's networks (Figure 3). Between these methods, a first distinction can be made between intracellular and extracellular techniques. Intracellular techniques allow the direct measurement of the voltage across the membrane. Therefore, this kind of electrophysiological measurement needs two measuring points: the first one inside the cellular cytoplasm and the second one outside the cell, requiring the breaking of the membrane.
Alternatively, utilizing an extracellular technique, one can place the first measuring point outside the cell, but very close to the membrane, and the second one, the reference, far away from the cell. When an action potential occurs, the intracellular and extracellular ionic concentrations are both modified by the membrane transport properties: the extracellular changes are localized near the membrane and currents entering or leaving a cell generate voltage signals at the electrode nearby. This results from a resistive drop in the medium between the reference electrode and the recording electrode.
Intracellular recording is more suitable for sensitive recording, but it needs breaks a part of plasma membrane to approach the cytosol directly. Therefore, intracellular recording allows to record signals from only few cells at time with high spatial resolution, ideally subcellular resolution, to reveal details of cellular signaling. Being a highly invasive method, it is very difficult to perform long-term or large-scale recording. On the contrary, extracellular recording is a non-invasive method, allowing recording extracellular signals simultaneously from hundreds of electrodes at the time to reveal network characteristics. It also supports long-term recording, but the weakness in signal strength and poor quality of the recorded signals are significant limitations of this method.

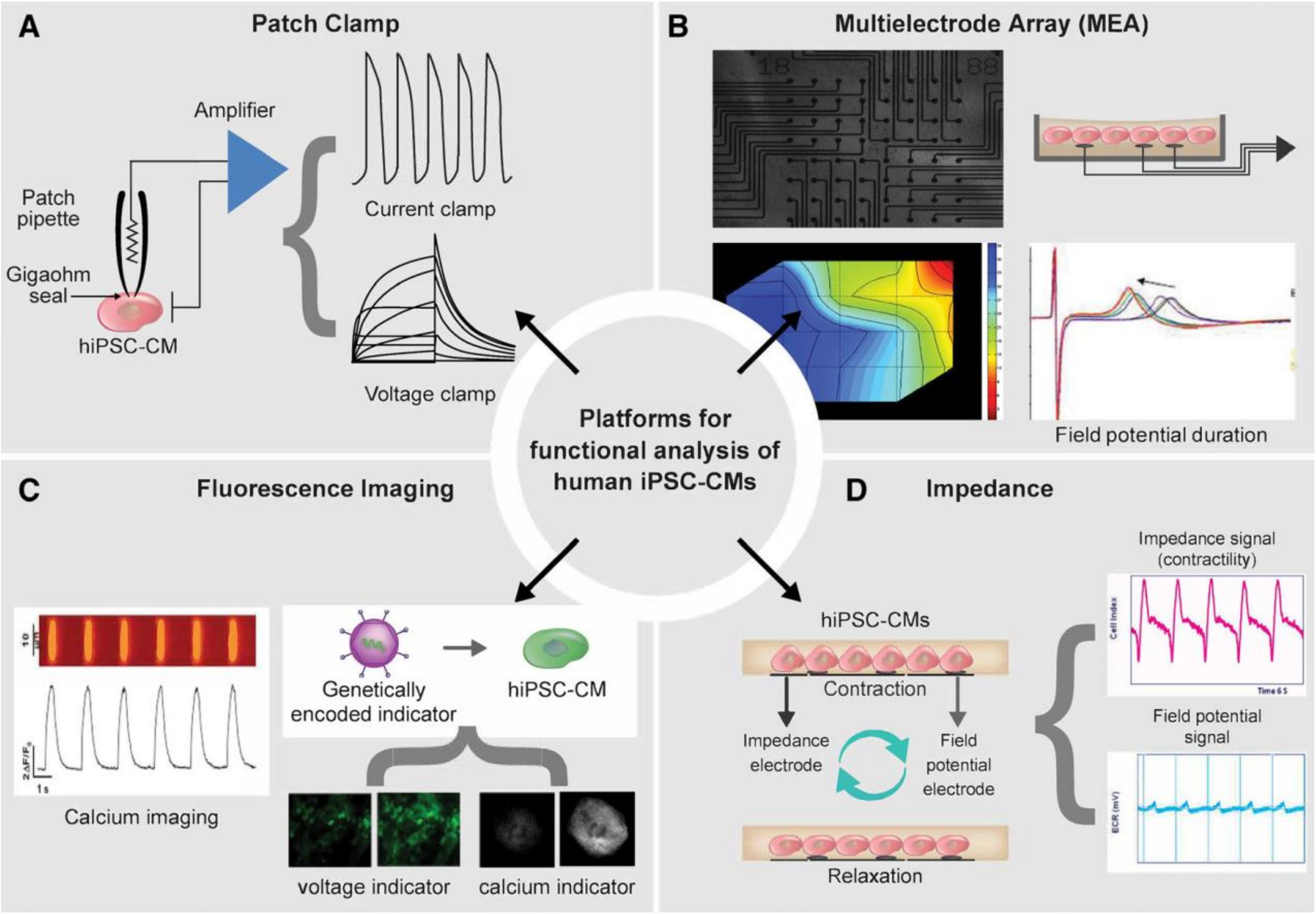

***Figure 3:*** *overview of electrophysiological measurements.*

# Technologies of Optical Imaging

Going beyond the recording of the activity from electroactive cell network through electrodes, one can exploit chemical dyes that react following ionic movements and dynamics that a cell undergoes during an action potential. Measurements of intracellular calcium in living cells was revolutionized by Tsien in 1980 [32], when he reported fluorescent derivates of the $Ca^{2+}$ chelator 1,2-bis(o aminophenoxy)ethane-N,N,N',N'-tetraacetic acid (BAPTA). This molecule binds to $Ca^{2+}$ reversibly and specifically, with little sensitivity to other cations, like Magnesium and protons.

Initially they were impermeable and the delivery required microinjection. Formulation as an acetoxymethyl (AM) ester overcame this issues and indicators like quin2 [33] were produced and rapidly applied to cardiomyocytes [34]. Quin2-AM can cross the plasma membrane and accumulate in isolated cardiomyocytes after brief incubation allowing Ca2+ transient recording. Calibration was achieved by selective cell permeabilization using either triton (to disrupt all cell membranes) or digitonin (to selectively disrupt the plasma membrane).

Quin-2 has single excitation and emission wavelengths; and thus is intensiometric. Therefore, overall fluorescence intensity changes can be affected by indicator concentration, a parameter that is difficult to control due to the passive diffusion mechanism of intracellular accumulation. Robust use of quin2 proved technically difficult due to its poor dynamic range, brightness, and inaccurate calibration. The second generation of intensiometric chemical Ca2+ dyes, such as fluo-3 [35] and fluo-4[36], were ~30 times brighter than quin-2, so lower dye concentrations are required for dynamic imaging in cardiomyocytes.

Despite these optimal features, calcium dyes have some drawbacks, related to the introduction of more variables that may change the native physiological function of the cell. Furthermore, Ca2+ dyes might affect also other intracellular components, cofactors, reducing enzyme's viability and limiting cell survival. (Smith *et al.* 2018).

## Impedance measurements

Functional assays based on cellular impedance (an indirect measure of cardiomyocyte contractility) offer a noninvasive, label-free, and high throughput analysis method [37].

The mechanical displacement of the cells during cardiomyocyte contraction is measured as variations in impedance (equivalent to resistance in a direct current circuit) that directly correlates with the beating frequency. These measurements use weak alternating current between the electrodes with tissue culture medium as the electrolyte. The electronic hardware monitors the voltage across the electrodes, and the impedance is calculated using the alternating current version of Ohm law where impedance (Z) rather than resistance (R) is calculated as Z=V/I. Monitoring of the impedance signal does not alter cellular physiology either in an excitatory, suppressive, or cytotoxic fashion [37]. Clearly, this method has great potential in ascertaining the beat rate/arrhythmogenicity of hiPSC-CMs. However caution should be taken to avoid over-interpreting the data. It is important to realize here that the changes in the beating pattern of the cells are recognized only because of a measurement of a minute change in resistance and not because of their direct electrical activity [38].

## Multi Electrode Array for Extracellular Recording

External micro-transducers are used to monitor and measure the extracellular electrophysiological activity, which occurs outside of the cells membrane. These micro-transducers allow for precise measure of the concentrations of ions in their vicinity, which changes according to the activity of the cell. During an action potential, ions, mostly sodium and potassium, travel across the cell membrane in and out of the cell. These moving ions generate a floating electric field, the so-called Localized Field Potential (LFP), which can directly influence the open-gate region of a field-effect transistor or which can be recorded by means of metal microelectrodes [39].
The LFP encompasses the spatiotemporal electrical activity of cell clusters attached to the electrode, thus, it is the superimposition of all ionic processes, ranging from fast action potentials to slowest fluctuations [40]. Since the biophysical processes underlying the generation of LFPs are well known, it is possible to reconstruct the corresponding AP pattern and to extract important physiological parameters. Figure 4 depicts the different phases of a typical ventricular AP pattern (top) and the corresponding FP measured by an external electrode. The LFP waveform contains a strong transient spike attributed to the Na+ influx and associated membrane depolarization, followed by a gentle incline based on the intracellular increase of Ca2+ level and ending with repolarization associated with K+ efflux [41].

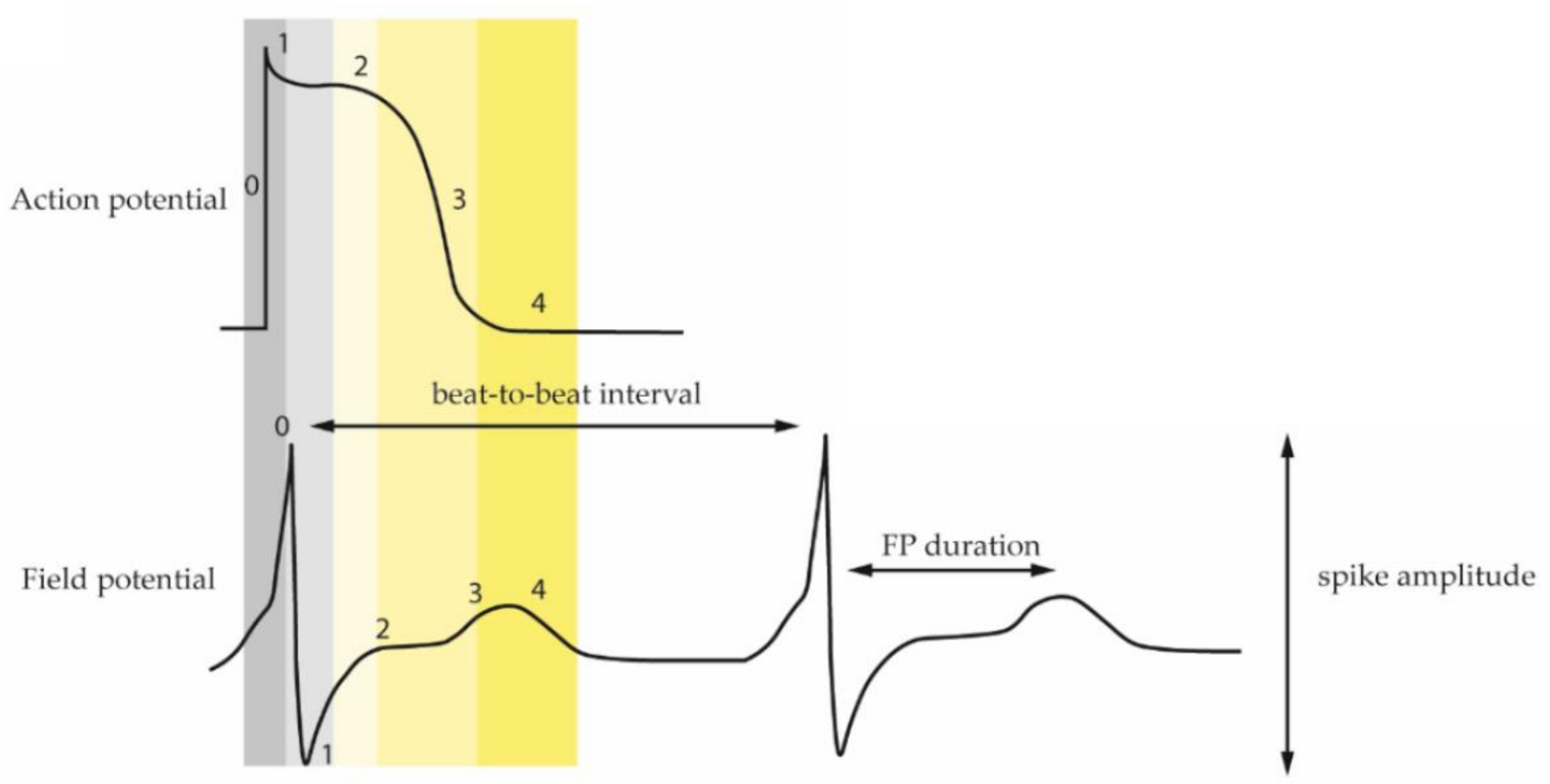

***Figure 4***: *Comparison of the different phases between recorded action potential and field potentials*.

Multielectrode Array (MEA) technology represents the gold standard for extracellular measurement and is extensively used for basic and applied electrophysiological research of neuronal- and cardiomyocyte-networks. A MEA is a two-dimensional grid of biocompatible micro-transducers embedded in an insulating surface, such that each micro-transducer is electrically independent of its neighbors. Each micro-transducer is connected to a differential amplifier, which measures voltage or current with respect to a common reference (Figure 5). Since cells can be maintained for up to several months, MEA technology is therefore suitable for long-term studies. Nevertheless, the ability to record extracellular signals alone is not enough to provide complete information about the activity of cell's network. Therefore, some MEA devices have bidirectional functionality, meaning that they can, in addition to recording activity, deliver either current or voltage stimuli, thereby eliciting a response in a population of electroactive cells. These devices can be used to stimulate single or small clusters of cells.

Thomas et al. described the first Micro-Electrode Arrays in 1972 [42]. The work aimed to record electrical activity from cells using a noninvasive method, allowing them to address questions about plasticity and electrical interaction among cells. The device consisted of platinized gold microelectrodes (two rows of 15 electrodes each, spaced 100 μm apart) embedded onto a glass substrate and passivated by photoresist. This device was able to record field potentials from spontaneous contracting sheets of cultured chick cardiomyocytes, but incapable to record activity from each single cell. Five years later, Guenter Gross and his collaborators proposed the idea of a MEA, without knowledge of the previous work [43]. They showed recordings from an isolated snail ganglion laid over the electrodes, with single action potentials having amplitudes up to 3 mV, depending upon the cell size.

The first successful recordings from single dissociated neurons using a MEA, instead, were reported by Pine in 1980. He succeeded in recording from a network of rat superior cervical ganglion neurons, cultured for up to three weeks over a MEA with 32 gold electrodes (two parallel lines of 16 electrodes each, 10 μm square and 250 μm apart), platinized and insulated with silicon dioxide [44]. He also used the same MEA for stimulating neurons with voltage pulses of 0.5 V and duration of 1 ms. These three results define a milestone for the following works, marking the beginning of in vitro network electrophysiology using MEA.

In conclusion, MEA are tools that allow for the monitoring of an area with dozens of recording sites reporting local field potentials. These local field potentials give information about clusters of cells close to a given electrode site such that experimental conditions could be correlated with changes in electrical behavior across the network. Nevertheless, much of the channel-specific information encoded in the action potential shape are lost in field potential measurements, therefore an intracellular recording approach could give a more detailed information about single cell action potential dynamic and electrical activity.

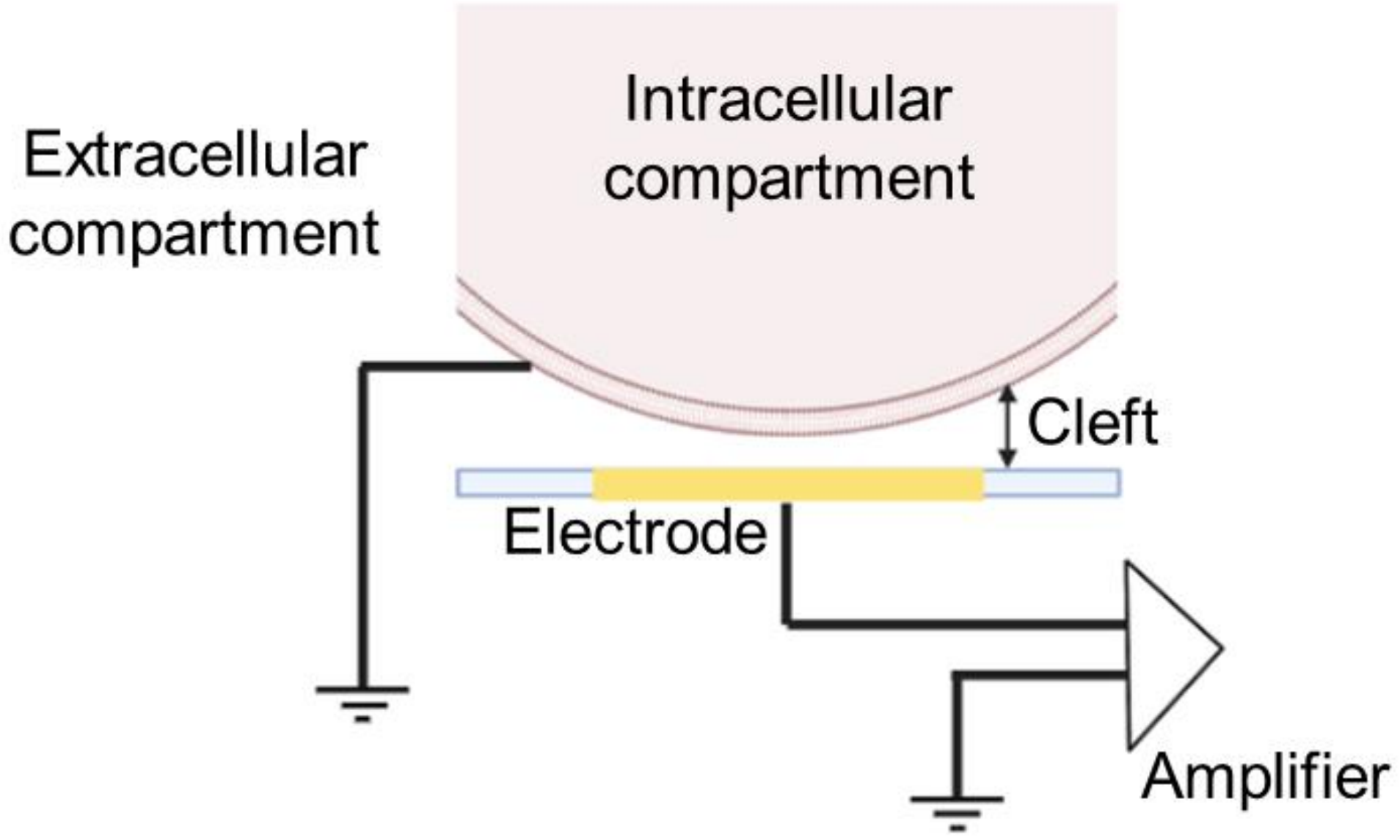

***Figure 5**: Sketch of extracellular recording configuration.*

## Patch Clamp Approach for Intracellular Recording

The clarification of human body functions and its associated pathophysiology pushed biologists and engineers on developing tools that span from whole tissues to single cells analysis [45–47].
These tools aim to reveal the condition of the cell's internal compartment, thus introducing immediately significant technological issues, infact human cells are regulated systems delineated by a complex and active lipid bilayer membrane and measuring inside cells require the insertion of a probe through this membrane.
Intracellular electrodes have a greater sensitivity and can measure the full spectrum of membrane potentials, including not only the action potential but also the much smaller post/pre-synaptic potentials. Traditionally, the patch clamp technique has been the dominant electrode-based method for intracellular electrophysiology. Erwin Neher and Bert Sakmann described for the first time how to record currents from single ion channels [48].
The conventional patch clamp experiment (Figure 6), which consists of manual electrophysiology, uses glass microelectrodes that press against the cell surface to form a tight interaction with a giga-ohm (GΩ) seal resistance (Rs) between the cell membrane and the rim of glass microelectrode (Obergrussberger *et al*, 2015). The current that flows across the plasma membrane can be directly measured by controlling the voltage applied to the membrane (voltage clamp configuration), or *viceversa* (current clamp configuration).

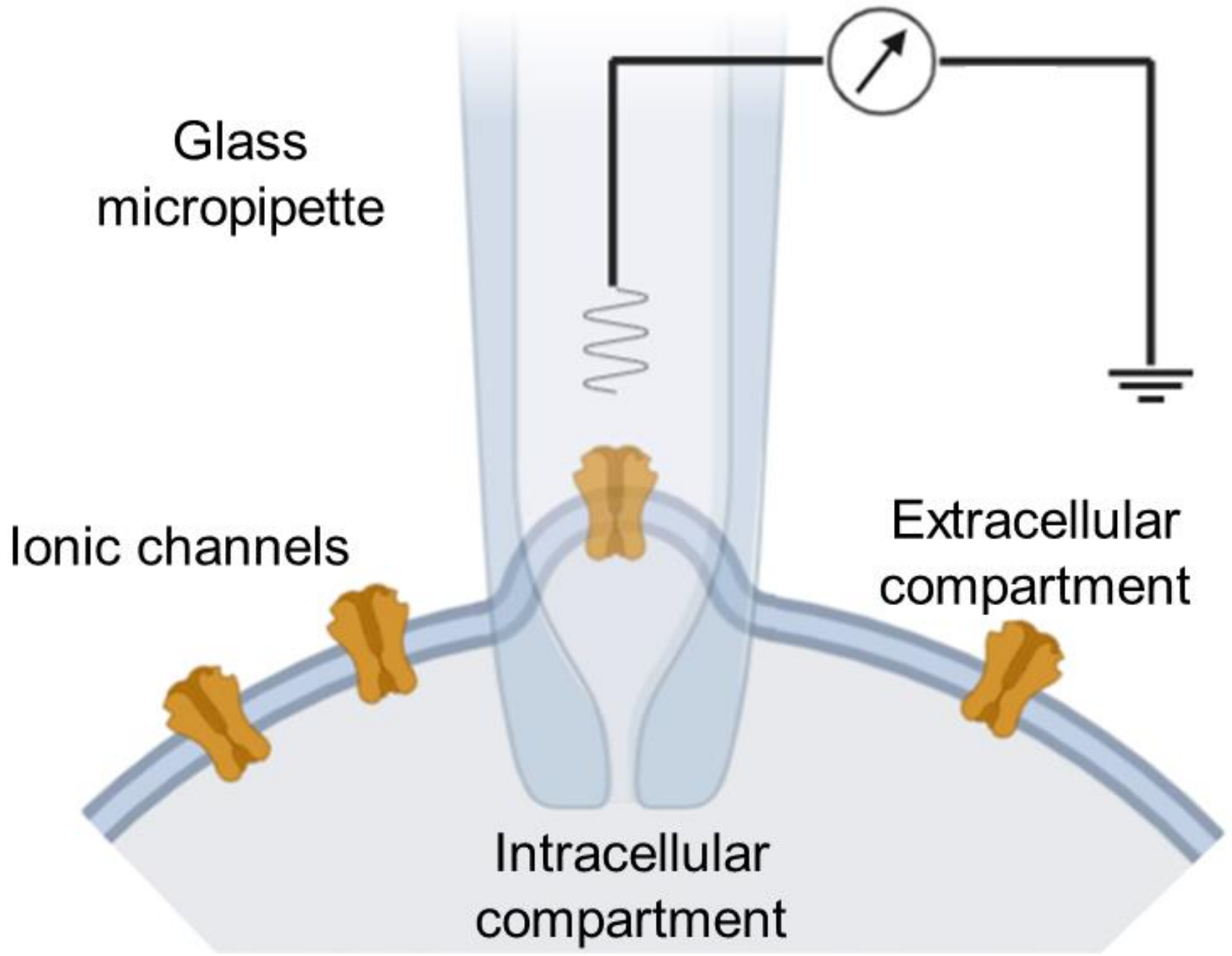

***Figure 6**: Sketch of patch-clamp configuration.*

This manual technique allows for a high degree of flexibility in recording configurations, including cell-attached patches and whole-cell patches.
The cell-attached method, or on-cell patch, is often used to study ligand-gated ion channels, channels that are modulated by metabotropic receptors, or neurotransmitters [50]. Compounds that will directly contact the external surface of the membrane are usually included in the pipette solution. This contact allows the concentration-response curves of the compound on ion channels can be accurately determined. The main disadvantages of this method are that only one compound concentration in a concentration-response curve can be measured per patch and the possibility of dialyzation of the cytosol after few minutes of recording.

The whole cell method records the currents of multiple channels at once through a voltage clamp and measures the membrane potential variation under the current clamp. The advantage of this method is that it allows better electrical access to the inside of a cell because the larger opening tip of the electrode provides lower resistance. The manual patch clamp has long been considered the gold standard for investigations of ion channel properties and compound activities because it usually generates high quality data, but the experimental procedures are complicated and time consuming. Therefore, high throughput screening for the compounds is very limited. To address this bottleneck, several automated patch clamp systems, capable to analyze hundreds of cells under variable experimental conditions, have been developed and introduced to the research market [51].

# Nanopillar Electrophysiology and Interaction with Cardiac Cells

In order to optimize the electrode-cell interface and capture the full spectrum of intracellular signals, the Rs should be increased (to prevent signals leakage) and Ra should be decreased (for maximize voltage transfer).

The larger is the cell-material surface, the better will be the signals recorded because Ra is inversely proportional to the surface area.

However, larger electrodes are not very useful due to the lack of single cell resolution. Therefore, the community started to push on the development of alternative solutions in order to increase the total surface of an electrode, without losing spatial resolution. The most exploited approach derived from nanotechnologies and nanostructures fabrication.

The development of lithography techniques capable of constructing nanoscale electrodes [52] and methods of controlled membrane poration has advanced the available technology to a new era in which intracellular potentials can be recorded using MEA architecture. The main advantage of these new platforms is the capability of recording both extracellular and intracellular signals during the same experiment. This approach could pave a way to several application that aim, for instance, to understand biological processes in the communication between single cells in network structures. In this field, the expansion in computer capacity and the miniaturization of electronics has paralleled the desire to measure the network activity of cells in order to create a new generation of devices that mimic the behavior of these cells.

The term “nanopillar” encompass the wide variety of 3D nano geometries and sizes, from sub-100 nm needle-like structures, to 1 µm mushroom-like structures. In general, these electrodes are out-of-plane vertical structures with features that have dimensions lower than mammalian cells. The nanopillar electrode technology and its various iterations share the common objective to record intracellular action potentials.

For intracellular nanoelectrode applications, pillar geometries in 3D configuration have been the most widely used. In fact, in comparison to traditional planar electrodes, vertical nanopillars promote good sealing and a smaller cleft in the nearby of the cells, resulting in higher Rs [39].
The ability of cells to tight well on 3D vertical nanostructures has been investigated in detail, since they are good candidates to improve and develop novel intracellular recording technique. Many prominent groups in the world that actively work on this field, presented methodical investigations of the cell-3D nanostructure interface using focused ion beam (FIB) , scanning electron microscopy (SEM) and fluorescence microscopy (Figure **7**) [53–56].
They all worked with the aim of characterizing the engulfment-like event of 3D nanopillars by cells and their membrane deformation after the interaction with the structures. In particular, so far it has been shown that cells engulf 3D nanostructures in the center of the cellular body, deforming their membranes in a process guided and stabilized by an actin network. The FIB-SEM and fluorescence co-localization studies reveal huge differences in how cell membranes behave to different substrate nanotopologies.
These studies demonstrated that the cell membrane deforms readily and wraps conformably around the surface topology of substrates with protruding structures generating a cleft of about 15nm.
In contrast, for substrates with invaginations, the membrane hardly deforms and does not contour the surface of the pore, generating a cleft of about 400nm) [53,57].
Several studies indicate that a cell can respond by a highly preserved cell biological mechanism, endocytosis, which is a mechanism that underlies the internalization of particles into the cells. In this case the cell completely engulfs protruding mushrooms shaped electrodes, increasing electrical coupling and enabling intracellular recordings [58–60].

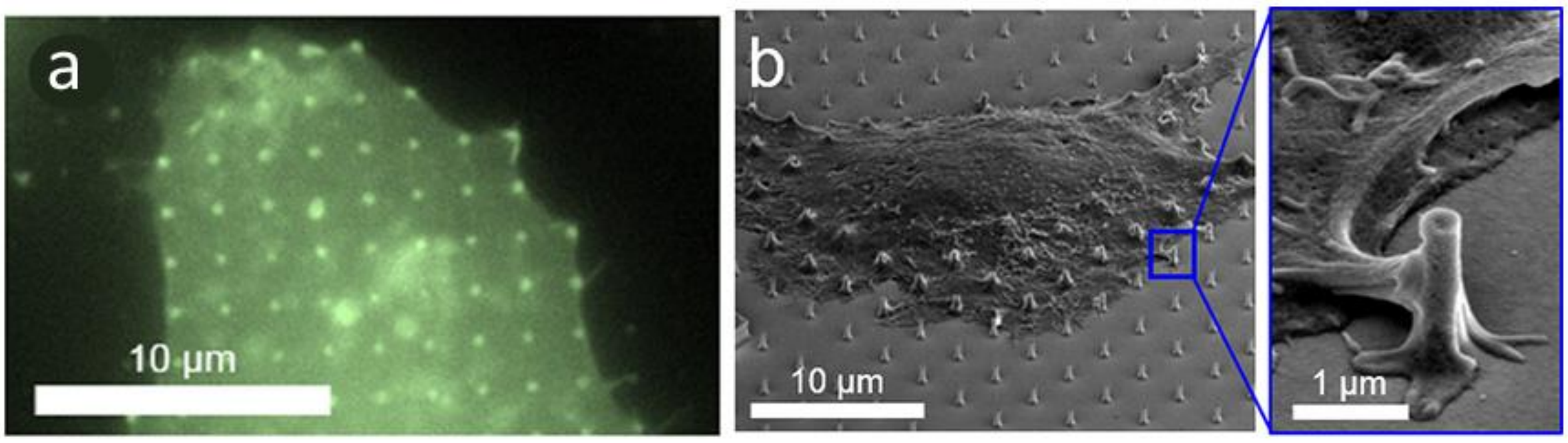

***Figure 7**: Imaging of the cell-material interface by fluorescence and FIB/SEM microscopy. Adapted with permission from (Santoro et al., 2017). Copyright (2019) American Chemical Society.*

# Methods for Cardiomyocytes Intracellular Access and Recording

When a cardiac cell, and in general a cell, is plated on a nanoscale electrode protruding from the surface, the membrane responds, deforming itself in order to wrap the electrode, excluding it from the interior of the cell. Many works have observed this membrane wrapping and many techniques have been exploited to study membrane deformations  and the process of gradual settling of live cells around vertical nanopillars [61]. In order to gain intracellular access, the electrode needs to bypass the cellular membrane and come into direct contact with the cytosol. In recent years, different approach have been analyzed (Table 1), from spontaneous to induced perforation of the membrane. In the following sections, I will discuss more in detail these topics, focusing on the different technique used to gain intracellular access of cardiomyocytes.

| Reference | Electrode Geometry | Access Method | Cells | Experiments |
|---|---|---|---|---|
| [62,63] | Kinked nanowire | Mechanical and lipid coating | Chicken cardiomyocytes | Recording |
| [9] | Vertical nanowire | Electroporation | Rat cardiomyocytes | Recording and stimulation |
| [64] | Silicon nanotube | Mechanical or spontaneous | Cardiomyocytes (embryonic chicken) | Recording |
| [10,65] | Vertical nanotube and porous metaelectrodes | Optoporation | HL-1 and rat cardiomyocyte sand hiPSCs-CM | Recording |
| (Dua*n et al*, 2012) | Vertical nanotube | Lipid coating | Chicken cardiomyocytes | Recording |
| [67] | Vertical nanotube | Electroporation | HL-1 | Recording |
| [68] | Nanovolcano | Spontanous | Rat cardiomyocytes | Recording |
| (Xi*e et al*, 2012) | Vertical nanowire | Electroporation | HL-1 | Recording |

***Table 1**: Summary of state-of-the-art approaches for intracellular electrophysiology on cardiomyocytes.*

# Biomimetic Approaches

A biomimetic approach aims to gain and maintain physical access to the cytosol with continuous flow of high signal-to-noise ratio potential signals. In general, a cell maintains homeostasis exploiting the cell membrane, which is a functional compartment enriched of protein channels and complexes that detect and bind molecules on a chemical basis. Exploiting the chemical feature of this boundary, scientists have had some successful chemical functionalization that dupe the cell, allowing nanostructures to penetrate the cell membrane for prolonged periods [62,66].
The group guided by C. Lieber for instance, fabricated free-standing probes made of a kinked silicon nanowire with an embedded field-effect transistor detector at the tip end. These probes can target specific cardiomyocytes or groups of cells monitoring real-time change in action potential oscillations. In order to promote a kind of biomimetic fusion between the probe and the membrane of the cells, the probe has been functionalized with a layer of phospholipids and gently placed in contact with the cell membrane. The recorded signals shown by the authors are high quality traces, without perturbing the electrical activity of the cells. The results are therefore consistent with biomimetic membrane fusion with the nanostructure [63,69].

Ideally, the interaction between a probe and the lipid bilayer could be also achieved by modifying the material's surface characteristics. The ability to specifically insert proteins into the bilayer core and form a strong interface, mimicking endogenous transmembrane complexes, could generate an "artificial membrane proteins". Melosh's laboratories reported an example of this approach. They presented a simple micro-fabricated architecture based on metallic multilayer probes that allows probe fusion into a lipid bilayer core and systematic control of lateral bilayer-material interactions. These "stealth" probes are designed to mimic two essential transmembrane protein characteristics: first, the transmembrane regions are mostly hydrophobic, with hydrophilic groups on either side, and second, the hydrophobic domain and the bilayer thicknesses should be proportional. Therefore, the probe replicates the nanometer-scale hydrophilic-hydrophobic-hydrophilic architecture of transmembrane proteins, allowing for a spontaneous insertion and anchoring within the lipid bilayer, forming a high-strength interface [70].

## Spontaneous Access

The hypothesis of spontaneous rupture of the cellular membrane in contact with 3D nanostructures is very fascinating. If this event could be controlled, its nature (repeatability and stability of the rupture) could determine the future design of nanostructured devices. A stable spontaneous rupture, in fact, could enable unprecedented advantages for measurements of intracellular phenomena. An unstable rupture, in contrast, would lead to reduced control of the device in switching from extracellular to intracellular recordings (i.e., in electrophysiology).

One of the first demonstrations regarding the idea of spontaneous access to the cytosol was published in 2007. In this work, Kim W. and colleagues presented a platform made of silicon nanowires with diameters of 90nm and length of few microns, therefore few orders of magnitude smaller in diameter than mammalian cells (a cell is on the order of 15 µm). The cells were cultured on this platform and the penetration of the SiNW array into individual cells was observed during the cell incubation. The cells survived up to several days on the nanowire substrates. The longevity of the cells was highly dependent on the diameter of SiNWs. Furthermore, successful maintenance of cardiac myocytes derived from embryonic stem cells on the wire array substrates was observed, and gene delivery using the SiNW array was demonstrated. In this approach, no external force was necessary for the penetration, owing to the small diameter and high aspect ratio of nanowires [71].

Recently, Desbiolles and collagues reported a novel microtechnology-based array made of nanovolcano structure electrodes for multisite intracellular electrophysiology. Proof of principle experiments demonstrate that the nanovolcano provide spontaneous intracellular access without application of chemical, physical or mechanical triggers [68].

Nevertheless, the idea of a spontaneous nanostructure-induced rupture of the cell membrane is an approach still controversial and debated [72].

Very recently, a work from Dipalo and colleagues (Dipalo, McGuire, *et al*, 2018) deeply investigated cellular membrane deformations on top of nanostructures with different geometries and dimension without detecting spontaneous penetration in any culture development stage. Only rare events have been observed with sharp nanopillars (diameter of 80nm) after 4 day in culture. However spontaneous rupture has transient dynamics and features with a full recovery of the membrane integrity after 3 day in culture.

# Electroporation

As depicted in Figure 2 the junctional membrane is defined as the resistance and capacitance of a membrane patch that faces the sensing electrode. The junctional membrane can be of very high resistance and low capacitance, because of the electrode geometry and the adhesion quality of the cells. This implies that only a few amount of the current generated across the cell's membrane flows through the junctional membrane. Reduction of the junctional membrane resistance would be very effective in improving the electrical coupling coefficient between a cell and an electrode.
Therefore, electroporation [74] is exploited to repeatedly gain temporary low impedance electrical access to the cytoplasm.
In bulk electroporation, which is employed in transfection applications, cells are suspended between two parallel electrodes and a large voltage (hundreds to thousands of volts) is applied to perforate the membrane.
Electroporation is wrongly termed both as electroporation or electropermeabilization. In fact, the former term refers only to the contribution to the increased permeability of the membrane owing to the formation of aqueous pores in its lipid bilayer, while the latter is more general and ascribes this increase to a broader range of (bio)physical and (bio)chemical mechanisms [75].

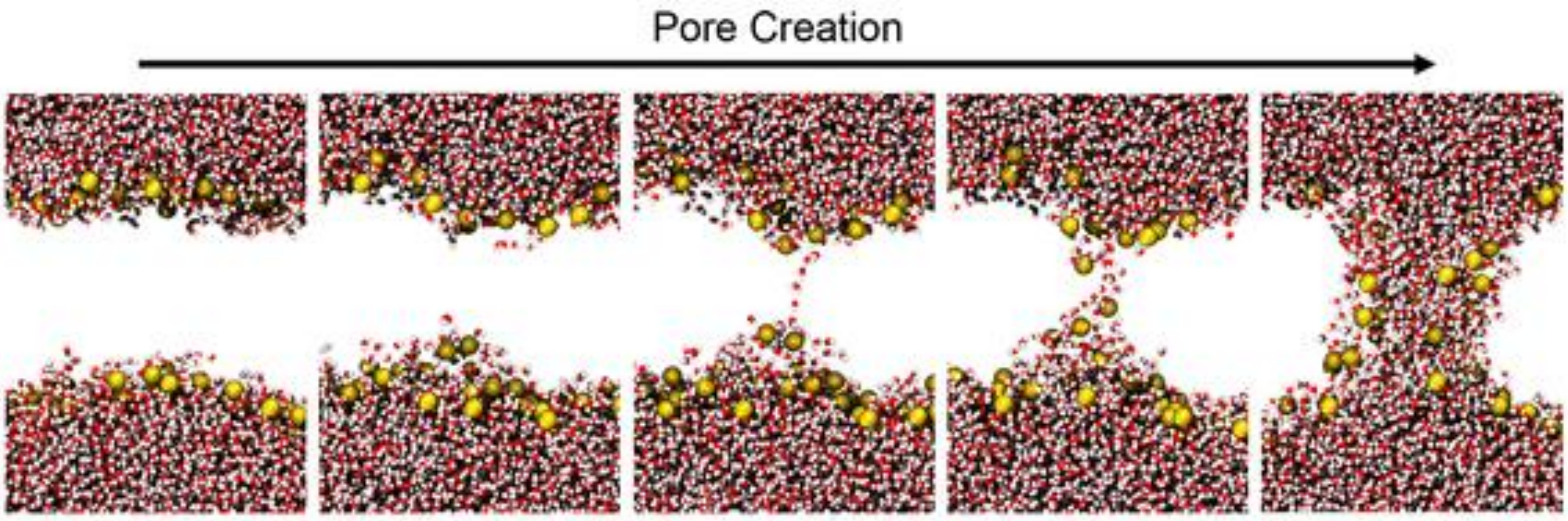

***Figure 8:*** *Phospholipid electro-induced pore life cycle. Structurally distinct steps in pore creation as observed in molecular dynamics simulations. Adapted from* [75].

Broadly speaking, an exposure to sufficiently strong external electric field can induce an increase in transmembrane voltage, far exceeding their resting range and causing structural changes to the membrane and changes to its constituent molecules [76]. From a theoretical point of view, the electropermeabilization is not a threshold event, in the sense that the process does not occur only if the increase in the field amplitude reaches a certain value. On the contrary, it depends on the cell type, exposure duration, and is influenced also by cell size and local membrane curvature [77,78]. Figure 8 shows that the pore generation process can be divided into five stages: the initiation of the permeable state, its expansion, stabilization with partial recovery, the resealing of the membrane, and finally gradual cessation of what are referred to as residual memory effects reflected in cells' altered physiological processes.
The application of an external electric field modifies the transmembrane voltage that locally generates an electric field that promotes water molecule accumulation. Such event generates the insertion of water molecule into the hydrophobic core of the bilayer (by intermolecular hydrogen bonds), allowing for the formation of a hydrophobic pore across the bilayer. Then, the lipids adjacent to the water molecule inside the pore start reorienting with their polar head-group toward these water molecules, stabilizing the pore into its hydrophilic state and allowing more water and other polar molecules, to enter. Once the electric field ceases, pore closure follows a reverse sequence of events [79].
The combination of large local electric field gradients and intimate cell membrane-electrode contact enables the use of small potentials to achieve poration and, thereby, minimize heat-induced cytotoxicity [80,81].
Cui's group, in 2012, introduced a novel nanoscale system based on vertical nanopillar electrodes, which can not only record both the extracellular and intracellular action potentials of cultured cell with excellent signal strength and quality, but also repeatedly switch between extracellular and intracellular recording by nanoscale electroporation and resealing processes (Figure 9**Figure 9**). The nanopillars can form tight junctions with mammalian cell membrane, promoting the active engulfment by means of protrusions from the cells. This suggests that tight engulfment results in good sealing at the interface. A transient electroporation drastically improves the quality of the nanopillar electrode-recorded signals by lowering the impedance between the electrode and the cell interior.

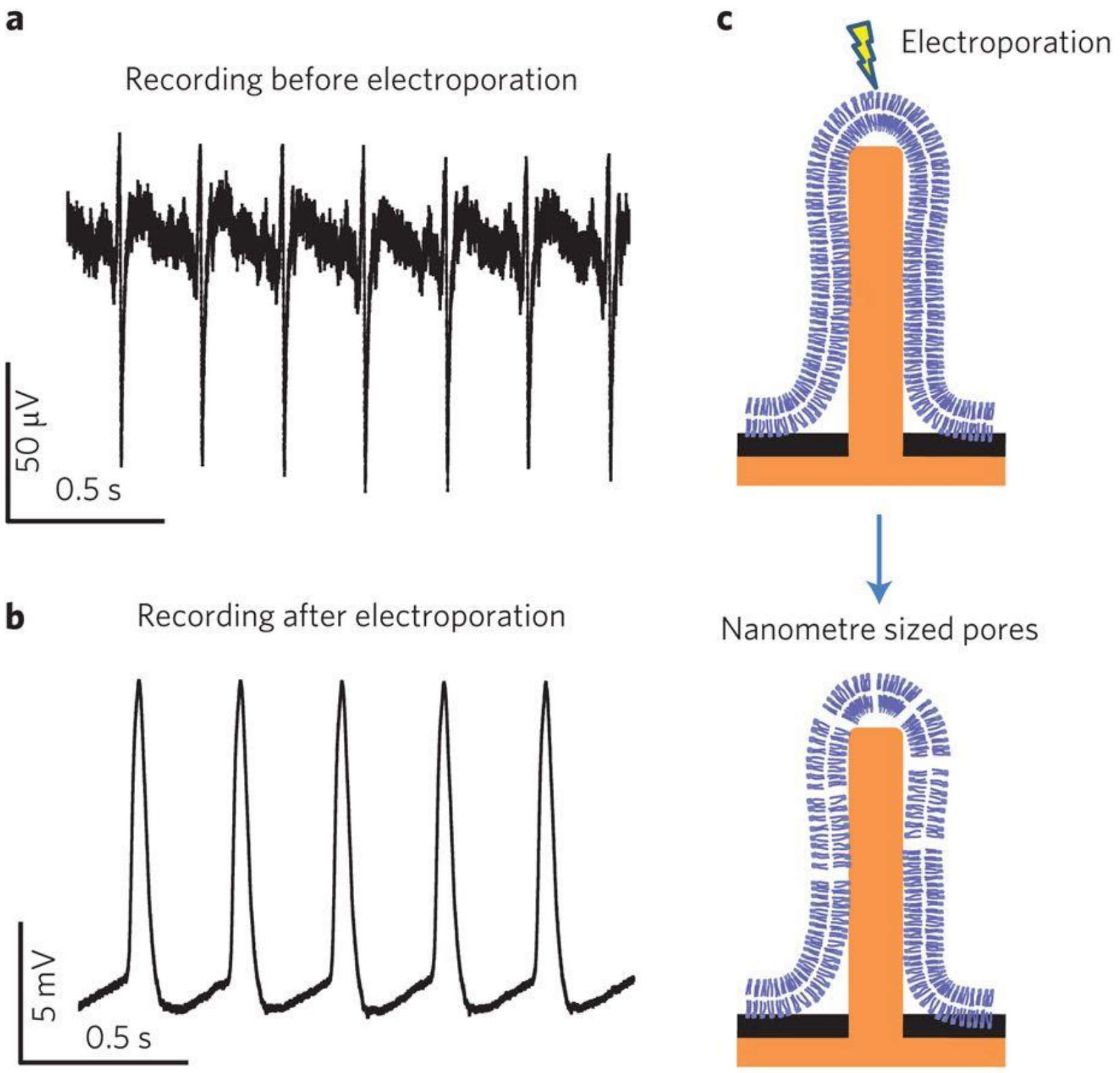

***Figure 9:*** *Recording action potentials of a single cardiac cell before and after electroporation. Reprinted by permission from [7].Copyright (2019) Springer Nature.*

A high electric field can induce nanometer-sized pores in the cell membrane, as in the established *in vitro* technique that used electroporation to introduce genetic material into cells. Due to the sharp shape of the nanoelectrodes and their strong coupling to the membrane, they can create a large electric field around the tip, with a small voltage (∼1−3 V) delivered in pulse trains of one or few seconds, to transiently and locally increase the permeability of the cell membranes (Xi*e et al*, 2012).

Lin et al., in 2014 used cardiomyocytes that actively engulfed vertical hollowed iridium oxide nanotubes, also protruding into the hollow center. The average diameter of the structure was of 181nm, with a height of about 500nm. Electrochemical studies revealed lower impedance than gold nanopillar and a higher charge storage capacity.

They show that this geometry increases the electrical coupling resulting in measuring much larger intracellular action potentials (up to 15mV). The nanotube electrodes afford much longer intracellular access and are minimally invasive, making it possible to achieve stable recording up to one hour in a single session and for few days of consecutive daily recordings. This is possible because of tight cell-nanotube interaction confirmed from co-localization staining of cytoskeleton protein that show actin filaments also in the hollow center of the nanotubes. A positive curvature of the membrane therefore promotes longer intracellular access duration [67].

Despite several successes, the process has yet to be characterized in detail, and whether nanoelectrode-based electroporation leads to a single pore or a group of pores is not very well established. The effective diameter of the pore has been estimated to be 10−20 nm for ∼180 nm diameter nanotubes [67] and 500−700 nm for ∼1.5 µm diameter mushrooms [58].

Very recently, De Angelis's group exploited the plasmonic properties of gold nanopillar to further reduce the voltage used in a "soft-electroporation" approach, avoiding collateral effects of electrical pulses on cell culture (Caprettin*i et al*, 2017).
Therefore, during nanoelectrode-based electroporation, a significant increase in signal amplitude is observed with respect to extracellular signals, right after electroporation pulses, indicating reduction of the access resistance for intracellular measurement. This access is transient (a few minutes with vertical nanopillars and up to 1 h with IrO2 nanotubes): as the cell membrane recovers, the perforations in the membrane become resealed, expelling the nanoelectrodes and causing the signal amplitude to decrease.
This method takes practical advantages of the presence of stimulating electronics that often accompany the recording system, but has the drawback of a recording blind period immediately after electroporation as the amplifiers recover from the influx of charges. Furthermore, electroporation may perturb spontaneous activity of the cells.
In the following section, optoporation will be described in details, considering all the advantages of this technique respect to electroporation.

# Optoporation

Laser-mediated plasma membrane poration is an alternative to the classical methods of intracellular access, receiving in the last years much attention. A variety of laser systems have been used for optical poration, including continuos-wave, pulsed picosecond ($10^{-12}$ s) and nanosecond ($10^{-9}$ s) lasers [82]. Of particular note are ultrashort pulsed lasers, the most widely used systems for optical poration [83]. In the femtosecond regime ($10^{-15}$ s), poration of the cell membrane is mediated by a process of multiphoton absorption by the membrane, which leads to the generation of free electrons [84]. Using femtosecond laser sources with high-repetition-rate (> 1 MHz), a low-density electron plasma is generated, which causes a photo-chemically induced degradation of the membrane. Notably, with this class of lasers, membrane poration is achieved at pulse energies that are lower than the optical breakdown threshold in water, defined as the threshold pulse energy to generate a cavitation bubble [84]. At energies exceeding this threshold, gas bubbles are formed. In contrast, poration with lower repetition rate (order of kHz) is performed at pulse energies at least one order of magnitude higher than with sources with MHz-scale repetition rates, with the drawback of the generation of small transient bubbles on the membrane.
To date, this laser technology has been used to porate different kinds of cells and allow for delivery of various molecules. Among them, the delivery of DNA, genetic material [85–89] and in general sub-100nm materials [90] has been long discussed and demonstrated. In the following section, I will discuss about the mechanisms of laser-induced optoporation.

## Direct Laser-induced Optoporation

In direct laser-induced optoporation, it is crucial to confine the laser energy on the cell membrane in order to obtain sufficient density for the generation of membrane pores.

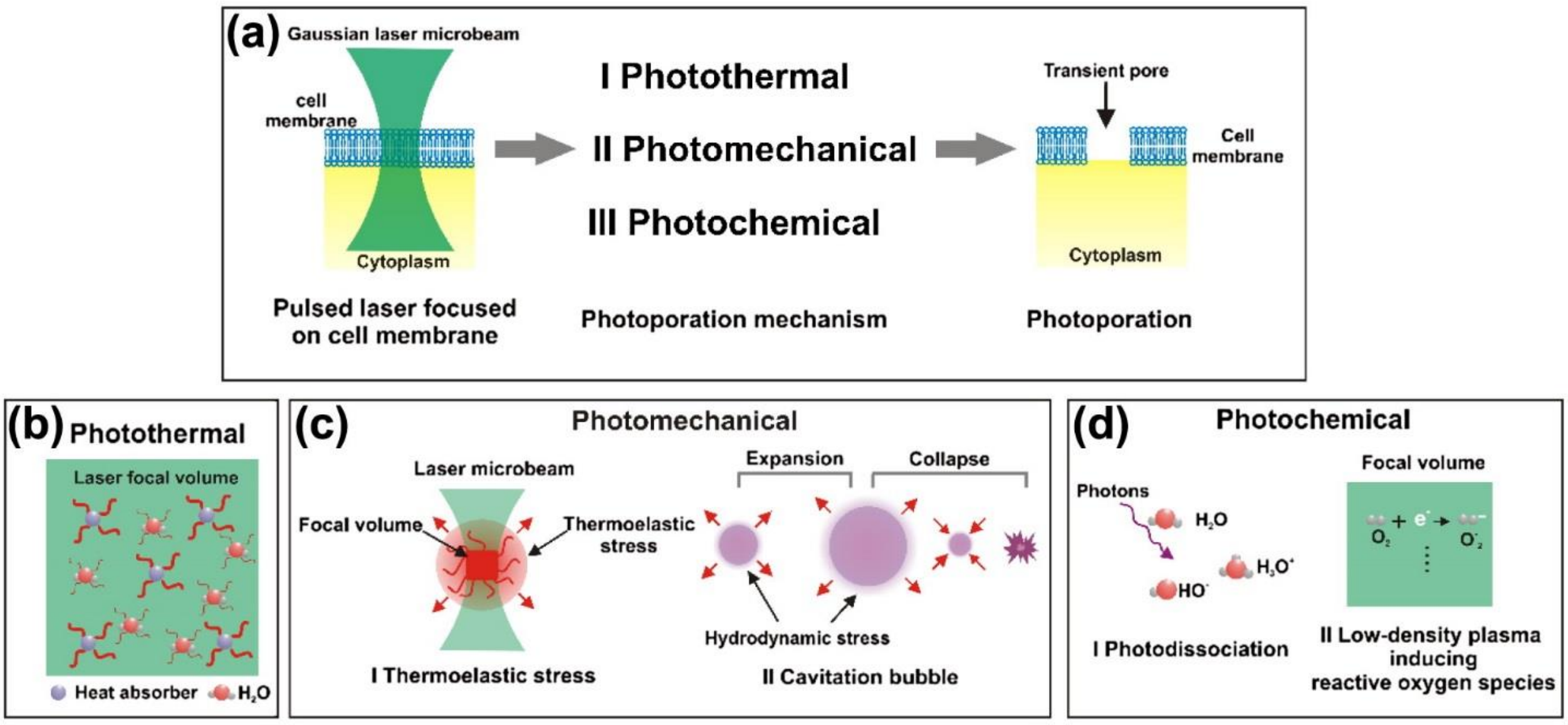

***Figure 10**: Main mechanisms contributing to direct laser-induced optoporation.* [91].

To do that, a laser beam is focalized in a very small spot, with a size typically of about 1–10 μm through a microscope's objective lens. As schematically shown in Figure 10, a variety of possible mechanisms may contribute to pore formation, including photo-thermal, photomechanical, and photochemical processes [84,92]. The contribution of each of these processes is related to pulse duration, wavelength and intensity (determined by laser energy and beam size).

The photothermal effect is related to a temperature increase following single photon absorption by molecules like water, proteins or DNA. Electronic transitions are allowed in molecules that absorb in the UV and visible range, while infrared regime is associated with vibrational transitions. Non-radiative relaxation to the ground state results in heat production that, locally, can increase the permeability of the cell membrane. This phenomenon is associated with a local phase transition of the lipid bilayer or by thermal denaturation of integral proteins [93]. However, it was reported that photothermal heating by single photon absorption alone is not sufficient to effectively form pores in cell membranes. This is mainly because water, lipids, and proteins have a relatively low absorption in the 350–1100 nm wavelength range [94].

Pores can also be generated by mechanical stress induced by acoustic waves or by phenomena such as cavitation bubbles. These bubbles are created when femtosecond laser pulses cause plasma formation following a multi-photons absorption process. Free electrons generated in the plasma thermalize triggering phenomena of pressure propagation that generate bubbles. The expanding bubble can lead to perforation of the cell membrane by hydrodynamic stress. Furthermore, when the bubble has expanded to its maximum size, the bubble collapses by the surrounding hydrostatic pressure, inducing liquid jets or shockwaves that can form pores in the cell membrane.

Photochemical reactions may also contribute to optoporation of cell membranes. When femtosecond laser pulses are used below the threshold for optical breakdown or bubble formation, reactive free electrons can be generated by multiphoton ionization of, for example, water molecules. This process results in highly reactive oxygen species (ROS) generation that can locally induce cell membrane damage [95].

Following optoporation, cells reseal the membrane pores in a matter of tens of seconds to a few minutes, depending on the pore size. The repair mechanism is based on Ca2+ influx that induces exocytosis of lysosomes for ‘patching’ of the pores and takes from seconds to minutes.
In order to enhance the performances of optoporation, very recently has been shown that nanomaterial can be used to improve the capability of optoporation of cell membranes at lower laser energies, as detailed in the following section.

## Optoporation Mediated by Plasmonic Nanopillar

Very recently, the exploitation of 3D plasmonic nanoelectrodes has allowed for the development of a particular kind of photoporation mediated by nanostructures. It promotes the opening of transient nanopores into the cell membrane without compromising the seal between the cell membrane and the nanoelectrode and with no side effects; this process is called plasmonic optoacoustic poration [96,97].
When plasmonic nanostructures in aqueous environment are excited with visible/near infrared laser pulses (fs or ps duration), they can emit electrons in free space. This process can be resumed in three steps: electromagnetic absorption into the plasmonic structure, electron ejection by multiphoton or field emission and ponderomotive acceleration by the plasmon-enhanced electromagnetic field.
For example, if vertical gold nanotubes in acqueous environement are shined by a pulsed laser source, it excites strong and confined electric fields at the nanotube ends by means of plasmonic enhancement. These excited surface plasmons rapidly decay in highly energetic electrons, called “hot electrons”. These electrons are not in thermal equilibrium with their environment and are known because they have unique features and are very effective in being injected into semiconductor materials.
Since the aqueous medium around the nanotubes can be approximated to a semiconductor, in the timescale of picoseconds, free-electron density can be generated by exciting electrons from valence to the conduction band of the water. This process requires more energy than the one to remove electrons from the gold nanostructure, therefore the plasmonic hot electrons emission is favored. A free electron promoted to the conduction band of water during a laser excitation, can absorb a photon and increase in kinetic energy when collide with heavy charged particles, like ions or nuclei. Then, a cascade collisions event take place, where these electrons with high kinetic energy excite other electrons and so on. The result of this process is that water molecule increase their kinetic momentum and are accelerated, thus generating nanoscale shockwaves that locally open transient nanopores into cell membrane [11].

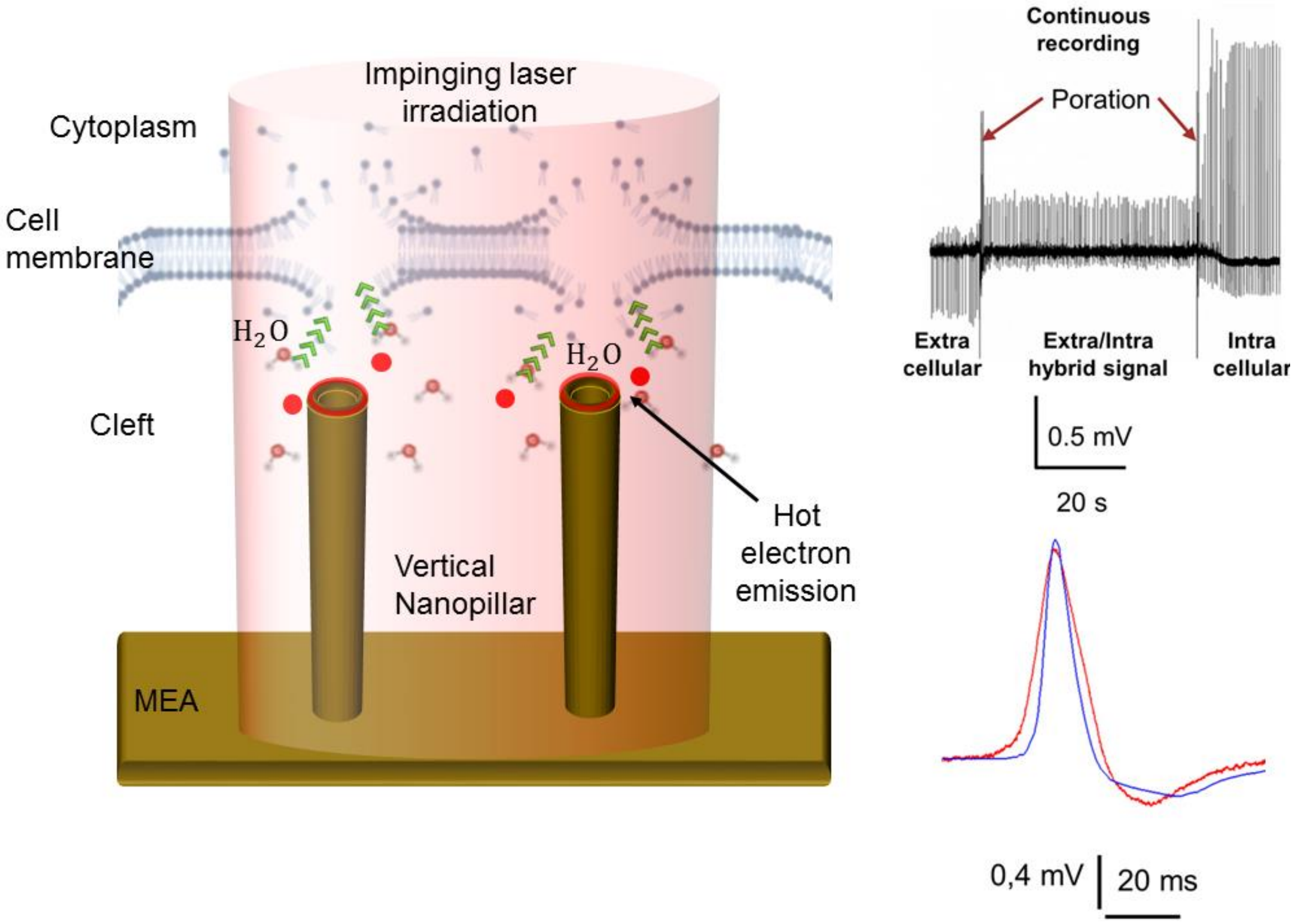

***Figure 11****: Sketch of optoacoustic poration of cell membrane with examples of intracellular recording.*

This process, coupled with other improvements, can be used to develop novel technologies and for acquiring intra- and extra-cellular information about mammalian neurons and cardiomyocytes.
In this context, Dipalo and colleagues recently introduced an approach of plasmonic optoporation for long term and stable continuous recordings of both intracellular and extracellular electrical activity in cultures of primary mammalian neurons and of cardiac-derived cells (Figure 11). These results have been achieved exploiting the combination of vertical nanoelectrodes with plasmonic optoporation. The former promotes a tight seal with the cell membrane that is essential for a high SNR. The latter is an extremely local process that enables a gentle and well-controlled insertion of the nanopillars in the cell membrane with no effects on the overall sealing [10].
Since plasmonic optoporation is completely decoupled from electrical processes, a first advantage in respect electroporation is the lack of recording blind time after the poration. In fact, after electroporation, a blind-time window of tens of seconds is typically present, due to a charge that accumulates on the MEA electrodes during the electrical pulses.
The comparison between electroporation and optoporation on vertical nanostructure has been deeply investigate recently (Dipalo *et al.*, 2019). They provided new details about membrane dynamics in case of these two different poration approaches both mediated by the same 3D nanostructures. The authors shed light on the poration approaches, correlating them to change in electrophysiological recordings and morphological imaging. They show that there are differences in terms of localization, efficiency and membrane reformation dynamic.

Despite this technology could pave the way for a novel platform more promising for electrophysiological recordings, it still has limitations that will be discussed in the following section.

# MEA Platforms for Electrophysiology

In the following sections, I will describe in detail structure, materials and size of the most important commercial devices used in electrophysiology for the study of electrogenic cells. I will start from commercial passive MEA, describing the single well and the multiwall configuration, up to the CMOS-MEA technology.

## Passive Planar MEA in Single Well Configuration

Conventional MEAs, as those depicted in Figure 12, have fixed wiring and are passive (i.e., no active circuit elements, such as amplifiers). Each electrode connects directly to a signal pad outside the array through a conductive feedlines. The pads are then connected to external equipment for signal acquisition and conditioning. Passive MEAs are easy to fabricate and many different substrates and electrode materials can be used. In order to record the extracellular signals, MEAs need a reference electrode in respect to which evaluating the potential. Some arrays, presenting an internal reference electrode, do not need to open the lid, protecting the culture (differently from what happens by using the external reference), which exposes the cell culture more easily to contaminations. For this reason, internal references are usually used during experiments on developmental studies in order to maintain the condition of the culture as healthy as possible during the whole developmental stage.

A glass/plastic ring is placed at the center of the array, and it allows to contain the culture medium; when placed in an incubator, therefore, the culture can survive for several weeks.

The current technology typically provides glass-embedded MEAs, consisting of 60–256 electrodes, 10–30 μm in diameter and spaced at 100–500 μm. MEAs consist of micro-fabricated electrodes embedded in a biocompatible insulation substrate (e.g., polyamide or silicon nitride/oxide) which prevents short circuits with the electrolyte bath, forming a sort of wired Petri dish. The electrodes, typically made of Au, Indium-Tin Oxide (ITO), Titanium Nitride (TiN), or black platinum, must be biocompatible, long-lasting, and should have low impedance. Individual sensing electrodes must be made small enough to match the size of the individual cell (good spatial resolution).

However, reduction of the surface area that accompanies the reduction of the sensing area size results in a significant increase in the impedance and consequently decreases the signal-to-noise ratio.

The challenge of manufacturing very small electrodes and, at the same time, keeping the impedance and the noise level down has been reached with fabrication processes based on the thin-film technology realized in clean-room facilities. Photolithography, i.e., the process of transferring geometric shapes from a mask to the surface of a silicon wafer, is used to make MEAs [98].

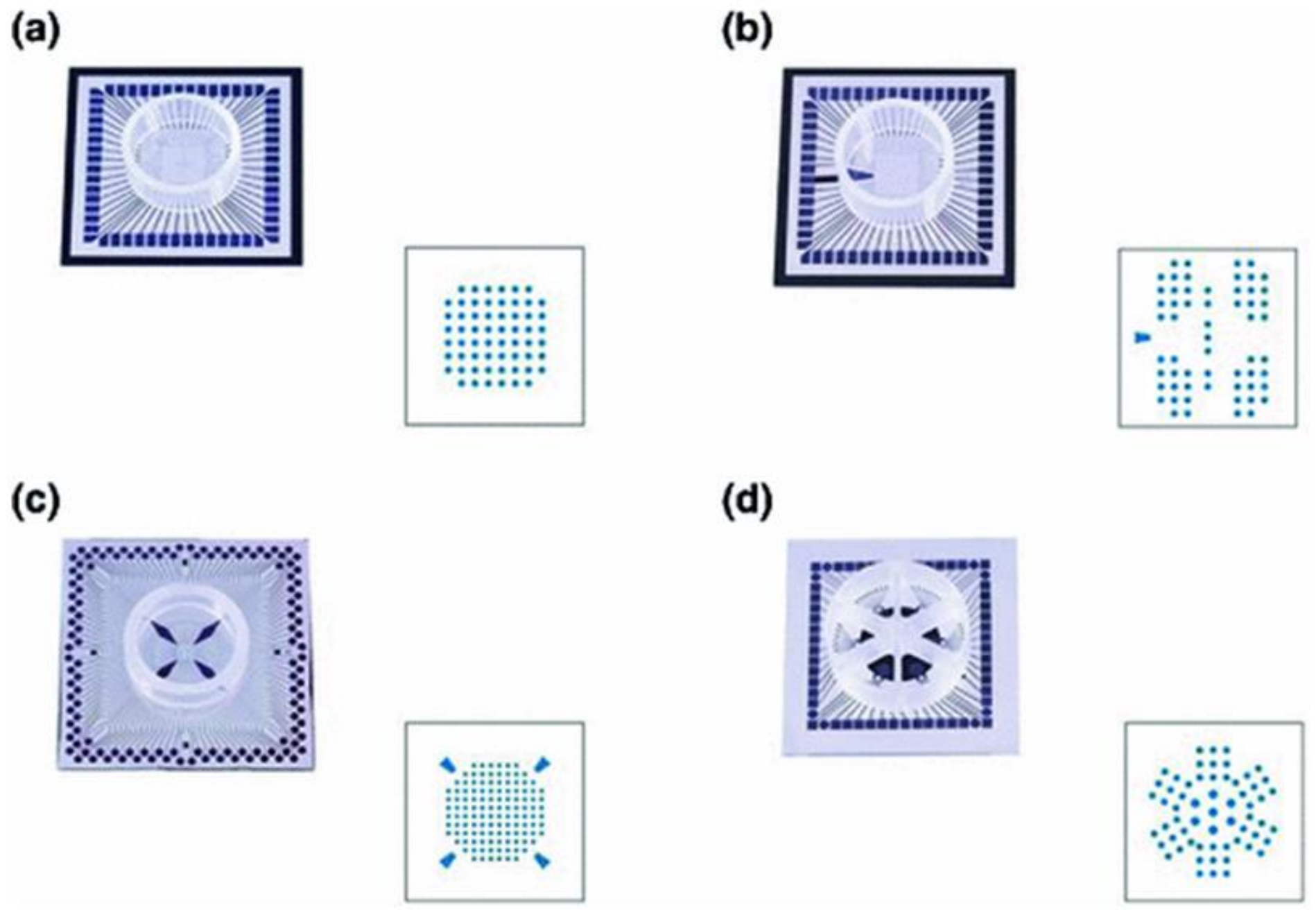

***Figure 12****: MEA devices and electrodes layouts.* ***a*** *Standard 60MEA.* ***b*** *60-4QMEA.* ***c*** *120MEA.* ***d*** *60-6WellsMEA.*

Cell lines or primary cell preparations are cultured directly on the MEA. Adhesion factors, such as protein coatings, promote the adhesion and proliferation of the cells on the surface of the device.
The rapid success met by MEAs in the neuroscience and cardiology research fields moved some electronic companies to develop commercial systems to perform electrophysiological measurements using MEA.
At present, there are on the market at least two complete acquisition systems based on MEAs: the MED System developed originally by Panasonic, developed and manufactured by Alpha MED scientific (www.med64.com, Osaka, Japan), and the MEA System by Multi Channel Systems (www.multichannelsystems.com, Reutlingen, Germany).
Despite these features, MEA in single well configuration have limitations in term of parallelization of the experiments in research labs where high-throughput applications are required. Therefore, some companies developed MEA in multiwell configuration.

# MEA in Multiwell Configuration

Axion Biosystems and MultiChannel Systems actually represent the leader companies in the production of MEAs for *in vitro* high-throughput applications (Figure 13**Figure 13**). These platforms consists of high-throughput configurations with 12 to 96 wells. Each well may have 12 or more low noise microelectrodes with few integrated ground electrodes. Each electrode, fabricated on transparent glass or opaque epoxy resin substrates, may consist of nanoporous Platinum or Gold, PEDOT or nanoporous titanium nitride. The arrangement of these electrodes into a grid extends the recording range across an area of few squared millimeters, providing concurrent access to both single-cell and network-level activity. These MEA are ideally suited for the investigation of electroactive cells on traditional plate readers and automated instrumentation, making them highly suitable for applications in pharmacological research and drug screening and development processes, where a high level of parallelization is required.

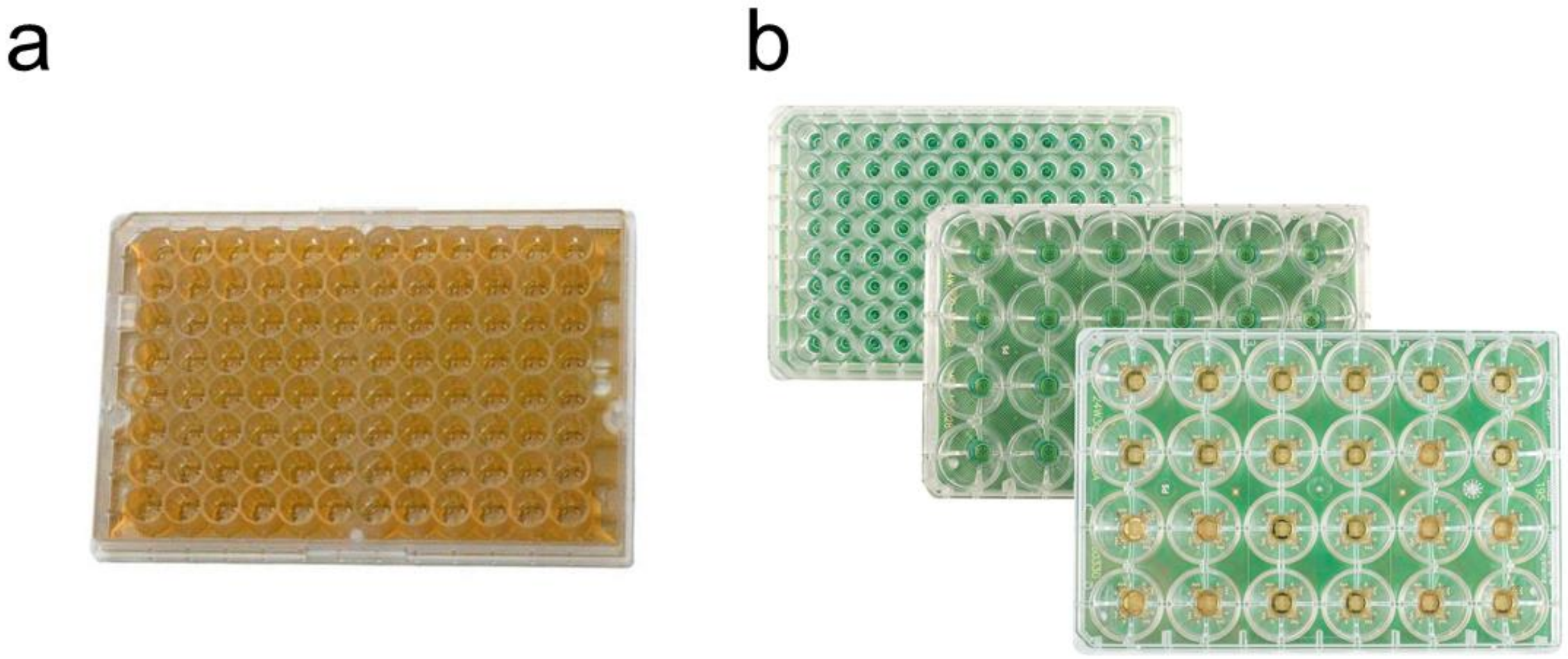

***Figure 13****: Examples of MEA in multiwall configuration. a Axion Biosystem. b Multichannel Systems.*

## Complementary Metal Oxide Semiconductor-MEA

CMOS is a powerful tool for biosensor improvement, equipped with a huge amount of transistor on-chip, which provide powerful signal processing. CMOS technology offers new possibilities for the realization of faster and more compact integrated circuits. This emerging technology enables the integration of multiple layers with vertical interconnections, providing potential performance, functionality improvements, and the possibility of technology scaling [99,100].

The term CMOS describes the physical structure of the devices: generally it has a top electrode (metal) placed on top of an oxide insulator, which is positioned on a semiconductor material [101].

Even though the increasing impact and relevant contributions aimed at upgrading MEA-based devices, the potentialities of this approach to map cell's signaling in large-scale networks at spatial resolution down to cellular resolutions are still not available. This improvement requires innovative approaches to integrate large and dense microelectrode arrays and to read-out signals at sufficient temporal and spatial resolution.

Indeed, currently available devices based on thin-film technology are limited both in spatial resolution (typically 100 µm) and in the number of integrated microelectrodes (range of hundreds) due to the external wiring, and to the amplification and signal conditioning circuitry.

With current technology, high resolution measurements are only possible from small active areas [102].

Berdondini's group introduced for the first time the concept of CMOS-MEA based on active pixels technology. The device was able to record from 4096 microelectrodes with high sampling rate, suitable for neuronal culture [103].

There are three primary advantages to using this technology. (1) Connectivity: on-chip multiplexing means that many electrodes can be addressed, allowing for measurements at high spatiotemporal resolution. (2) Signal quality: signal processing circuitry can be located directly below the electrode, facilitating the detection of weak signals through immediate signal conditioning and digitization. (3) Ease of use: many functions can be implemented via user-friendly software that communicates with on-chip logic units through a digital interface.

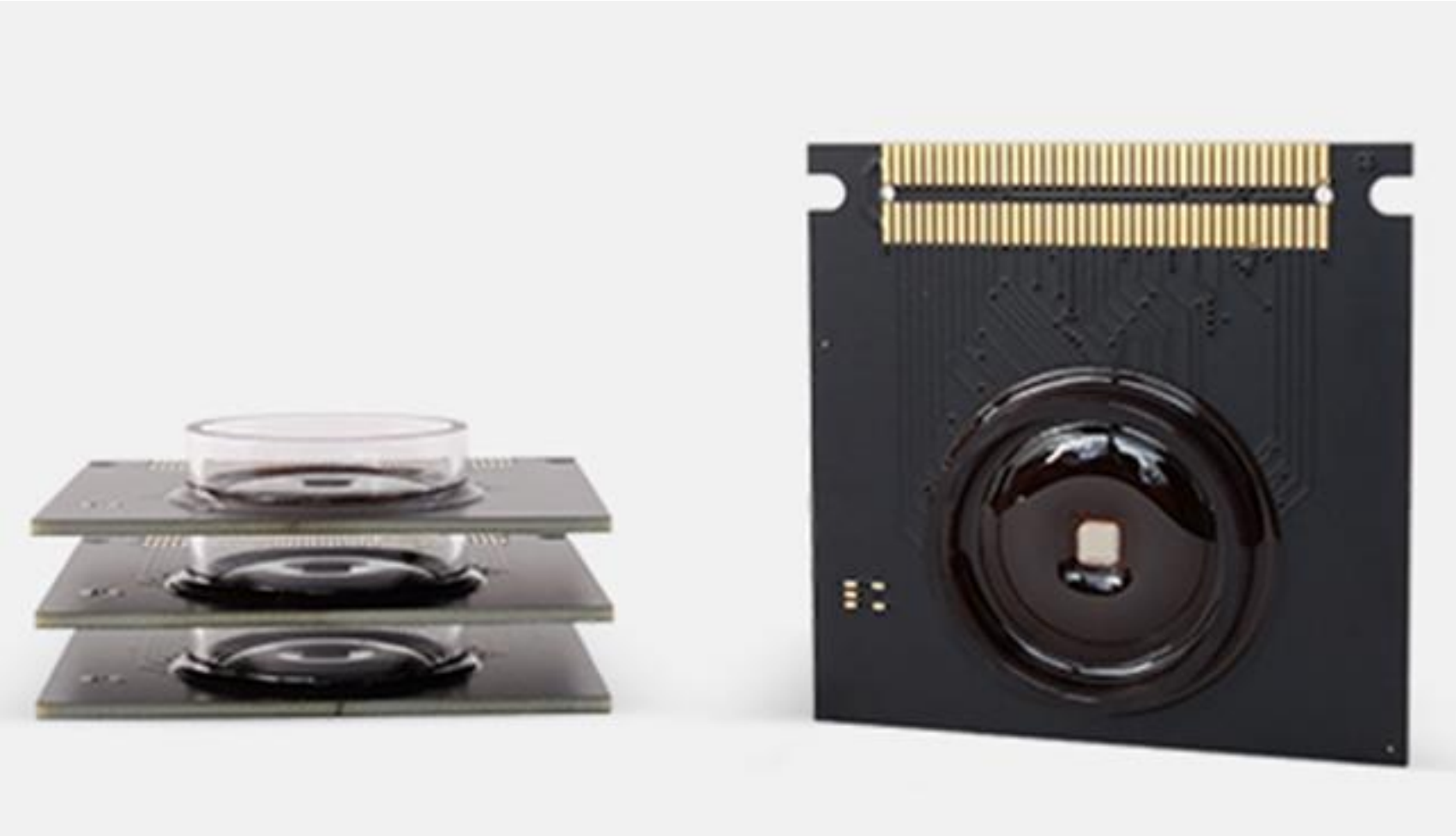

***Figure 14****: Example of commercially available CMOS-MEA device.*

Moreover, the use of CMOS technology allows for the realization of small system chips with large numbers of electrodes; arrays with thousands of electrodes on 6 mm by 6 mm chips have been reported (Figure 14***Figure 14***), with high sampling rates for each channel [103–105]. High-density microelectrode arrays realized in standard CMOS technology offer the potential to perform recordings at single-cell or even subcellular resolution [106].

When designing such systems, a compromise between noise performance and spatial resolution must be made. To achieve high spatial resolution, small electrodes are preferred; unfortunately, thermal noise is inversely correlated to electrode size. A further impediment is that, because of spatial constraints, filters and amplifiers with large noise-reducing capacitors cannot be used.

Other issues that must be considered with CMOS MEAs are the packaging of the chip and the risk of corrosion of electronic circuits by aqueous salt solutions, so the chips themselves need to be protected from the biological environment.

# MEA applications: from Scalable Recording to Cardiotoxicity and Pharmacology

In 2014, the US department of health and human services estimated that nearly 1 million patients showed adverse drug reactions each year. Among these, drug-induced arrhythmias are the leading cause (WHO, 2019). In fact, small molecules and biologic drugs that may address different target, affect human cardiomyocyte biology, thereby significantly reducing cardiac function.

Drug induced cardiotoxicity is a major problem, even occurring after the introduction of the drug on the market. In fact, very often the effect of a compound on heart health rise after the treatment, for example by modify the expression of ion channel or other features of the cardiomyocyte. These of course make the drug useless and forces a withdrawal from the market with huge money and work time waste.

In recent years, furthermore, the need of efficient cardio-pharmacological and cardio-toxicological *in-vitro* testing is increasing, as there are new directives to restrict animal use for laboratory tests. Therefore, there is a great need for systems that recapitulate and model human biology, with scalable and reproducible features, and preferably from an inexhaustible source.

New experimental strategies based on alternative methods, in which the use of time and materials is reduced, are required. Furthermore, using animal hearts as a model, is also not sufficient due to interspecies differences in electrophysiological properties and different responding behavior to drugs [107].

hiPSCs-CM may have this potential, nevertheless their relatively immature phenotype might affect their drug responses. Remains to be checked if drug responses on these cells recapitulate drug responses of the adult myocardium.

The Comprehensive *in vitro* Proarrhythmia Assay (CiPA) initiative was originated for drug pro-arrhythmic potential assessment in order to analyze several known drugs and substances and how they affect the cardiac system.

For the categorization, therefore, the CiPA initiative recommends assays that are mechanistically based on *in vitro* assays. To date, one of the most promising tools for pharmacological tests is the Micro-Electrode Array (MEA). MEA technology has been recognized as a standard experimental approach for *in vitro* long-term electrophysiological investigations on cardiac cells and tissue. Furthermore, the use of hiPSC-CMs on MEA for drug testing is a promising tool due to their large-scale production circumventing the lack of a source for human adult cells (Figure 15).

Since the inception of the CiPA initiative, several cell lines (including self-generated and commercially available cell lines like iCell Cardiomyocytes (Fuji), Pluricytes (Pluriomics), Cor4u (Ncardia), Axol Bioscience, I HCm (Cell applications), ASC (Applied Stem Cell, ix Cells Biotechnologies), CDI (Cellular Dynamics International), Cellartis (Clontech, Takara), ReproCardio (ReproCELL) and ACCEGEN (immortalized from patients or transdifferentiated from hSC) have been used to analyze the impact of compound administration on cardiomyocyte electrophysiology.

Recently, a lot of literature has been published reported studies about the effect of several compounds on cardiac cells. Data on drug-induced APs modification have already been studied using different readout, such voltage sensitive dyes [5], patch clamp [108] and MEA [4,109,110]. Despite these attempts, all these readout encompasses many advantages, but also disadvantages. The most important is that none of these approaches give complete information about waveforms of the APs on long-term period. The only one that allow for single APs detection, patch clamp approach, is a high invasive technique that prevent analysis over many hours/days. Therefore still, remain the need to develop a novel tool for addressing these needs.

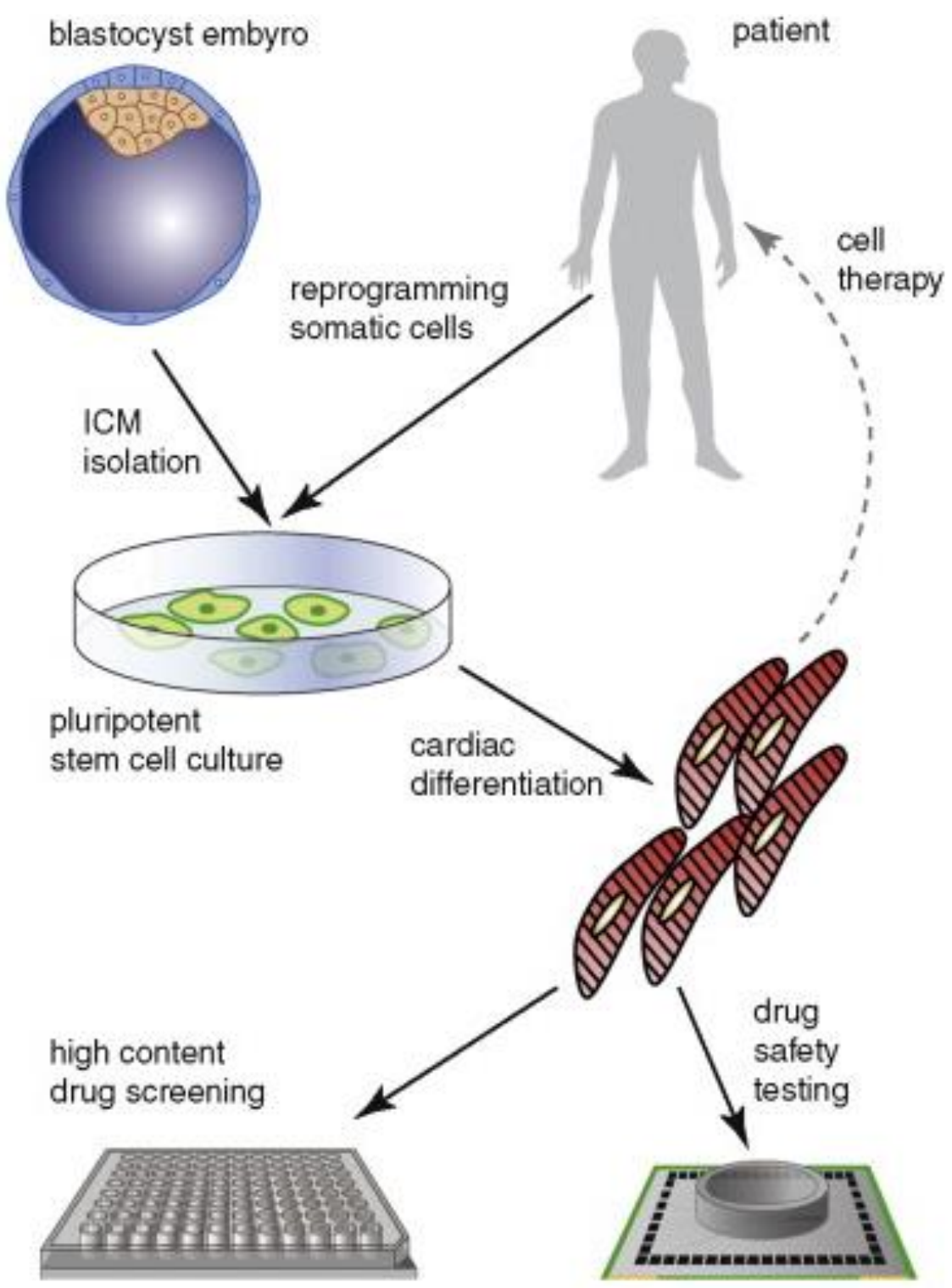

***Figure 15**: Applications of cardiomyocytes from pluripotent stem cells. Reprint with permission from (Braam, Passier and Mummery, 2009).*

# Aim and Objectives of the Thesis

Cardiotoxicity is one of the leading causes for failure during drug development process as well as after a compound has been approved for therapy. Between 1990 and 2013, a total of 81 drugs had to be taken off of the market in the USA, Europe and Asia, due to treatment-emergent effects, with 16 of the compounds producing cardiac arrhythmias. Moreover, since 1997, cardiotoxicity-related withdrawal has dramatically increased from 5.1 to 33.3 %.
In order to establish whether a potential molecule can became a drug, the discovery process at an early stage uses processes based on high-throughput screenings, which involve groups of candidates comprising more than 100,000 different molecules divided in libraries. These screenings are followed by strict toxicity analyses, which filter out approximately 60–70 % of all candidates due to unwanted effects, also in the late stages of drug development [111]. Only one in 5000–10,000 screened chemicals reaches the market, with an associated cost of $1–2 billion for development and a 12- to 15-year commitment. The economic risk in the development process can be reduced if compounds with therapeutic but toxic properties are recognized early. Moreover, cardiac adverse side effects have been the major causes for the withdrawal of drugs from the market. [112].
Therefore, preclinical cardiotoxicity screening is useful to detect unsafe compounds early in the drug discovery process to regulate candidate selection, reducing later-stage attrition, and risk to participants in clinical studies, development time and costs. (Figure 16).

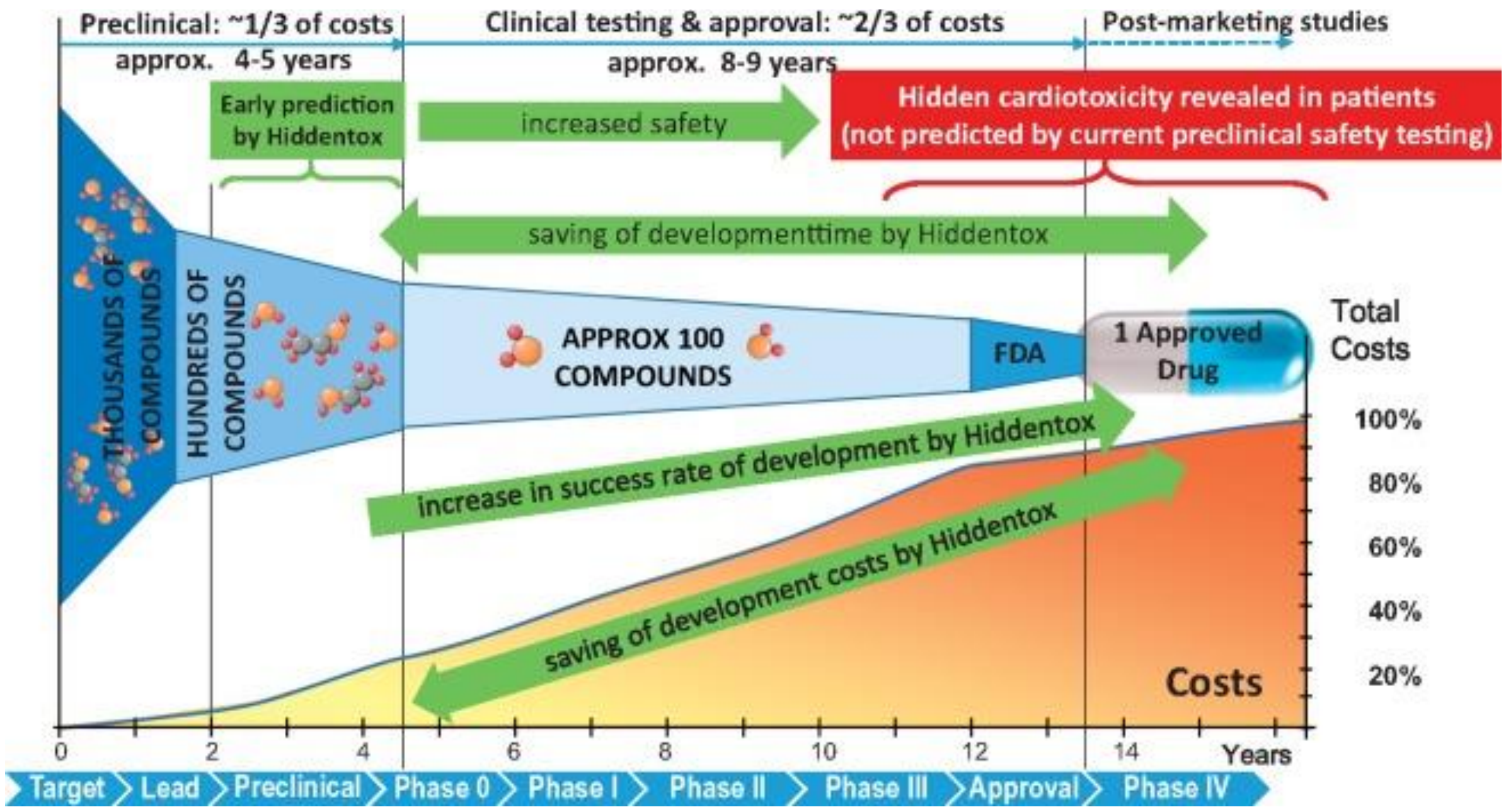

***Figure 16**: Benefits of pre-clinical prediction of hidden cardiotoxicity by pre-clinical testing platform and drug development process* [113].

So far, screening assay involve simple *in vitro* systems and surrogate markers for cardiac liabilities in animal or animal-derived tissues and organs. A major limitation of the current range of *in vitro* and *ex vivo* models of non-human species in use today is that none can be utilized as a single test system across both structural and functional cardiotoxicity assays to generate a cohesive risk assessment [114]. Recent advances in methods for large-scale production of cardiomyocytes (CM) from human induced pluripotent stem cells (hiPSCs) offer an opportunity to provide a human model system that is compatible with high throughput screenings. Being a more relevant nonclinical model for *in vitro* cardiotoxicity testing, these cellular models have the potential to bridge analytical platforms, thus providing an integrated model for probing many possible causes and mechanisms of cardiotoxicity. Hundreds of reports in literature demonstrated the utility of hiPSCs-CM across a wide range of analytical platforms and readout. These instruments include electrophysiology [115], optical imaging [116], contractility studies [117], metabolic studies [118].

Despite several approaches, the gold standard for drug screening remains the manual patch clamp. This technique, while extremely sensitive, is also immensely time-consuming and labor-intensive, which limits its usefulness for rapid-throughput cardiotoxicity screening. In the last decade, platforms for automated patch-clamping have been developed, and the technology has made impressive advances. The major benefit of automated patch clamping is the ability to make measurements in parallel, allowing multiple recordings to be made simultaneously.

Despite many advantages, automated patch clamp still works with specific and stable cell line that express a single ion channel type in large quantities. These cells used in the automatic techniques usually produce well-behaved patch clamp characteristics such as membrane seal, stability, and sufficient amplitude of currents. The primary cells, differentiated cells derived from iPSCs/ESCs or transiently transfected cells with low expression levels of ion channels may not feasibly be used in most of the systems on the market. Therefore, conventional patch clamp technique is still not completely replaceable by these automated patch clamp systems due to its unique features of high data quality and flexibility with cell types (Schee*l et al*, 2014; Li *et al*, 2016).

Multi Electrode array (MEA) platforms allows for high-resolution and non-invasive electrophysiological recordings and are also used to track drug-induced effects on hiPSCs-CM[121–124]. The great advantage of the MEA platform is that it offers the opportunity to conduct experiments with intact cellular networks.

The MEA assay is an electrophysiology-based technique that uses microelectrodes to measure fluctuations in extracellular field potential (FP), generated from spontaneously beating cardiomyocytes, and it is correlated with action potentials, as well as with the QT interval measured on the clinical electrocardiogram [125,126] . The resulting FP waveform is analogous to an in vivo ECG trace, which measures the change in voltage on the body surface generated by ionic currents that flow through the heart. By measuring the drug-mediated impact on multiple features of the FP signal in the hiPSC-CM MEA assay, the cardiac liability and mechanism of action of a compound can be predicted [127].

In 2013, the Comprehensive I*n vitro* ProArrhythmia (CiPA) (http://cipaproject.org/) initiative was proposed, with the aim to assess cardiac safety based on a mechanistic approach.

This initiative introduced and validated the use of hiPSCs-CM and the micro-electrode array as a non-invasive and label-free assay for detecting cellular arrhythmias. Although some characteristics can differ from a type of hiPSC-CMs to another one, it is a promising way to overcome the limitations of the existing methodologies used for preclinical safety evaluation of pharmaceutical compounds and is now considered as a reliable cardiotoxicity assay.

Since MEAs provide information about electrical activity from clusters of cells, as described in previous section of this thesis, it is impossible to recognize a single cell response after a drug treatment. Consequently, MEA cannot satisfy completely the requirements for analyze ion channel associated currents.
There is therefore an interest in finding more accurate and more predictive new methods for avoiding late stage drug attritions and side effects. Specifically, for heart-related issues, the market needs innovative solutions to surpass the MEA state-of-the-art performance and for accurately detecting cardiotoxicity.
The CiPA initiative could be, therefore, addressed with a MEA platform capable also of intracellular access and examination of intracellular APs instead of FPs.
A very recent advancement in intracellular recording that reached the market of MEA is represented by the LEAP (Local Extracellular Action Potentials) technology from Axion Biosystem (https://www.axionbiosystems.com/LEAP). This novel approach allows for recording of detailed information on multiwell MEA plates for high-throughput applications. However, the Axion Biosystems induction protocol cannot be repeated reliably on the same cells over several hours and days, with limited predictive power.
In the last years, several efforts have been done to develop robust approaches for gaining intracellular access by combining MEA with engineering solution for intracellular recordings.
Two important features are necessary to achieve high-fidelity network-level recordings using electrode-based tools: scalable intracellular electrodes to couple to the cells and scalable electronics to record each electrode's electrophysiological signal.
The development of better microfabrication techniques, mainly driven by the Complementary Metal Oxide Semiconductor (CMOS) industry, have enabled scaling up the number of recording electrodes while further enabling the integration of intracellular measurement electronics on the same substrate as the electrodes (Abbott *et al.*, 2017). Here, they combine 3D nanoscale electrodes integrated on complementary metal-oxide–semiconductor (CMOS) circuits to realize a high-fidelity parallel intracellular recording at the network level. However, the complex design of nanostructures and intrinsic limits of electroporation still limits the technique and hinder the translation of these platforms in pharmaceutics and drug development fields.
Optoporation could represent a step toward scalable technologies for both network and single cell electrophysiology since the use of laser open the way for high-throughput approaches and automation of processes [128].
Therefore, exploiting the features of plasmonic optoporation, recently introduced by De Angelis's laboratories (Dipalo *et al*, 2017), I worked on to the development of the concept of planar meta-electrodes that, combined with plasmonic optoporation, allow for recording of intracellular signals from hiPSCs-CM on high density commercial complementary CMOS-MEAs from 3Brain. Since the approach does not require any rework or 3D post-processing of the devices and does not employ more complex electronics, it overcomes and fixes the abovementioned issues of electroporation on CMOS-MEA coupled with 3D nanoelectrodes, making an important step toward the recording of electrical signals from whole networks of cell on commercial devices.
Although this approach is extremely promising for high-content recordings and for specific assays that are beyond the capabilities of standard MEA or patch-clamp, the use of integrated silicon technology could obstruct the translation to the pharmaceutical industry standards and the fast adoption from the *in vitro* electrophysiology community.

Therefore, I also worked on overcoming these limitations by providing evidence that optoporation can be used on the majority of electrode types used today in commercial MEA acquisition systems, from the research labs to the pharmaceutical industry. In particular I focalized on MEA in single well configuration from MultiChannel Systems and MEA in multi-well configuration from Axion Biosystems.

# A Novel Platform for *in vitro* Electrophysiology

# Results and Methodological Approaches

## Meta-electrodes

A cell that interfaces with an out-of-plane structure, such as with a 3D nanopillar, makes a tight physical coupling as shown in detailed in the previous section. Exploiting chemical or physical mechanisms, intracellular recording from single cell and whole network has been introduced previously in this thesis, highlighting the limits.
The combination of 3D nano-electrodes and laser optoporation has been presented as novel approach to gain intracellular access and record action potentials from cardiomyocytes and primary neurons [10]. However, the requirement of 3D nanofabrication methods limits the practical implementation and spreading of these techniques for many reasons. For example, the fabrication of 3D nanostructure is time consuming and has high costs, requiring clean room facilities; furthermore, these processes are not well developed on large area substrates. Therefore, these platforms are not suitable for recording from whole network of cells in commercial applications.
In order to overcome these limitations, the first step conducted in this work has been the establishment of cell culture on alternative materials free of high-costs and time-consuming fabrication steps, which represent a valid alternative for the implementation of scalable techniques for industrial and research applications.

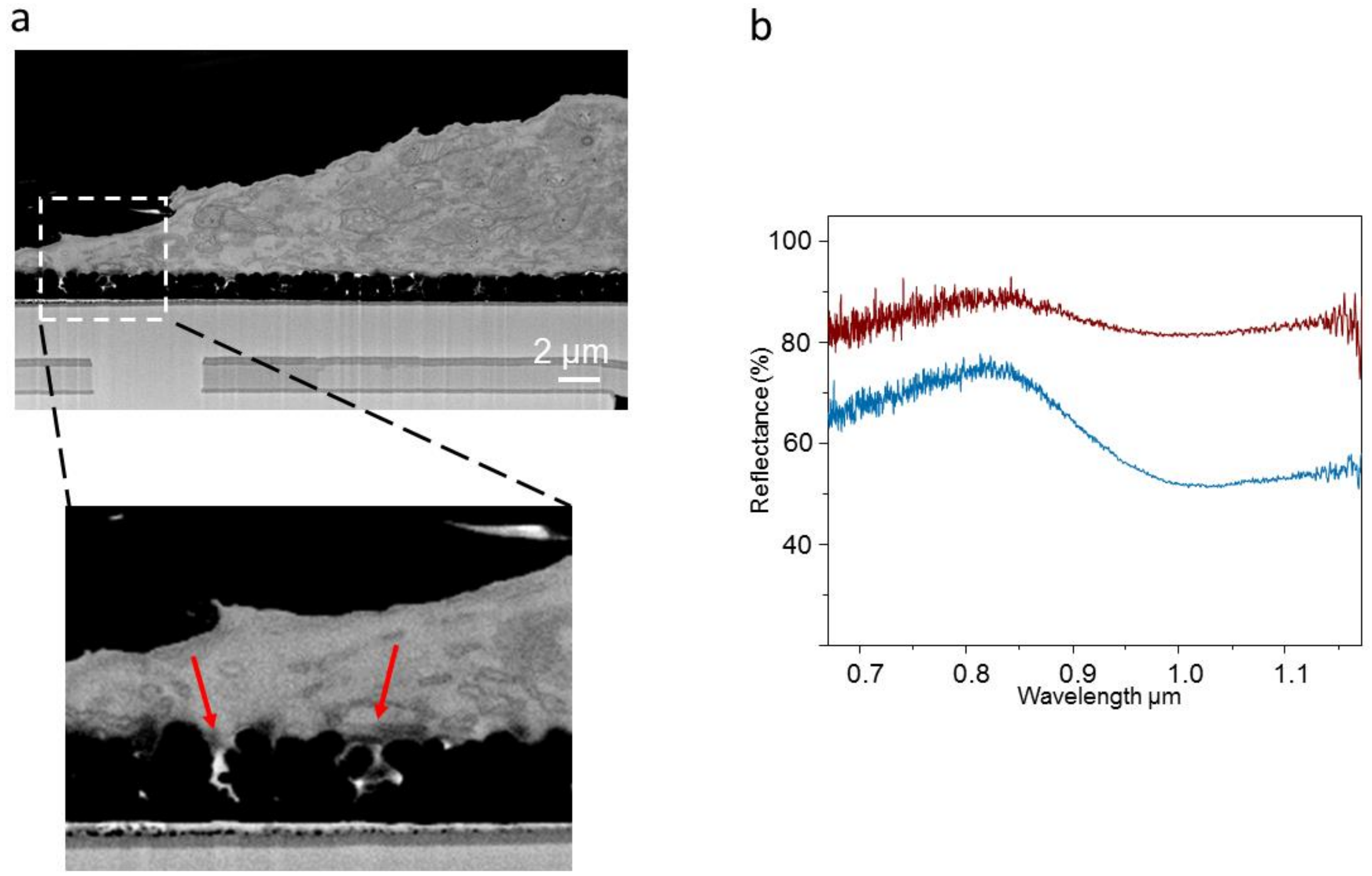

***Figure 17**: Optical and biological behaviour of meta-electrodes a. SEM cross-sectional images of a cardiomyocyte cultured and fixed on meta-electrodes. b. Porous (blue trace) and rough (red trace) platinum reflectance in the NIR range.*

Material surfaces with porous and rough features at the micro- and nanoscale are well known in the field of electrophysiology as optimal promoters of tight junctions and electrical coupling. The typical porous configuration shows, at the nanoscale, empty nanogaps between the material bumps. As demonstrated for 3D vertical nanopillar, the cell is able to settle on top of porous surfaces, establishing a strong connection with the material's pores, and a tight adhesion over the whole cell/surface contact as depicted in Figure 17a (red arrows in the magnified view).
Here, a cross section of a cardiac cell cultured on the surface of porous platinum has been reported in order to prove the optimal physical and biological coupling established with the material. Most importantly, the figure shows how the cell membrane establishes a coupling inside the nanopores, by interlacing with the material surface. Porous or rough surfaces, therefore, enhance the process of cell adhesion and, eventually, generate optimal conditions for engulfment dynamic and mechanical wrapping of nanostructure as also demonstrated by [61].
On the other side, past studies demonstrated also that porous or rough metal surface have the capability of arbitrarily controlling the light at the nanoscale (Yu and Capasso, 2014).
Both the combination of good biocompatibility and electrical features suitable for the development of novel platforms for electrical recording, makes platinum an optimal candidate for laser optoporation.
In order to prove this, a first evaluation of its optical behavior has been conducted on porous and rough platinum thin films. A reflectance measure has been performed with a FTIR spectrometer from Thermofisher, considering the near-infrared (NIR) range of wavelengths. The reflectance spectra have been acquired in respect to a perfectly reflective gold layer. Figure 17b shows that porous platinum surface shows strong scattering and adsorption in the NIR region where the laser used for optoporation is located.
In particular, porous platinum thin film (red trace) with a pore size of 50–200 nm behave like a broadband absorber in the Near Infrared (NIR) region of the electromagnetic spectrum, with an absorption about 50% higher than a rough platinum surface (blue trace) at these wavelength.
Therefore, under suitable conditions, porous platinum surfaces can very efficiently mimic both the biological and optical behaviours of 3D antennas. [129], being compatible with laser optoporation.
Nanoporous or rough metallic materials are exploited as covering layer of the electrode in the MEA productions. CMOS-MEA from the Swiss company 3Brain AG for example are produced with nanoporous platinum as metal of choice for electrical recording from electrogenic cells. Due to the abovementioned results, it is reasonable to conclude that MEA electrode covered by porous metallic films can very efficiently mimic both the biological and optical behaviors of 3D nanoantennas. From now on, these class of electrodes will be define as meta-electrode for the rest of this thesis.
Furthermore, when a porous metal thin film is excited with light in the NIR regime, this incident radiation is adsorbed by the porous surface and the electric field associated to the electromagnetic wave is concentrated and strongly amplified inside the nanogaps of the material, thus forming optical "hot spots" [130], as shown in Figure 18.
Here, I report details on the surface morphology and optical behavior of three different class of commercial meta-electrodes: Figure 18a reports details about platinum porous electrode from 3Brain. Figure 18b reports details about gold electrode from Axion Biosystems. Figure 18c reports details about Titanium Nitride (TiN) electrode from Multi Channel Systems. All SEM images were acquired with a FEI Nanolab 600 dual beam system. Imaging was performed in immersion mode, setting the voltage at 3 kV and the current at 0.20 nA. The FIB cross sections were performed operating with a 0.23 nA Ga ion beam accelerated at 30 kV.

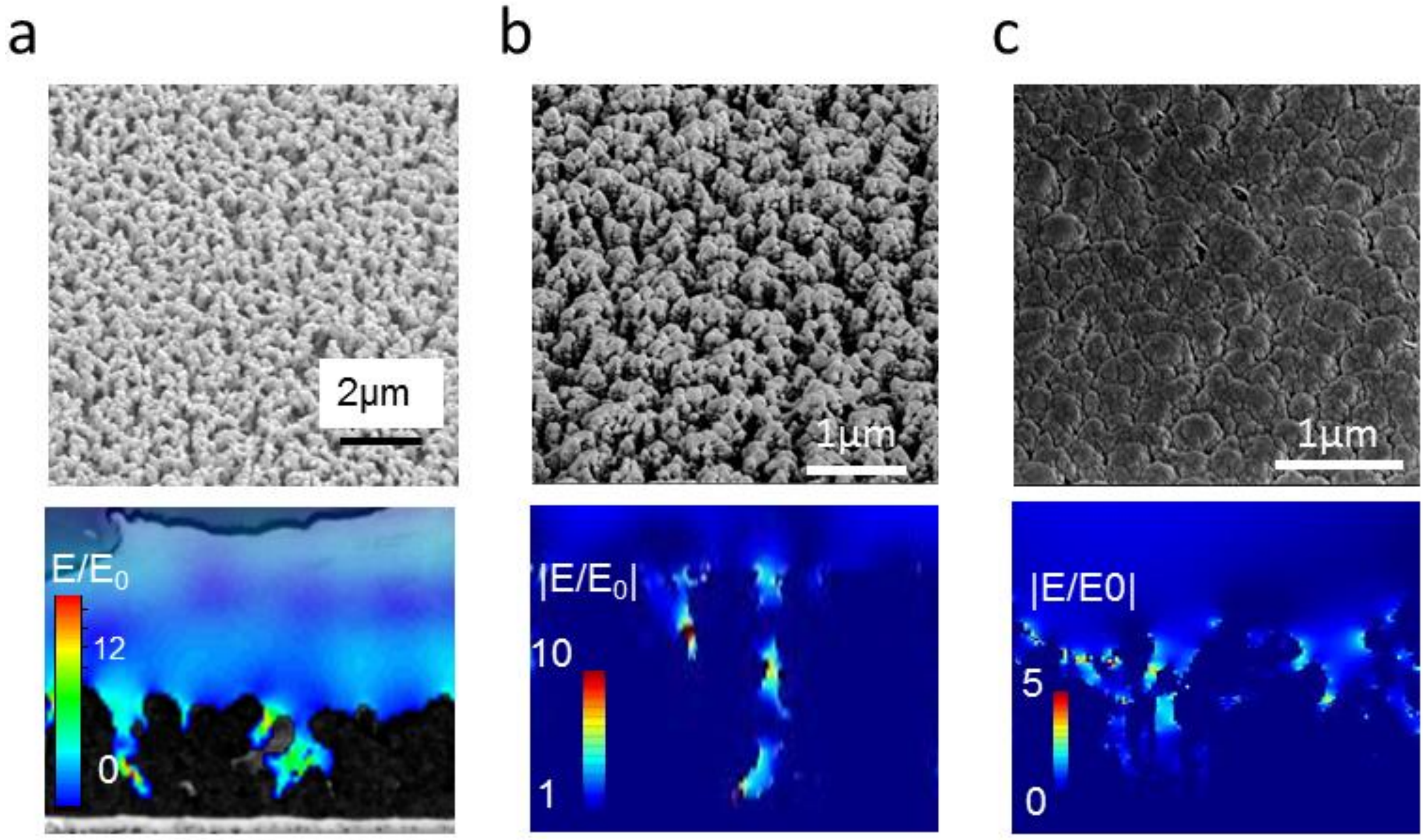

***Figure 18**: Surface morphology of porous and rough commercial electrodes and related 2D simulation of electromagnetic enhancement. a. Electrode from 3Brain. b. Electrode from Axion Biosystems. c. Electrode from MultiChannel Systems.*

To study the optical behavior of these electrodes, 2D electromagnetic simulation of the electric field distribution at the interface between electrodes surface and water was performed. Finite Element Method implemented in the RF Module of Comsol Multiphysics was used to calculate the near-field enhancement of the porous materials. A SEM image of a vertical cross-section of the meta-electrodes was used as structure in a 2D simulation where the incident field was a linearly polarized plane-wave at $\lambda$ = 1064 nm. Perfect matching layer condition has been put at the boundaries of the simulation region.

The acquisitions show the high porosity surface of the three different meta-electrodes, with deep nanogaps between the bumps of the material. Notably, between the three different configurations, Gold electrode from Axion Biosystems (Figure 18b) presents disordered vertical sharp nanostructures structures rather than a planar nanoporous surface.

After cross-sectional SEM imaging of the meta-electrodes, the morphological profiles of each of them has been extracted, associating the corresponding dielectric constants of water to the empty gaps and of Platinum, Gold or TiN to the electrodes. Then, we used these configurations for simulating the optical response under laser excitation at 1064 nm. As shown here, the laser radiation produces hot spots in all the materials, concentrating the electric field inside the nanosized features of the electrodes.

Notably, the optical hot spots are overlapped with the sites of tight junction at the electrode-plasmatic membrane interface. Therefore, the membrane is located inside the sites of higher amplification of the optical energy delivered by the incident light. This particular feature is the key parameter for optoporation, explained in the next section.
For TiN electrodes, given the different dielectric constants in respect to gold and platinum, the plasmonic enhancement results to be much lower than that obtained on gold or platinum electrodes with similar morphology. However, TiN has a lower work function than gold and can generate a higher number of hot electrons under light excitation in the visible and near-infrared region [131]. Since the hot electrons are responsible for generating the nano-shockwave that leads to cell poration, TiN remains a promising candidate for intracellular recording by means of laser optoporation of the cell membrane.

# Optoporation and APs recordings of hiPSCS-CM Cultured on CMOS-MEA from 3Brain

Commercial high density CMOS-MEAs available from 3Brain offer 4,096 square porous Platinum electrodes 21 × 21 $\mu m^2$ in size and with a 42 µm centre-to-centre pitch; they have been used in the past to record hippocampal neurons [103], cardiomyocytes and retina [132]. Considering the typical size of cardiomyocytes, which are larger than other electrogenic cells, such as neurons, the electrode size of the 3Brain device offers a single-cell resolution recording.
All the experiments have been conducted with human cardiomyocytes derived from induced Pluripotent Stem Cells (hiPSC-CM). HiPSC-derived cardiomyocytes have been purchased from Ncardia (Cor.4U (Ax-B-HC02-1M)).
Cor.4U cardiomyocytes are fully functional hiPSC-derived cardiomyocytes obtained through the *in vitro* differentiation of transgenic hiPSCs and puromycin selection technology for the resulting cardiomyocytes (Ncardia). HiPSCs were thawed and pre-cultured in a cell culture T25 flask coated with 1:100 fibronectin in Dulbecco's Phosphate Buffered Saline (DPBS) and grown over night in Cor.4U complete medium (Ncardia) before seeding on MEA devices.
This procedure allows the removal of dead cells prior to seeding and results in a better assay performance. CMOS-MEA were sterilized with Ethanol 70% for 30 minutes and then washed 3 times with sterile water. They were subsequently treated to promote the tight adhesion of cells. The substrates were coated with Geltrex ready-to-use solution (Thermo Fischer) and then incubated for 30 min at 37 °C in a humidified environment to promote the tight adhesion of the cells. Subsequently, the solution was totally removed and the cells were rapidly plated without the coating being allowed to dry. Cor.4U were plated at a density of 70000 $cm^2$ and grown with Cor.4U complete medium.
The cardiomyocytes were grown on the devices at 37°C, 5% CO2, 95% air for four days until they reached confluence, showing spontaneous beating activity. At the fifth day, first, five minutes of unperturbed extracellular activity was recorded to characterize the culture. Then laser pulse trains were applied on the surface of the CMOS-MEA electrodes and were used to porate the cardiomyocytes.

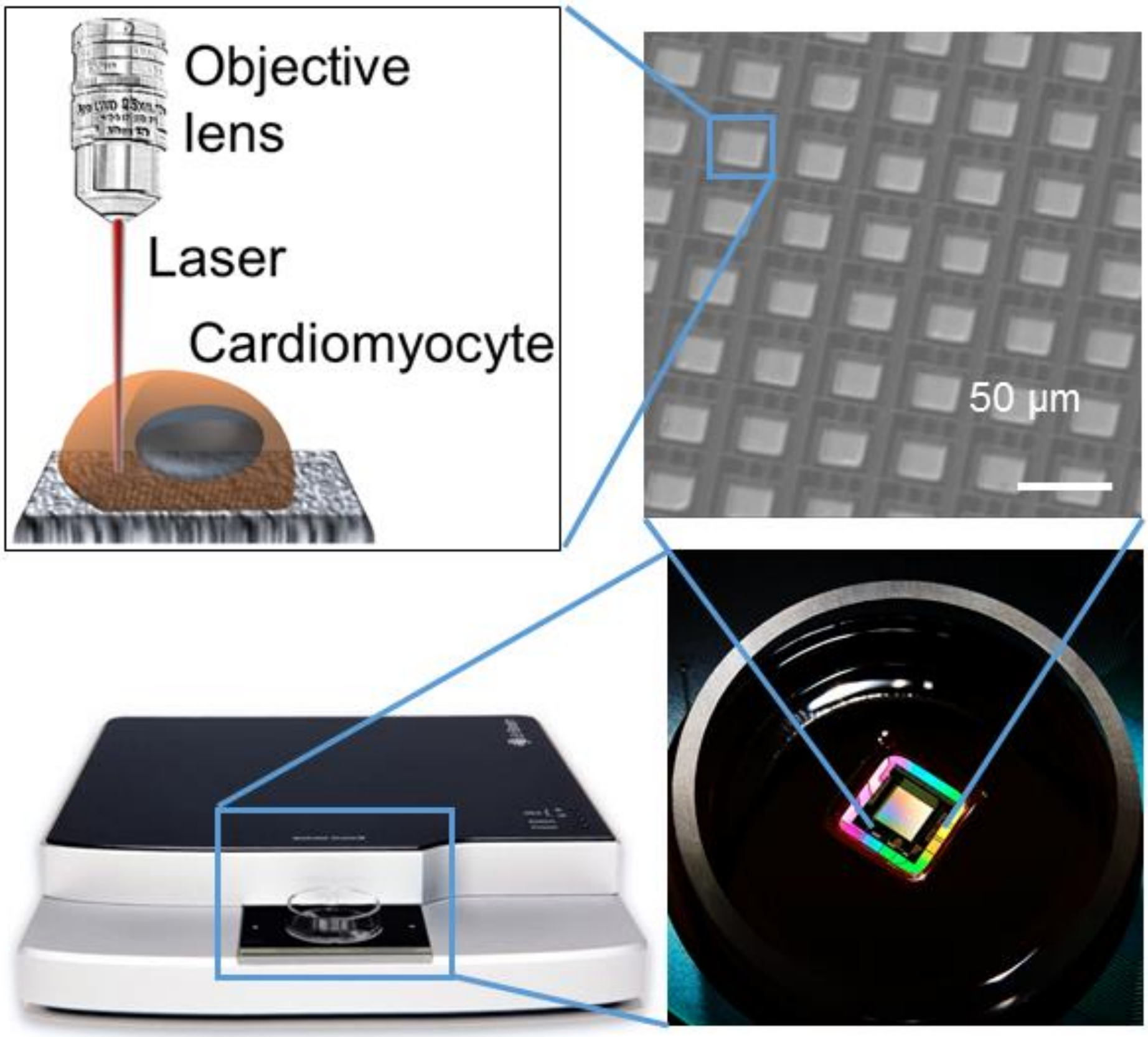

***Figure 19**: Overview of optoporation. An upright microscope is coupled to a NIR ultrafast laser source. The laser is focalized by an objective lens on the surface electrode of a CMOS MEA. The microscope accomodate the 3Brain apparatus for electrical recording.*

For the laser poration, Figure 19 reports a schematic overview of the whole process. The first harmonic (λ = 1064 nm) of a Nd:YAG (neodymium:yttrium–aluminium–garnet) solid-state laser (Plecter Duo (Coherent)) with an 8 ps pulse width and 80 MHz repetition rate was used as the radiation source for the optoporation experiments, with an average power of approximately 1mW. The exploited train pulses is therefore made of sequences of temporal windows with a duration of 12,5 ns and in this time, only one event of laser pulse take place with a duration of 8 ps. The total train time is 20 ms.
The laser was coupled to a modified upright microscope (Eclipse FN1 (Nikon)) able to accommodate the acquisition system from 3Brain directly on the microscope stage. A 60X immersion-mode objective lens (NA = 1 working distance = 2.8 mm) was used during the experiments to observe the cells on the devices and to focus the NIR laser used for poration.
In each experiment, laser optoporation was performed on several electrodes, by manually pointing the laser on the electrode and then recording intracellular APs. Successful recording experiments has been conducted on 20 different culture preparations, with a success rate of about 90%.
Examples of extracellular and intracellular recordings of hiPSCs before and after laser poration are shown in Figure 20**Figure 20**. In Figure 20a, the extracellular signals show an extremely high signal-to-noise ratio, with a total amplitude of approximately 4 mV on average: this result is in line with the maximum input–referred signal amplitude of the electrodes of 4 mV allowed by the original optimization of the 3Brain system to record high quality extracellular signals thanks to the tight adhesion of the cells on the nanoporous platinum electrodes. Figure 20b-c depicts intracellular spikes after optoporation. Here the signals reach an amplitude of approximately 4–5 mV on average.

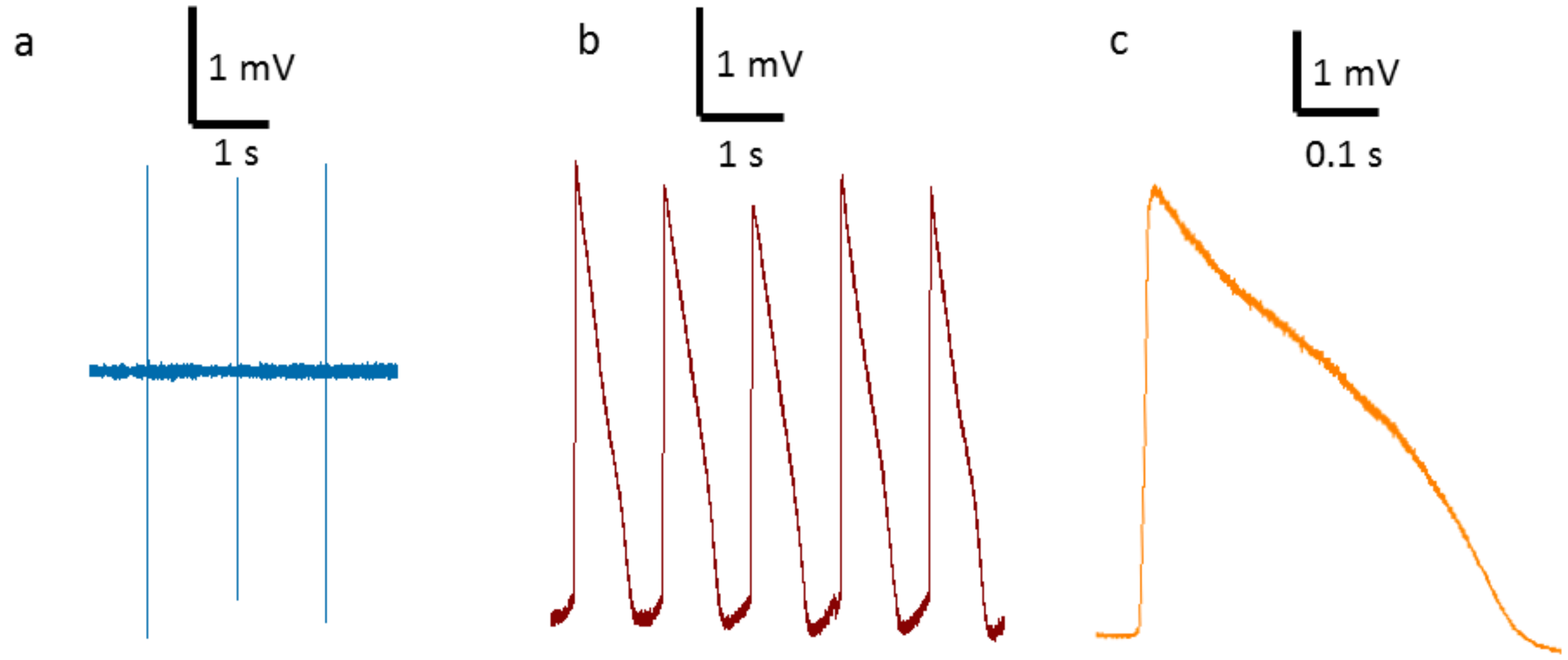

***Figure 20**: Examples of extracelluar (a) and intracellular (b-c) signals from hiPSCs-CM cultureed on top of porous electrode of a CMOS-MEA from 3Brain.*

The spikes present a fully positive phase, as expected from the action potentials recorded from the intracellular compartment of cardiac cells, as it was discussed in the previous chapter of this thesis. In addition to the waveform polarity inversion, a clearly different spike shape is observed, which closely resembles an action potential recorded with a patch clamp. This is more evident with the magnified spike shown in Figure 20c. Here, the total duration of the action potential is around 200–300 ms, which is the typical intracellular action potential duration of cardiac cells. In contrast, the extracellular spike duration is approximately 5 ms.

# Network-level APs recordings from cardiomyocytes on commercial 3 Brain CMOS-MEA

Park's laboratories first carried out network recording from hundreds of cells implementing a protocol of massive electroporation on a CMOS MEA platform with 3D nanopillars; despite very good results, this pioneering work still suffers of both electroporation and nanofabrication drawbaks.

As discussed above, meta-electrodes array coupled with non-invasive approach of intracellular recording, can pave the way to mass electrophysiological recording on devices already available on the market avoiding post processing fabrications.

Figure 21 reports the results of network-level optoporation and intracellular recordings on the whole CMOS-MEA array of a 3Brain device.

To achieve massive poration, a fast scan of the laser over the whole array has been performed, using the motorized stage mounted under the CMOS acquisition system (sketched in the left part of the figure).

The right part shows traces from 24 different electrode porated during the laser scan. Due to the limited amplitude allow from the device used for the recording, the electrode are capable of high-quality intracellular recording with average amplitude of maximum 3-4 mV. Most important, the recording of the electrical activity of the whole cell culture was continuously enabled during the laser scan of the array, that is crucial for gain uninterrupted monitoring of the APs. Notably, optoporation does not affect the behavior of the electrode in the nearby of those subjected to optoporation, thus implying high selectivity and resolution.

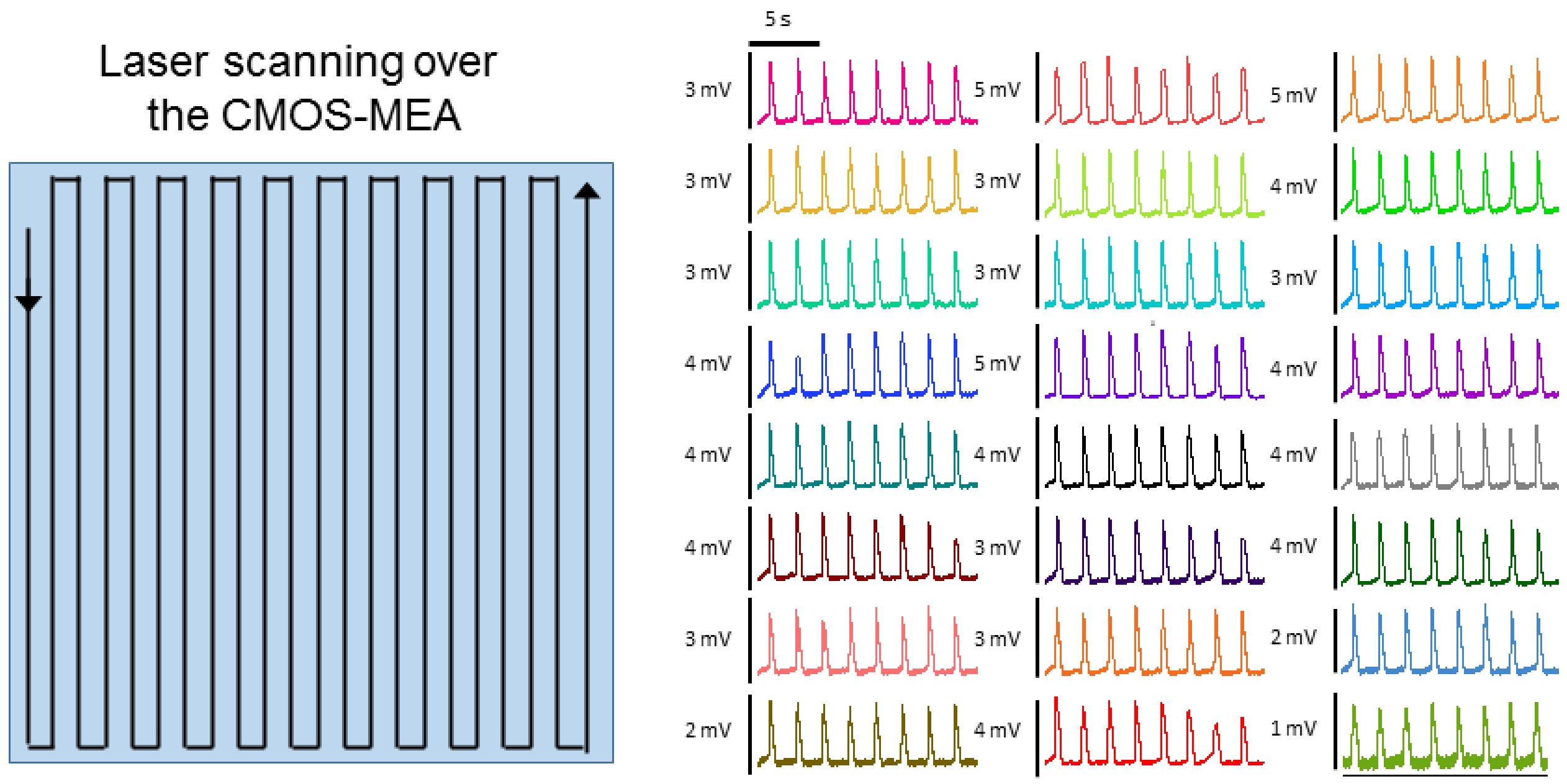

***Figure 21**: Optoporation for mass electrophysiology from hiPSCs-CM.*

# Optoporation and APs recordings from hiPSCS-CM cultured on meta-electrodes from Axion Biosystems Multiwell MEA

Axion BioSystems provides MEAs in multiwell configuration with high number of electrodes per well, unparalleled access to cellular electrical network information and a transparent plate bottom for culture visualization and assay multiplexing. Axion's MEA plates are characterized by low-noise signals, and retain the ability to be read over days, weeks, or months. MEA plates are available in several multi-well formats. All these features make these platforms suitable for high throughput applications in drug development and chemical research due to high compatibility in terms of parallelization and reliability.

Each well of a 12-well Axion MEA contains 64 low-noise working electrodes (8 x 8 configuration) and 4 integrated ground electrodes. The electrodes are covered with gold and embedded in a glass substrate of 300 µm. As reported in Figure 22, the surface of meta-electrode in the Axion's products show a particular fractal structure with tree-like structures with an height of approximately 1 µm that resemble the structure of a fern, with nanoscale rugosity. These features reflect more or less the structure of a porous meta-electrode, therefore it's reasonable to have similar optical and biological behavior.

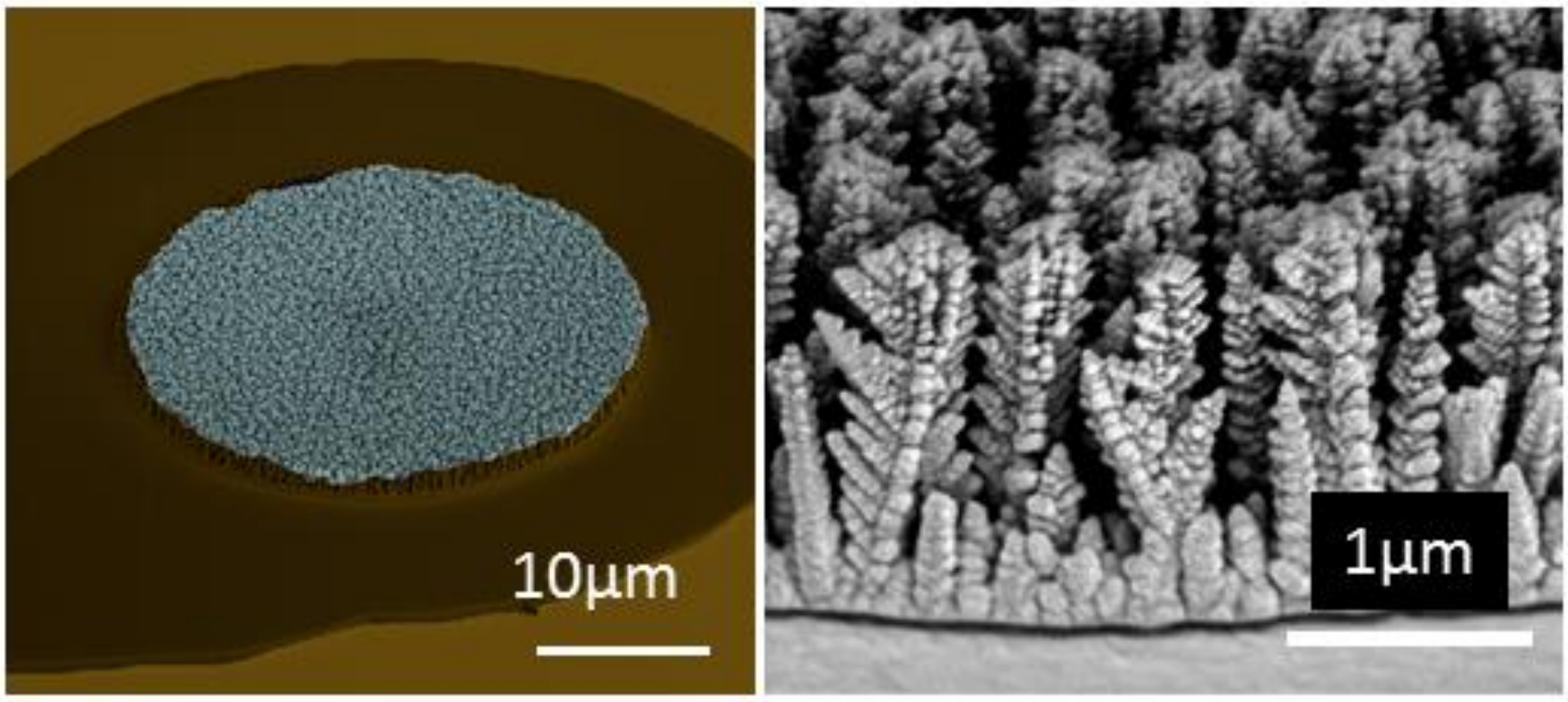

***Figure 22**: SEM acquisition of surface morphology of fractal-like gold electrodes (left) and SEM cross section of the same electrode (right).*

Axion Biosystems 12-well plates with fractal-like gold electrodes (Figure 22) were used as substrates for culturing hiPSC-CMs. The Axion Biosystems' MAESTRO system has been used to acquire signals from the cardiomyocytes. The MAESTRO system is capable of recording simultaneously from 768 electrodes (64 electrodes in each well for 12-well plates).

Before the cell culture, Axion 12-wells plates were sterilized with a 30 min ultraviolet exposure. The same coating, cell culture conditions and laser configurations described in the previous section have been adopted for these experiments. Successful recording experiments has been conducted on 30 different culture preparations, with a success rate of about 90%. After the cardiomyocytes monolayer reached spontaneous beating, extracellular recordings have been conducted. The recordings on Axion 12-wells plates, conducted with the Axion BioSystems' Maestro system, were done at a fixed sampling rate of 12.5 KHz. The cell medium from each sample was changed approximately one hour before the experiments and all the data sets were analyzed by AxIS software.

Figure 23 provides extracellular spikes recorded from hiPSC-CM on top of a fractal-like gold electrode from Axion Biosystems 12-well plates. The figure shows that the recorded field potential exhibits three signatures of extracellular recording: a spike with a shape that corresponds to the first derivative of the intracellular action potential ($-\frac{\partial V(t)}{\partial t}$), a duration of about 5-10 ms and an average amplitude of 1–2 mV.

These electrical measurements suggest that tight junction between fractal-like surface electrode and the cell membrane results in good coupling at the interface and therefore promotes the high value of extracellular signal's amplitude.

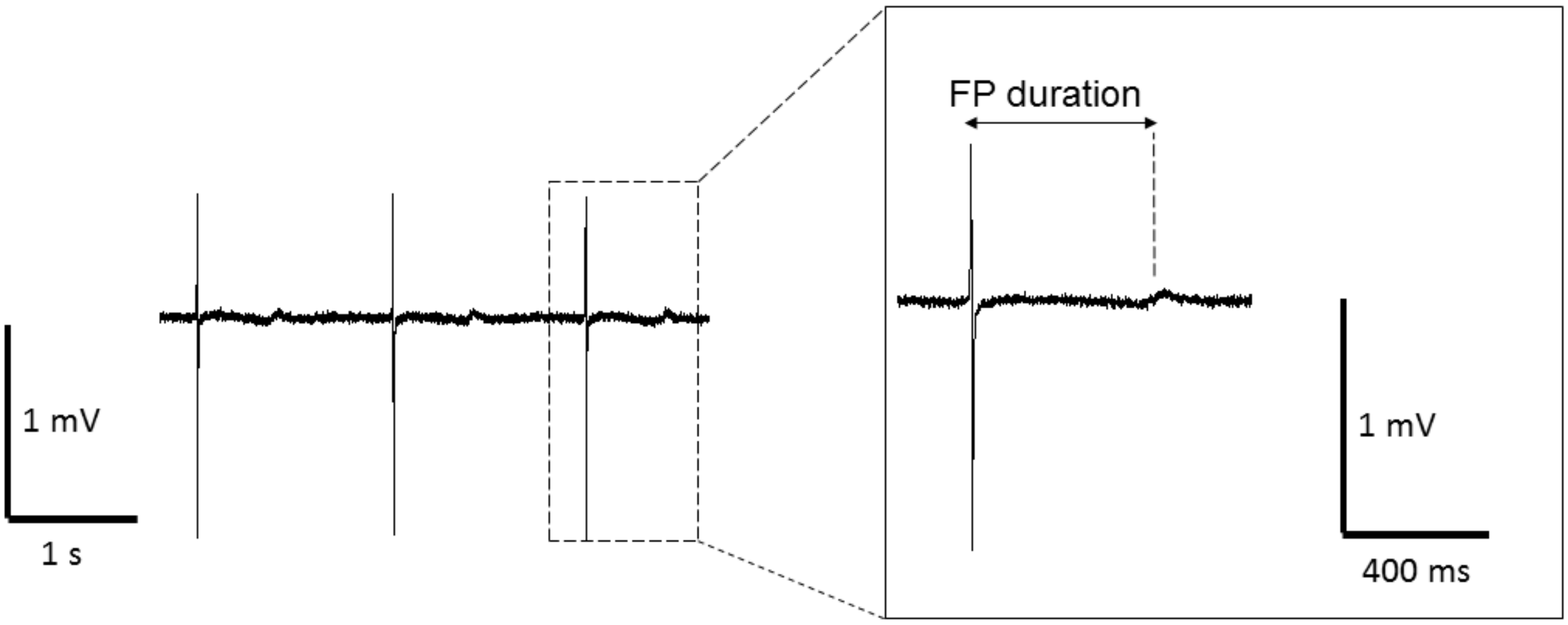

***Figure 23****: Extracellular recording from hiPSCs-CM cultured on multiwell MEA.*

These signals show also a very clear T wave corresponding to the low-amplitude positive peak present after each spike. The time between the spike and this positive peak represents the FP duration, corresponding to the time between depolarization and repolarization of the cardiac cells in the heart during a mechanical beat (as shown in the inset).

Despite high quality FPs have been recorded, these information do not fully characterize electrophysiological properties like AP repolarization periods, upstroke velocity and contribution of different ionic currents to the AP, which need to be derived from intracellular recordings.

Because of the fractal-like structure of the electrode's material, it is interesting to study the response of the electrode to optoporation in order to record intracellular access and record AP-like signals from hiPSCs-CM.

After the application of optoporation with an average laser power of 1 mW, as described in the previous section, an instantaneous switch from extra to intracellular APs is recorded.

Figure 24**Figure 24** provides evidence that high-quality intracellular recordings can be recorded on fractal-like gold electrodes used on multiwell MEA plates from Axion Biosystems. Immediately after the optoporation, the recorded signal amplitude increase to 9-10 mV. The noise level is similar to that of extracellular recording levels, but the signal-to-noise ratio undergoes a large increase.

In addition to this increase in the signal-to-noise ratio, our recorded intracellular signals have the following APs attributes: triangular shape and a total duration of about 300 ms, as depicted in the magnified view of the figure.

Furthermore, all the steps that characterize the cardiac action potential are easily recognized due to high quality traces that depend on the strength of the physical coupling between cells and the fractal surface. These traces have the shape and features characteristic of the intracellular action potential of cardiomyocyte cells [133], including fast depolarization at the beginning of the peak, a plateau region, fast repolarization and hyperpolarization, and a return to baseline.

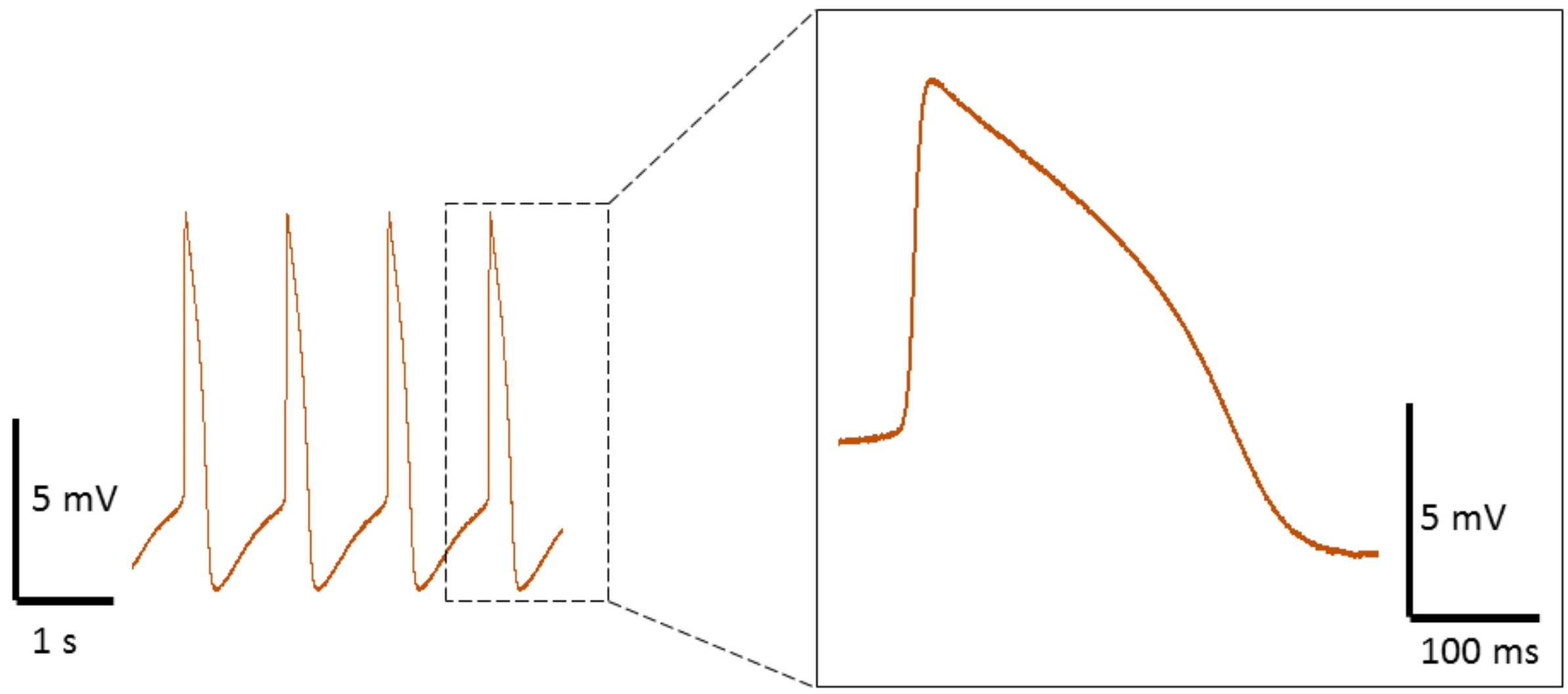

***Figure 24****: High quality intracellular APs from hiPSC-MC cultured on Axion Multiwell.*

To test the low invasiveness of our method, other experiments have been conducted where, after a first phase of intracellular recording, an attempt of repeated poration of same cell has been tried by exciting with the laser the same electrode approximately 30 minutes later. The results in Figure 25**Errore. L'origine riferimento non è stata trovata.Errore. L'origine riferimento non è stata trovata.** show that, due to the extremely controlled and localized pore opening, we are able to porate cardiomyocytes on the same electrode multiple times. The long-term coupling and the high sensitivity allow for the detection of the dynamic temporal evolution of the signals from intracellular APs to extracellular field potentials.

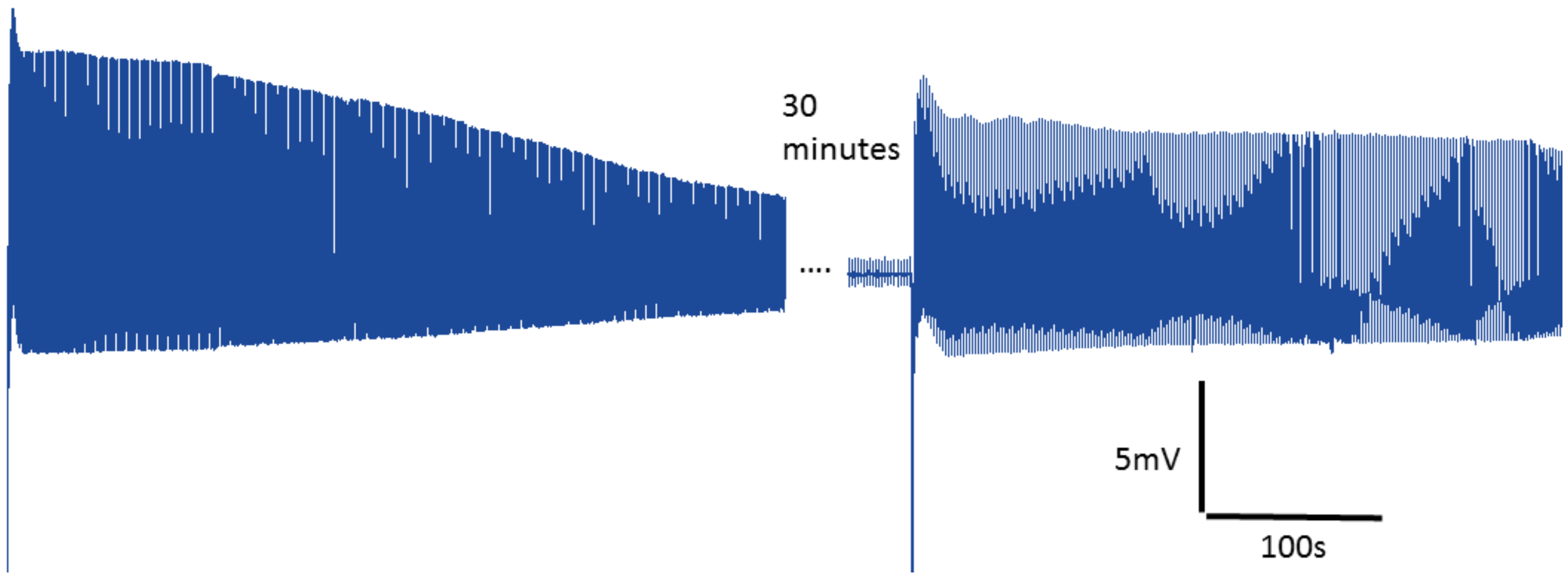

***Figure 25****: Intracellular recordings of cardiomyocytes from the same electrode after repeated optoacoustic poration.*

By translating the Axion MAESTRO acquisition system under the optical setup with a 2-axes translation stage, we were able to apply optoacoustic poration to the electrodes of different wells, obtaining intracellular recordings on several wells on the same multiwell MEA plate, as sketched in Figure 26a. This offers the possibility to test toxicity effects in parallel in diverse conditions (variable concentrations, different molecules…).
The highly detailed recording after optoporation also allows for the examining the effects of ion-channel drugs on hiPSCs-CM action potentials, with the specific aim of testing the capability of optoporation in conditions closer to those applied in drug screening assays.
Dofetilide, a class III antiarrhythmic drug and selective hERG channel blocker that has been approved for the treatment of atrial fibrillation, has been used as pivotal drug in order to check the capability of the system. This drug promote a prolongation of the repolarization duration, with an enhancement of Action Potential Duration (APD).
As control experiments, optoporation has been conducted on few cells to record action potentials in the absence of drugs. Then, the drug has been incubated for 10 minutes and another optoporation was applied to record the action potentials of the drug-treated cells.
Figure 26b**Figure 26** illustrates intracellular APs before (black trace) and after (blue trace) the administration of Dofetilide (100nM) from hiPSCs-CM cultured on the disordered fractal-like gold electrodes of the Axion Biosystems 12-wells MEA plate.
Intracellular recording revealed the expected changes in the shape and duration of the action potentials.
Optoporation therefore allows for recording highly defined intracellular signals that display the expected prolongation of the AP after the treatment with Dofetilide.
It is reasonable to conclude that the non-invasive and massive recording features of this novel platform can open the way for high throughput applications in pharmacological industrial applications, where high parallelization and low time and money consuming experiments are required for the development of new drugs.

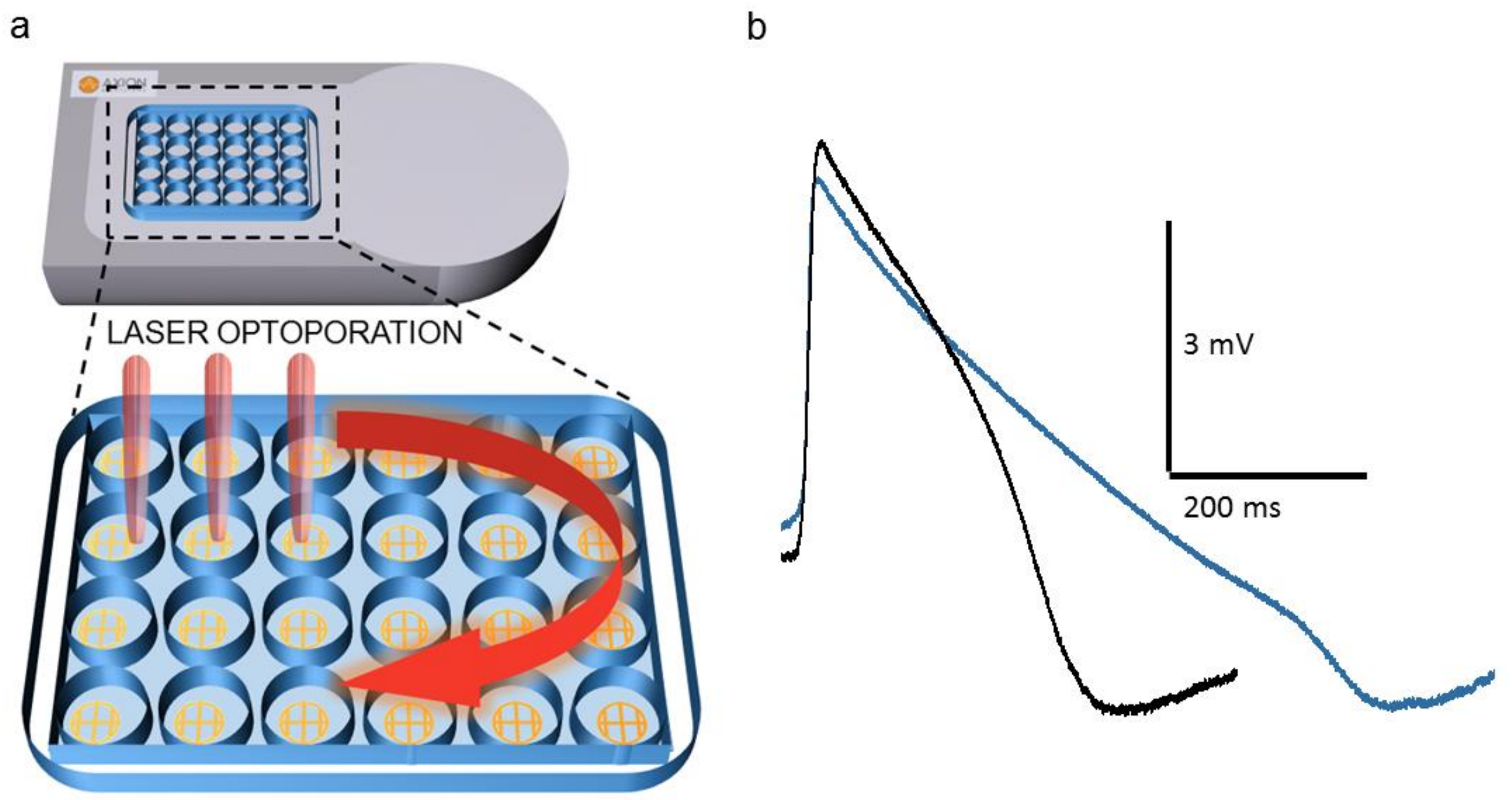

***Figure 26****: (a) Sketch of serial optoporation on multiwall MEA. (b) Intracellular recordings on disordered fractal-like gold electrodes on multiwell MEA from Axion Biosystems: with Dofetilide (blue) and control (black).*

# Optoporation and APs recordings from hiPSCS-CM cultured on meta-electrodes from MultiChannel Systems

MEA from MultiChannel Systems used in this thesis are an arrangement of 60 Titanium Nitride (TiN) electrodes allowing the targeting of several sites in parallel for extracellular recording and stimulation. Cell lines or primary cell preparations can be cultivated directly on the MEA. Freshly prepared slices can also be used for acute recordings, or can be cultivated as organotypic cultures (OTC) on the MEA. Excitable or electrogenic cells and tissues can be used for extracellular recording *in vitro*, for example, central or peripheral neurons, cardiomyocytes, whole-heart preparations, or retina.

A standard MEA biosensor has a square recording area of about 700 µm.

In this area, the electrodes of 10-30 µm are aligned in an 8 x 8 grid, as shown in the left panel of Figure 27, with inter-electrode distances of 100-200 µm (this distance depending from the size of the electrodes). The size of the electrode used in this thesis are 10 or 30 µm. The biological sample can be positioned directly on the recording area and the MEA serves as a culture and perfusion chamber as well.

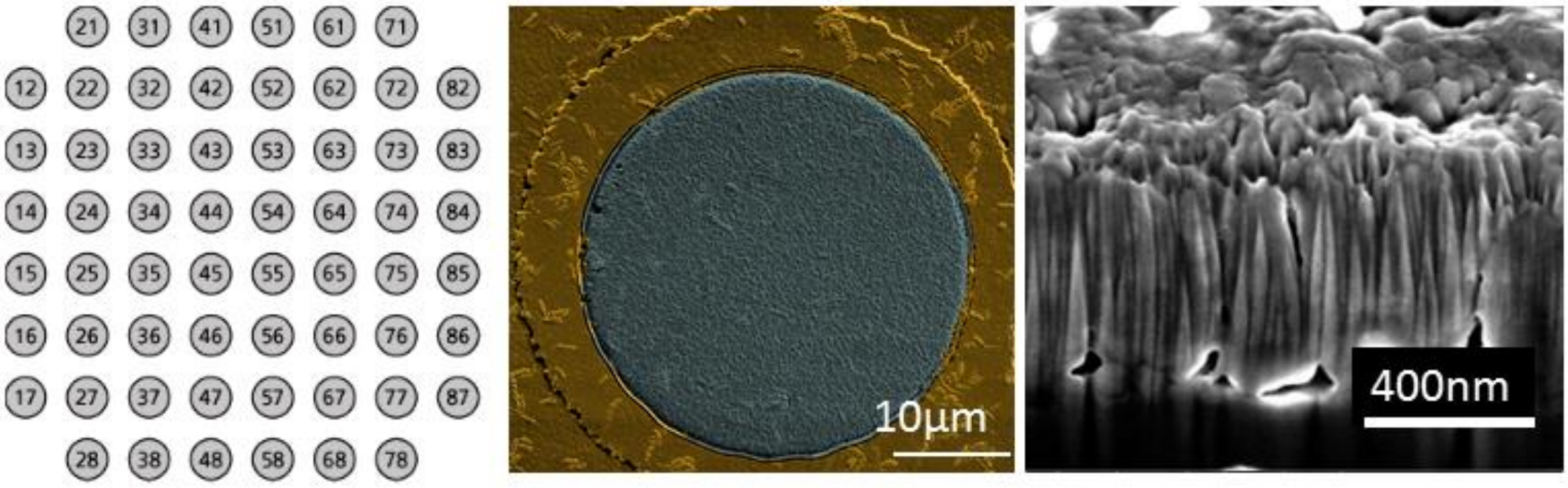

***Figure 27**: Electrode configuration on MCS-MEA (left) SEM acquisition of surface morphology of TiN electrodes (middlet) and cross section of the same electrode (right).*

A very rough electrode surface result in a large surface area that allows the formation of electrodes with an excellent signal to noise ratio without compromising on the spatial resolution. Central and right panel of Figure 27 show very well the morphology of the electrodes. As for the other two classes of meta-electrodes described in the previous sections, also here, a very rough surface is the main feature of the surface; therefore the hypothesis is that also rough TiN electrodes from MCS MEA can behave like meta-electrodes.

Furthermore, TiN is a very stable material and it represents a robust solution adopted by MCS to produce MEA for electrophysiological research and pharmacological industry. All MEAs with TiN electrodes have long life and can be reused several times if handled with care (about 30 times). The electrodes are embedded in a carrier material, usually glass. Standard tracks made of titanium (Ti) or indium tin oxide (ITO) are electrically isolated with silicon nitride (SiN). Standard contact pads are made of titanium nitride (TiN) or indium tin oxide (ITO). ITO contact pads and tracks are transparent, for a clear view of the specimen under the microscope.

The surface properties of a culture substrate are well known to have a decisive effect on the attachment, proliferation and differentiation of the cells. They can provide adhesion, growth-promoting, migration or differentiation cues. Pretreatment of the MCS MEA surface is therefore important to achieve reproducible results.

All of these materials tend to become hydrophobic during storage, preventing the attachment of the cells and pretreatment of the culture surface with permissive molecules is highly required. The first step in preparing the MEAs is therefore to ensure that the surfaces are hydrophilic for coating and cell adhesion.

The MEA surface has been exposed to a gas plasma discharge (100 W, 100% oxygen, 1 minute), which will make the surface polar and thus more hydrophilic. Note that the effect wears off after some days. The treatment gives a very clean (unless there is thick contamination) and sterile surface that can be coated readily with water-soluble molecules.

After that, standard MEAs can be sterilized with standard methods for cell culture materials using either rinsing with 70 % alcohol**,** UV-light (about half an hour depending on the intensity), vapor autoclavation, or dry-heat sterilization up to a temperature maximum of 125 °C.

Coating of MEAs with various materials is used for improving the attachment and growth of cell cultures or cultured slices. In this thesis, Fibronectin (from bovine plasma) is the selected biological coating, because of its high compatibility for heart tissues. The adhesion tends to be very stable, which allows longer cultivation times. A stock solution of 1 mg/ml fibronectin in distilled water or PBS (phosphate buffered saline) has been prepared and stored at 4 °C. The stock solution is diluted with water or PBS to a final concentration of 10 µg/ml before use. The MEA surface has been covered with 10 µl of this solution and incubate at 37 °C for at least 3 h (avoiding the evaporation using humid chamber close to the MEA).

Our MEA have been rinsed, after each experiment, with distilled water first, and then apply 1% Terg-A-Zyme solution (Sigma) for several hours, followed by accurate washing with distilled water. In the end, the MEA is dried before another use.

In order to assess the functionality of optoporation on MCS MEA, Pluricytes (fron Ncardia) have been adopted as cellular model. In fact, as also suggested by CiPA initiative, a novel platform for cardiotoxicity assessment should give the same results with different cardiomyocyte lineage. Pluricyte cardiomyocytes are fully functional hiPSCs-derived ventricular cardiomyocytes that are particularly suitable for electrophysiology-based MEA assays for predictive safety pharmacology and toxicity testing. The cells are thawed in humid bath at 37°C and diluted at the specific concentration of 30000 cells/ 10 µl. Fibronectin coating is removed, after 3 hours, by aspiration from the electrode array and the cells are directly plated on the MEA surface. After one hour in incubator, 1 ml is added inside the culture chamber without disturbing the cells and the samples are feed every 2-3 days for 8 days. After that, a spontaneous beating monolayer of cardiomyocyte covers the entire electrode array, as depicted in Figure 28**Figure 28** where four representative electrode of a MEA are covered by well spread and adhered cardiomyocytes.

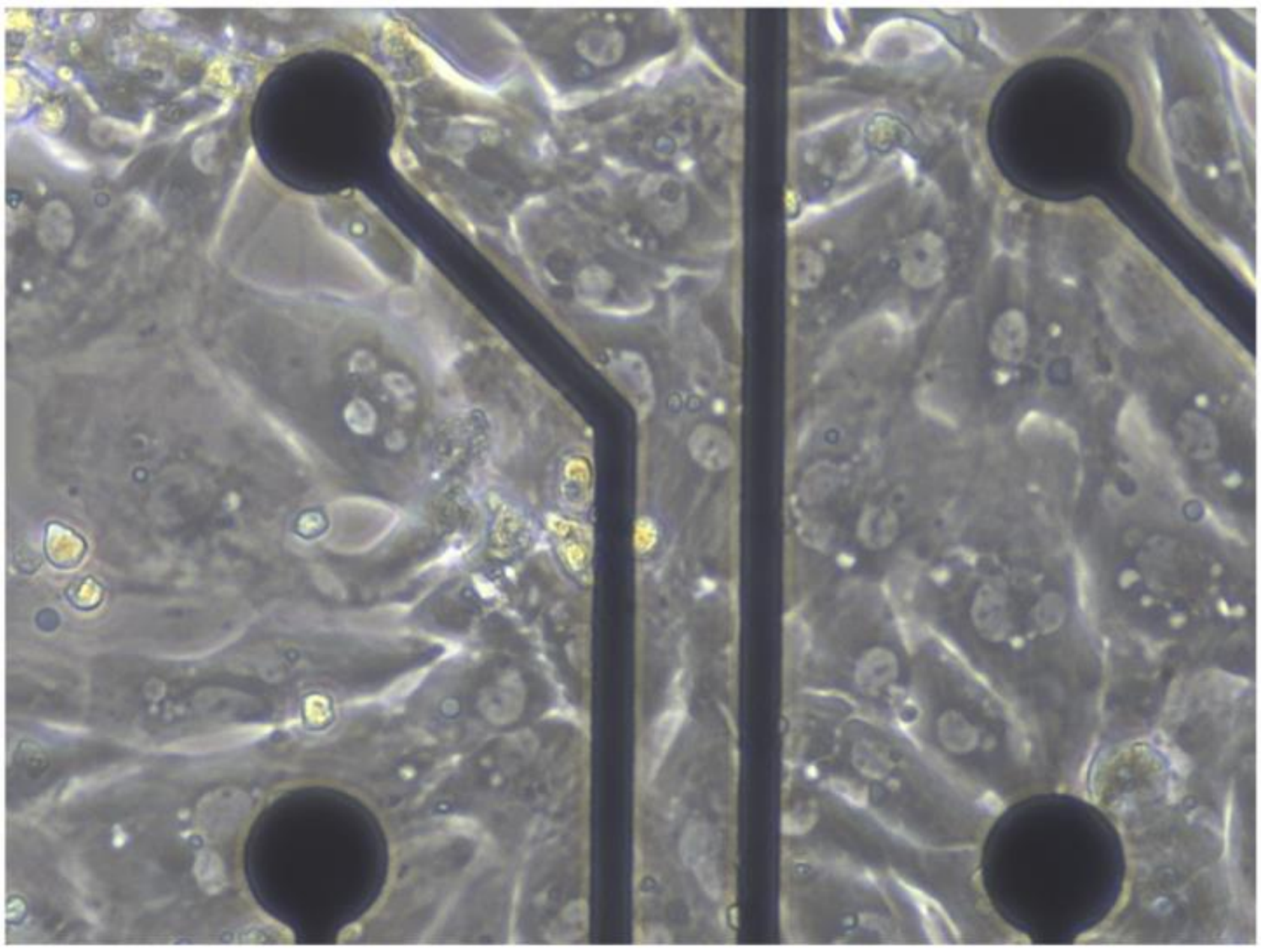

***Figure 28**: MCS MEA covered by a monolayer of beating Pluricytes.*

The recordings and optoporation experiments on MCS MEA have been performed at 37 °C outside the incubator by a custom-made MEA acquisition system, which could record 24 out of 60 channels at 7 KHz sampling rate. The same configuration of laser source, microscopy and laser spot focalization have been used as described above.

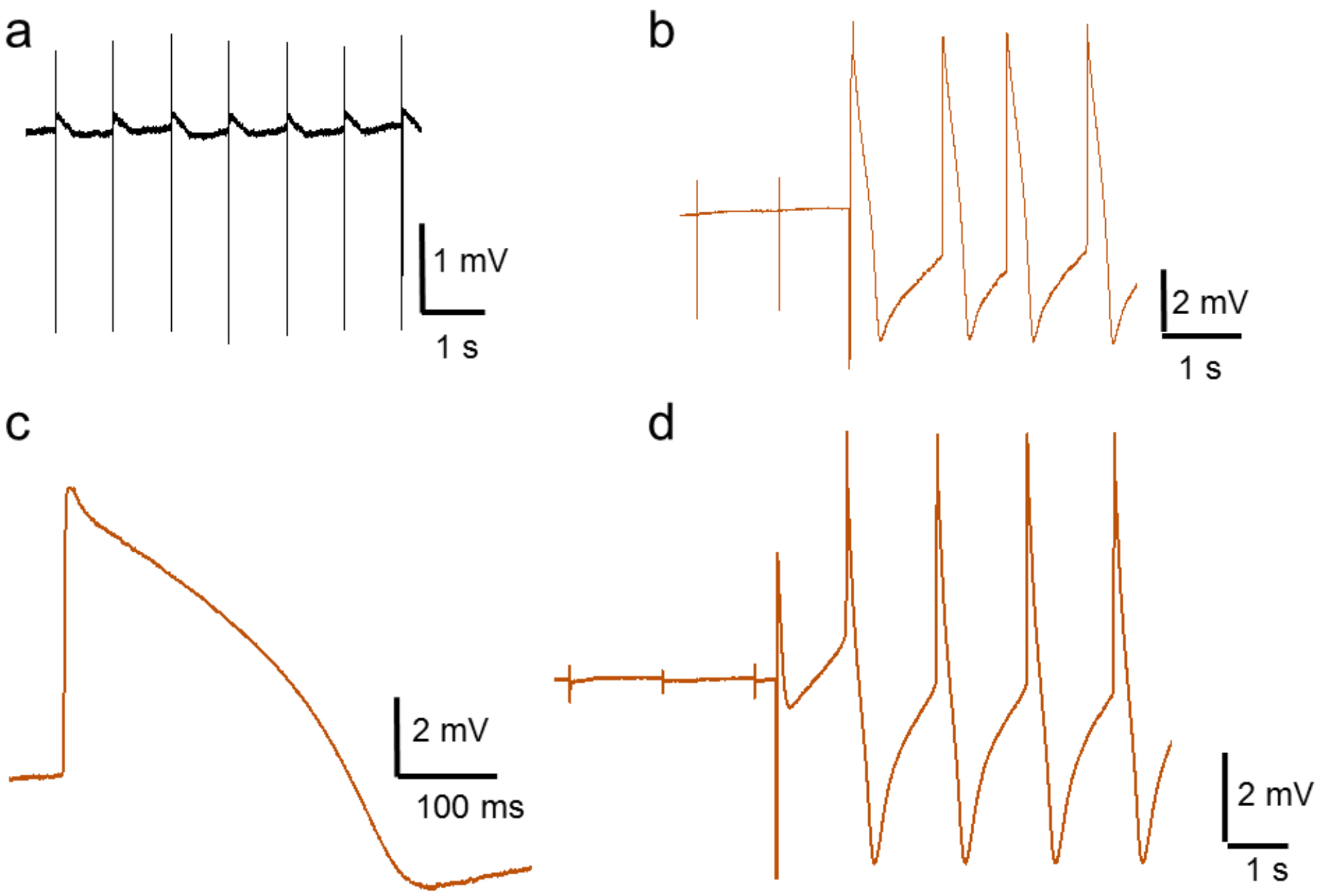

***Figure 29****: Signals recorded from hiPSCs-CM on MCS MEA. a. Extracellular signals. b. Switch from extra to intra cellular recording. c. Intracellular action potential. d. Example of saturated signals.*

First, five minutes of extracellular activity was recorded to assess the spontaneous beating.
Figure 29a shows an extracellular recording; the signals show very high signals-to-noise ratio, with an amplitude of approximately 3 $mV_{pp.}$ To achieve intracellular coupling, an ultra-fast-pulsed 1064 nm laser (pulse train duration = 20 ms, average power = 1 mW) was used to excite the TiN electrodes on the MCS MEA with diameter of 30 µm.
Figure 29b-c show that, after laser excitation, signal shapes that fit with intracellular cardiac APs are recorded, showing a fully positive phase followed by a repolarization step. The average amplitude of the intracellular signals is approximately 6 mV, characterized by a time duration approximately of 300-400 ms, reflecting the average values of intracellular APs of hiPSC cardiac cells, as shown in previous chapters.
Since the high quality recording on MCS MEA with 30 µm electrodes, an attempt of optoporation on MCS MEA with 10 µm electrodes have been conducted. In this case, the smaller electrode should provide a better sealing with the cardiomyocytes, thus providing intracellular recordings with high amplitude. As shown in Figure 29d in fact, the traces completely saturated the 10 mV range of the INTAN chip amplifier.

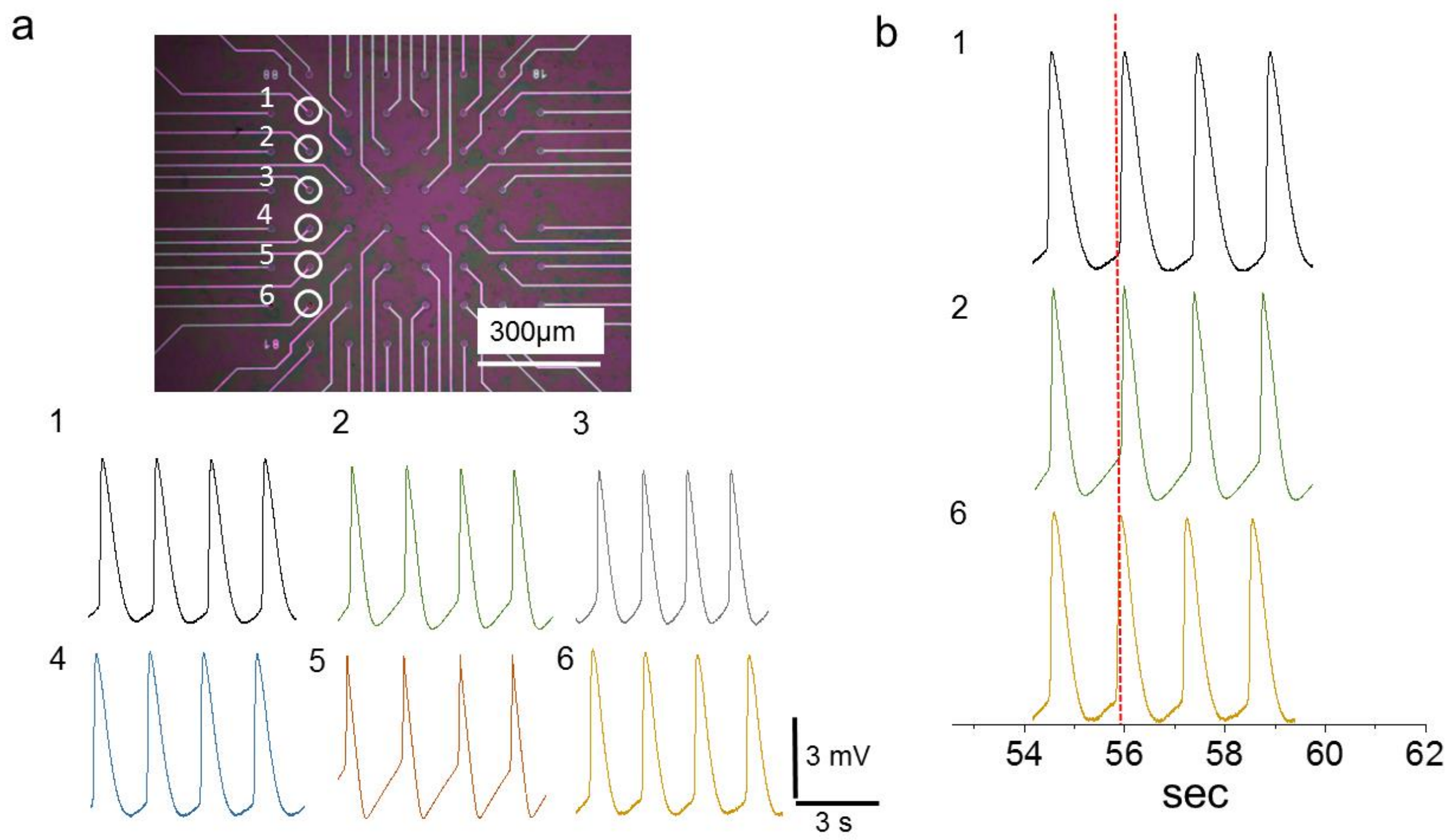

***Figure 30**: Evidences of low invasiveness of optoporation. a. Sequential multi-site optoporation on MCS MEA. b. Synchronized intracellular recordings from three electrodes after optoporation.*

These results highlight and confirm two important considerations. First, nanoporous TiN-based electrodes coupled to optoporation can effectively represent a valid alternative to past intracellular techniques for intracellular recordings. Despite the poor plasmonic enhancement shown in the previous chapter, the average power required for optoporation on TiN (1 mW) is comparable with that used on nanoporous platinum, as described in the above section. This result is related to the work function of TiN; being lower than other materials, it make easy the hot electron emission after laser excitation with low power. Although in different context, Naldoni and collagues reported, through experimental and theoretical evidences, that the improved performance achieved with TiN is due to the high absorption efficiency in a broader spectral region than that of Au or other plasmonic materials, leading to broadband hot electron generation and emission [131].

The second advantages of the TiN-based electrodes, as compared to common plasmonic materials, are that transition metal nitrides are robust, inexpensive, and, therefore, MCS MEAs coupled with optoporation represent a robust and efficient solution for the generation of a novel platform for intracellular electrophysiology and cardiac APs recording.

To show the reproducibility of the process and thus the compatibility with network-scale recordings, we depict sequential multisite optoacoustic porations on six different electrodes of the same MEA. As shown in **Errore. L'origine riferimento non è stata trovata.**, MCS MEA has been exploited in a multisite sequential poration experiment of hiPSCs-CM from six different electrodes.

After each optoporation, high quality intracellular signals have been recorded from each electrode, without perturbing the activity of the whole network. The bottom panel shows the intracellular APs from the six electrodes with average amplitude of 5 mV.

The process, as also highlighted in the right panel, does not affect the beating frequency of the porated cardiomyocytes. Here, a temporal view of few seconds shows that cell positioned on electrode 1 and 2 (spaced 100 µm) are still synchronized (red dot line) with electrode 6 that is far from these two of about 500 µm.

These data show that optoporation is suitable for reliable intracellular recordings on large cultures using materials with poor plasmonic enhancement, such as TiN meta-electrodes.

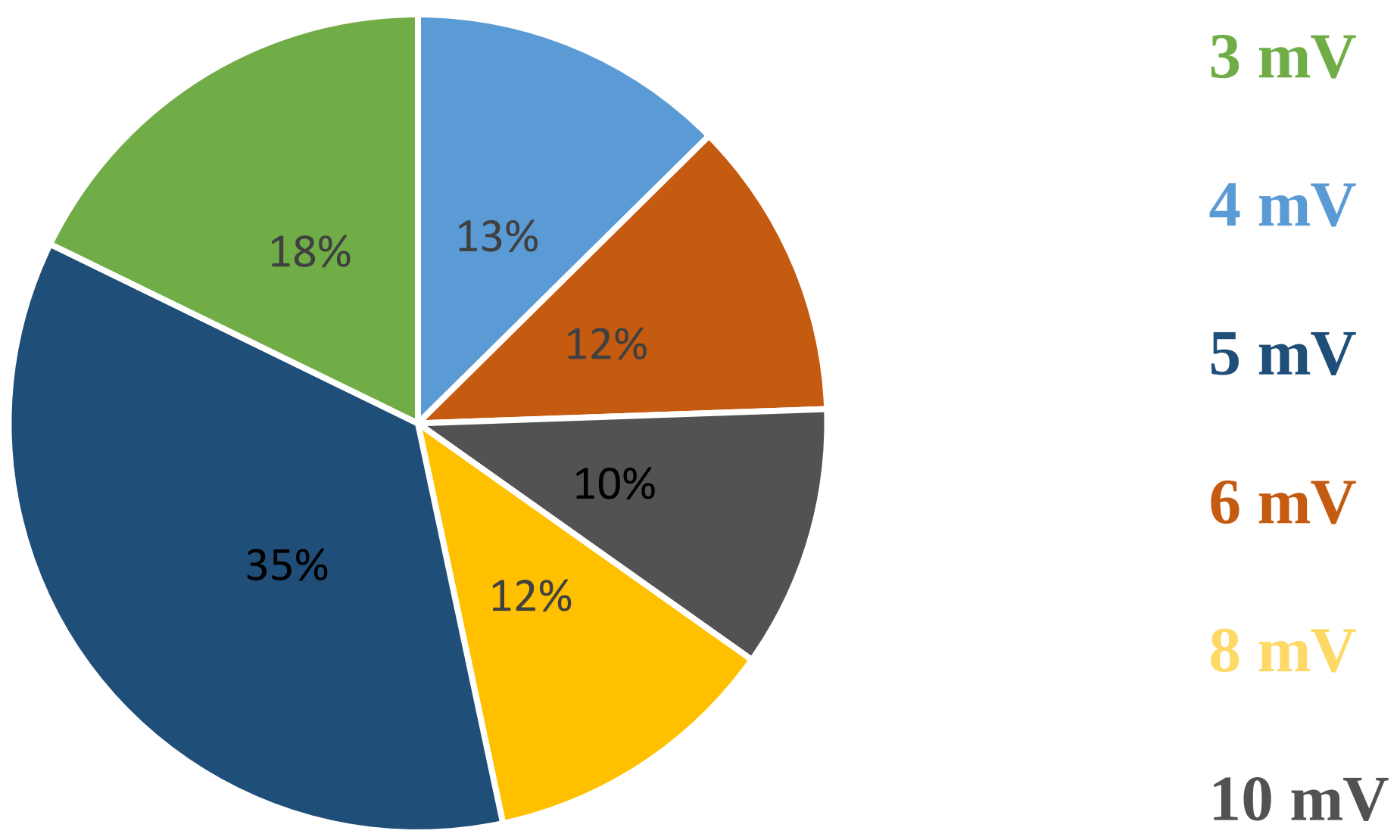

***Figure 31**: Quantitative amplitude classification of intracellular signals on overall experiments.*

In order to quantify the success rate of the platform, all the experiments of intracellular optoporation and recording have been classified. Figure 31 reports in detail, the amount, in percentage, of intracellular recording experiments after optoporation conducted on MCS-MEA devices, where the amplitude of the recording signals has been evaluated as quality factor. On the overall experiments, the 35% of the signals, show amplitude of 5 mV, the average value of a representative experiment.

The second and third portion of successful experiments, respectively about 18% and 13%, has a value of 3 mV and 4 mV, comparable with most of the state of the art literature about intracellular recording from human derived cardiomyocytes. Lastly, a couple of slices of the diagram are occupied by signals with amplitude of 6 and 8 mV, which represent a minor population of exemplary signals. Finally about the 10% of the experiments recorded intracellular signals with amplitude of 10 mV or more, according with the previous discussed adhesion strength of the cell membrane on electrodes smaller than 30 µm.

# Optoporation on MCS MEA for Acute Drugs Cardiotoxicity

The highly detailed recording by the TiN meta-electrodes after optoporation also allows for the examination of the effect of ion-channel drugs on hiPSCs-CM action potentials. We tested three different chemical compounds for  demonstrating this capability as an example of potential drug screening applications . All the drugs tested affect the intracellular signals' waveform and duration, inducing for example prolongation or reduction of APs. All the drugs were bought from Sigma.
The powder of each drug has been solubilized in DMSO, an organic solvent highly used in drug dilutions and stocking, at specific stock concentration and then stored as aliquots at -20°C. Before each experiment, a stock vial was thawed and equilibrate at room temperature for at least 30 min and then diluted in serum-free medium at specific concentration.
After that, a cycle of optoporation has been performed as reference experiment and then the medium culture was added with a volume of drug in such a way that the final desired concentration was reached. Subsequently, another cycle of optoporation has been performed in order to record drug-induced changes in APs.
In Figure 32 are depicted three examples of intracellular recording before (black traces) and after drug treatments (blue traces).
We measured the drug-induced AP deformation after the treatment with Nifidipine, a common calcium channel blocker that reduces the AP duration in a dose-dependent manner [110]; the effect starts at 10 nM and saturate at 1 µM.
Figure 32a shows clearly the effect of Nifidipine (60nM) on the shape of the intracellular signals after the treatment.

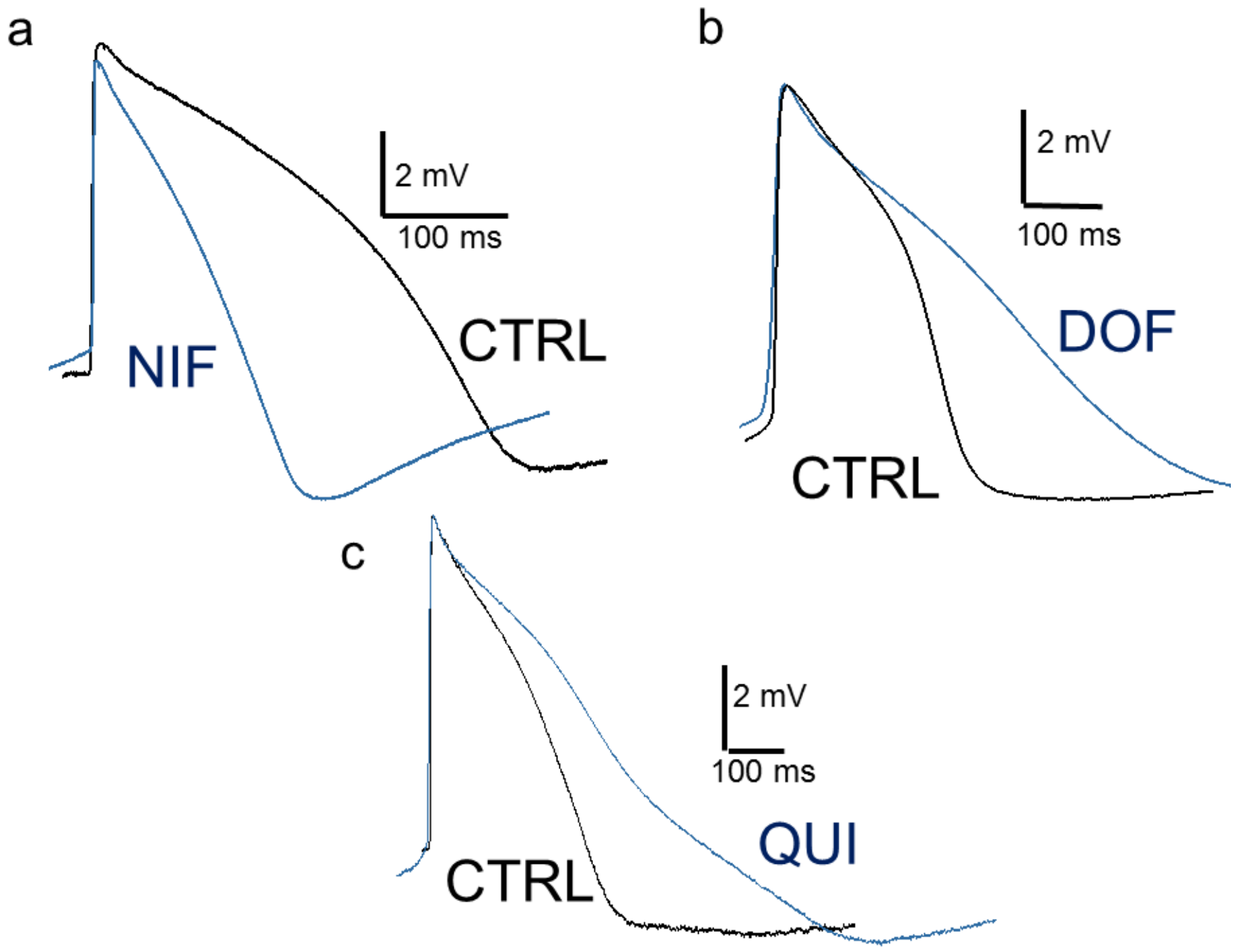

***Figure 32**: Optoporation on MCS MEA for acute cardiotoxicity. a. Effect of Nifidipine 60 nM. b. Effect of dofetilide 100 nM. c. Effect of Quinidine 4 µM.*

The second tested molecule, Dofetilide, a class III antiarrhythmic drug mostly used for the treatment of atrial fibrillation [110]. As shown in Figure 32b, it induces an AP duration prolongation compared to the CTRL waveform.
The third molecule used to further confirm the capabilities of optoporation has been Quinidine. Quinidine ion channel interactions are known to be complex. In addition to its blocking effect on hERG, quinidine blocks inward calcium current (ICa) and the rapid (IKr) and slow (IKs) components of the delayed potassium rectifier current. Quinidine induces a prolongation on AP duration as shown in Figure 32c.
The effect of Nifidipine on the beating frequency has been also investigated, as shown in Figure 33. In Figure 33a there are 10 seconds of intracellular recording before (black traces) and after (blue traces) the treatment with a concentration of 60 nM.
The increase of the beating rate is clarified in the histogramin Figure 33b. After the treatment with the drug, the number of Beat Per Minute (BPM) in fact increases up to about 110. Concomitant with decreased APD, we observed a higher beating frequency but no arrhythmic behavior, as also reported in literature [110].

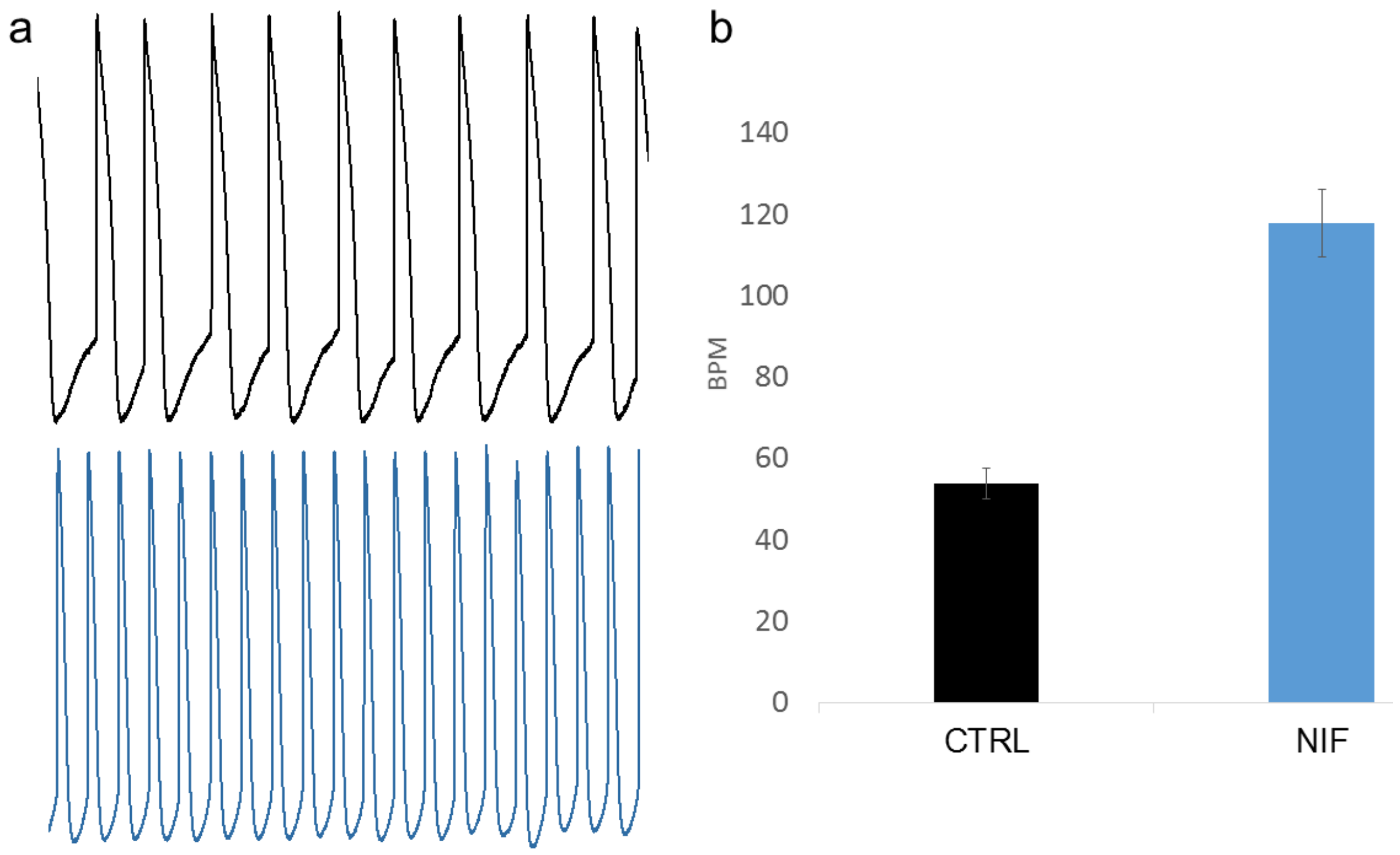

***Figure 33**: Nifidipine effect on beating rate. a. Intracellular traces from CTRL (black) and Nifidipine treated (blue) hiPSCs-CM cultured on two different electrode. b. Data analysis of the beating rate between treated cell (blue) and CTRL (black) not teated (p-value<0.005).*

The capability of the platform that has been developed can therefore benefit several area of cardiology, such as fundamental studies of signal propagation and tissue-based drug screening for cardiac diseases. For the measurements of simple variables of cardiac dynamics, such as beat frequency and conduction velocity, one can operate in the extracellular mode as already enabled by the standard MEA features.

In addition to these extracellular mode, however, the coupling of MEA technology with optoporation allows operating in the intracellular mode that determine with high precision, the duration and shape of membrane potential as well as their propagation as AP. These features, added to the advantages of meta-electrodes shown in the previous chapter, represent unique benefit over traditional MEA in studying network dynamic and drug effect.

# Long-term Optoporation of hiPSCs-CM on MCS MEA

The most important drawback that afflicts patch clamp and, generally, intracellular recording mediated by electrical pulses, is related to the damages and the stress induced to the cells during the experiment. After recordings with patch clamp, cells do not survive and cannot be measured again in later stages. This of course afflicts cardiotoxicity assessment and reliability of drug screening test using patch clamp. Although well spread in research laboratories and academia, patch clamp can be used following long term exposure of cells to a molecule prior the analysis. Nevertheless, this approach prevents the analysis of the same cells over multiple days and therefore limits the characterization of *in vitro* chronic activity of a molecule.

Coupling MEA with low invasive intracellular recording, could provide the advantage of long term recording to assess the kinetics of alterations in the action potential of the same cell during the treatment, that would represent a step forward a high resolution cardiotoxicity assay.

In this context, optoporation of the same cardiomyocyte (on the same electrode) on TiN electrodes of MCS-MEAs in three successive days-in-vitro (DIVs) has been conducted, in order to establish the performances in terms of cell health and recovery after each poration event. The experiments have been performed in order to monitor the same cardiomyocytes for up to 120 hours (from 3DIV to 8DIV). In this experiment, a first optoporation of three electrodes has been done at 3DIV. After the recording of intracellular signals, the cells are placed back in to the incubator at 37°C and then another cycles of optoporation have been repeated on the same electrodes at 6DIV, 7DIV and 8DIV.

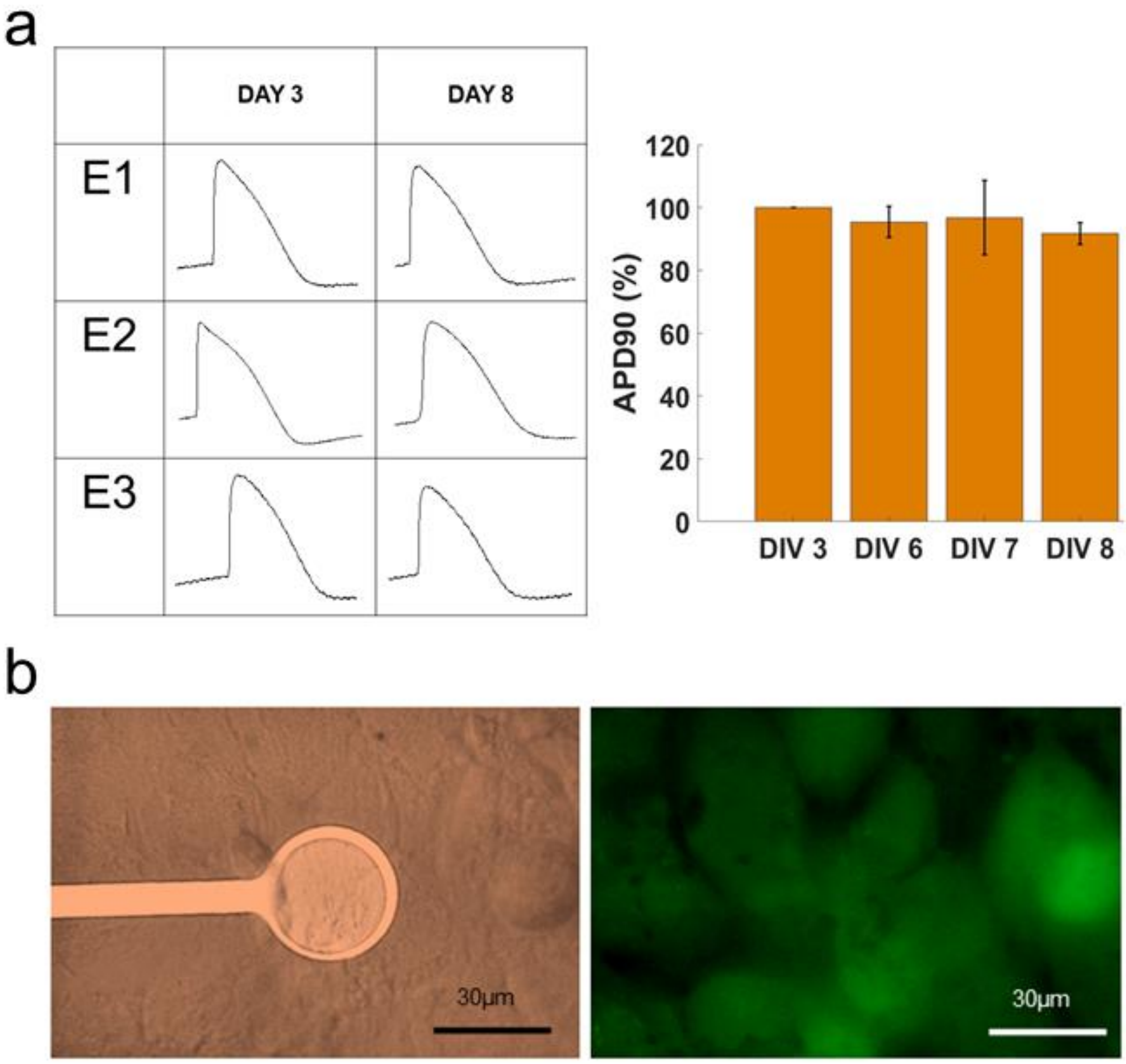

***Figure 34**: Long term experiment of optoporation oh hiPSCs-CM on MCS MEA. a. Intracellular recording from hiPSCs-CM culture on three different electrodes ( E1, E2, E3) of MCS MEA at DIV3 and DIV8 and. analysis of Action Potential Duration at 90% of repolarization (APD90) up to 120 hours in culture.*
*b. LIVE/DEAD assay on one representative electrode after long-term experiment of optoporation.*

In Figure 34a, on the left, intracellular recordings from three exemplary electrodes E1, E2 and E3 at 3DIV and 8DIV are reported. There are evidence of very good agreement of the intracellular signals recorded at 3DIV and at 8DIV in terms of amplitude, shape and noise level.
This demonstrates both the very good health of the cells and their good adhesion on the electrode after optoporation events over 5 days. In the histograms on the right, it is reported a plot of the Action Potential Duration (APD) at 90% (APD90) of the repolarization phase. It’s a typical key parameter used in cardiac electrophysiology to characterize the features of an AP. The data show the percentage variations of APD90 at 6, 7 and 8 DIV in respect to the first poration at 3DIV that represents the reference value (100%).
Despite minor fluctuations, which may be also attributed to electrophysiological maturation of the cells over time <sup>134,135</sup><sup>131,132</sup><sup>131,132</sup><sup>130,131</sup><sup>130,131</sup><sup>129,130</sup><sup>127,128</sup><sup>126,127</sup><sup>126,127</sup><sup>126,127</sup><sup>126,127</sup><sup>125,126</sup><sup>125,126</sup><sup>124,125</sup><sup>124,125</sup><sup>123,124</sup><sup>122,123</sup><sup>122,123</sup><sup>121,122</sup><sup>121,122</sup><sup>120,121</sup><sup>120,121</sup><sup>120,121</sup><sup>120,121</sup>(Braam, Passier and Mummery, 2009; Robertson, Tran and George, 2013)(Braam, Passier and Mummery, 2009; Robertson, Tran and George, 2013), the recorded APs show stable shapes over time and over four poration events. The long-term experiments have been performed on several electrodes distributed over more preparations.
After the last measurement, the cell culture has been stained with a live imaging technique. LIVE/DEAD kit has been exploited for the qualitative evaluation of cell health after the long term optoporation experiments. In Figure 34b **Errore. L'origine riferimento non è stata trovata.** , a bright field image of a monolayer of hiPSCs-CM is and the LIVE/DEAD fluorescence assay confirmed that the cells preserved good health, as shown also by the green fluorescence generate from intracellular enzymatic activity that metabolized Calcein AM into a fluorescent green dye (PrI data not shown due to lack of signal).

# Optoporation on MCS MEA for cardiotoxicity assessment

Long-term action potential (AP) studies from cardiomyocytes preparations are currently lacking, thus limiting our understanding of cardiac physiology in development and in disease. Furthermore, prospective studies have shown that some treatments for cancer are cardiotoxic [136]. The heart damage that they cause can manifest itself as arrhythmia, arterial hypertension, thromboembolism, angina pectoris, myocardial infarction, or heart failure.
It has been observed that potentially lethal complications can arise as late as 40 years after treatment of the original cancer [137]. So far, no attempts of long term APs recordings from cardiomyocytes treated with chronic drug injections have been successfully realized in this direction.
Therefore, building on the results shown before, a multiday administration of two molecules and subsequently optoporation has been performed to recapitulate drug-induced physiological responses and changes in the shape of the APs over several days.

Specifically, we analyzed the effect of Dofetilide 100nM and Nifidipine 60nM. In order to detect drug induced AP's changes, the same protocol described in the previous sections has been adopted. At 3DIV, through optoacoustic poration, we recorded intracellular APs from the cardiomyocytes. Subsequently, we applied Dofetilide 100nM or Nifidipine 60nM to the culture, recording the intracellular signals that reflected a shape modification due to the drugs; as expected, the recorded APs after drugs have an elongated shape in case of Dofetilide and a shortened duration in case of Nifidipine, as also discussed above in previous sections.

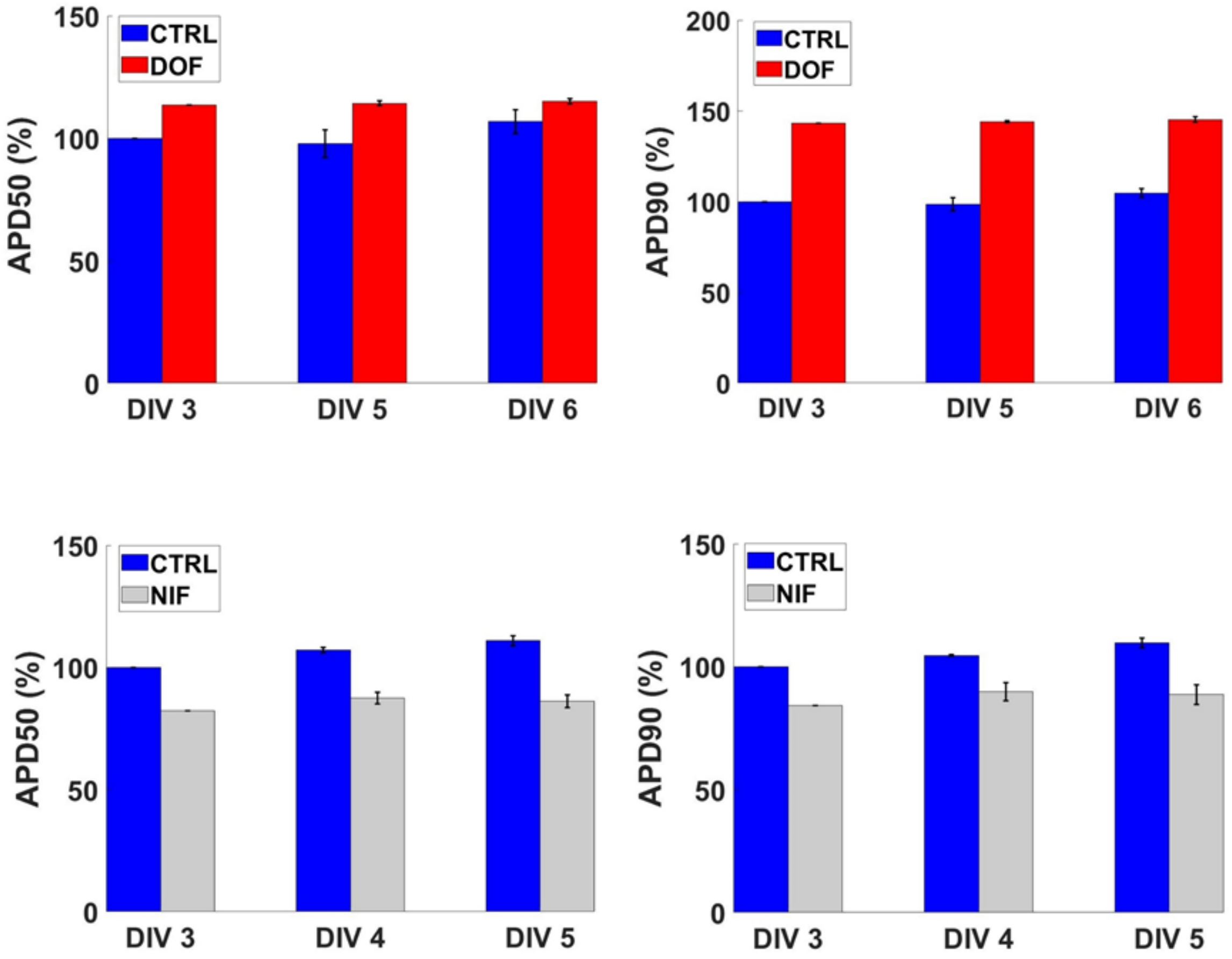

***Figure 35**: Long-term intracellular recordings and chronic detection of drugs on cardiomyocytes. On top intracellular APs recorded from the same electrodes at 3, 5, 6 DIV before (blue curves) and after (red curves) the treatment with Dofetilide 100 nM. On the bottom intracellular APs recorded from the same electrodes at 3, 4, 5 DIV before (blue curves) and after (grey curves) the treatment with Nifedipine 60 nM.*

After measurements, the cell medium was replaced for removing the drugs, and the MEAs were placed again in the incubator. We then repeated the recordings in successive days on the same electrodes, measuring again APs before and after drugs administration. The analysis of the Action Potential Duration at 50% and 90% of repolarization (respectively APD50 and APD90), expressed as percentage of variation in respect to the values at 3DIV, reports considerable results over multiple days. The data reported in Figure 35 show in fact very stable intracellular signals over the days of the experiments, as also demonstrated before. In the upper part of the figure, results about long-term treatment of hiPSCs-CM are reported. Dofetilide-induced APD 50 at day 3 is used as reference, in term of percentage, for a comparison with percentage values of APD 50 at day 5 and 6. Despite few differences related to biological dynamic abovementioned, the intracellular signals recorded show the reliability on long-term performance of the platform developed in this thesis. Furthermore, also APD90 has been detailed in the figure. The same experiments have been conducted in order to validate the results with Nifidipine (lower part of the figure). Also here, very satisfactory results have been observed.

These studies show that MEA systems with optoporation are promising platforms for long-term measurements of the APs of cardiac constructs to track their electrical activity. Of course, a system based on *in-vitro* constructs is not able to model alterations that happened in decades or longer period, but could provide an efficient instrument to assess maladaptive remodeling caused by drugs or genetic pathologies.

Furthermore, an extensive data analysis has been provided about cardiac pharmacology such as APD50 and APD90 in order to monitor the response of the cells to chronic drug treatment.

MEA coupled with optoporation therefore can fill a large technological gap between high resolution intracellular recordings and long term electrical analysis of cardiomyocyets, thus providing a breakthrough in the field of pharmacology by coupling the advantages of both optoporation and metaelectrodes. It is now possible to monitor cardiac activity from single cells and large ensembles by maintaining stable recordings over extended time periods (hours to days). Such attributes, when combined together, are necessary as they open up new opportunities to investigate advanced cardiac physiological phenomena such as syncytium formation, maturation, plasticity and dysfunction.

# Conclusion

Screening for drug adverse cardiac effects faces the challenge that most informative methods use animal or *ex vivo* models, which are distinctly low throughput, costly, and require large compound quantities. It is, therefore, desirable to identify approaches that combine strong prediction power for cardio-tolerability with experimental convenience.

The need for more reliable prediction of cardiotoxicity has prompted the development of several new *in vitro* technologies to help the detection of adverse cardiac effects in the early stages of the drug development.

In 2013, the *Comprehensive in Vitro Proarrhythmia Assay* (CiPA) emphasized the need of more integrated *in vitro* and *in silico* approaches for cardiotoxicity assays, thus suggesting the use of human induced pluripotent stem cell-derived cardiomyocytes (hiPSC-CMs) coupled to Multi Electrode Array (MEA). Although some phenotypic features can differ from a type of hiPSC-CMs to another one, the use of cardiomyocytes has been proposed as a promising way to overcome the limitations of the existing methodologies used for preclinical safety evaluation of pharmaceutical compounds, being now considered as a reliable cardiotoxicity assay.

On the other hand, MEA itself have many limitations that prevent correct and sharp analysis of cardiotoxic effect of drugs on specific ion-channels and ions exchange mechanisms.

Therefore, many research groups started to push on the development of platforms for detailed acquisition of intracellular electrical signals from hiPSCs-CM on MEA devices. The most exploited approach so far has been the delivery of electrical pulses in order to permeabilize the cell's membrane and gain intracellular access, recording attenuated intracellular action potentials from cardiomyocytes. Despite recent developments showed in the introduction chapter of this thesis, this technique suffer of many defects.

In order to overcome these drawbacks, this work of thesis proposed a novel tool for intracellular recording of sharp attenuated APs from hiPSCS-CM by means of optoporation. The tool has been optimized for cells cultured both on high density CMOS-MEA for whole network recording and on commercial MEA in single and multiwell configuration for research applications and high throughput applications in pharmacological industries.

The key point of the whole work has been the development of tools with high throughput, versatile and reliable features, by integrating ultra-fast pulsed laser excitation on unmodified commercial MEA systems used worldwide in several research laboratories and pharmaceutical companies, thus overcoming abovementioned issues of MEA decorated with 3D nanostructures.

Therefore, I first described the behavior of meta-electrodes, namely thin film of metals with porous or rough surface without post fabrication processes that behave like broadband absorber in the NIR range of the electromagnetic spectrum and are able to interface very well with biological cells. Since meta-electrodes are widely used in MEA technology as recording electrode from many companies, it is clear to conclude that these substrates can be alternative candidate materials for electrophysiological applications.

Therefore, an accurate study regarding the capability of optoporation has been conducted used hiPSCs-CM, which represent today a reference cellular model for pharmacological and toxicological studies, cultured on top of meta-electrode of large families of MEA electrode configurations already commercially offered.

Three different classes of meta-electrodes have been exploited: high-density CMOS-MEA produced by 3Brain, standard MEAs from Multi Channels Systems and multiwell MEA plates from Axion Biosystems.

After a deep investigation, all these three platforms, coupled with optoporation, have positively performed in terms of quality of intracellular signals, reliability, and fast translational potential.

In particular, the results showed in this thesis report recordings of high quality intracellular APs on porous platinum electrodes from 3Brain CMOS-MEA. These findings are significant because of the low invasiveness of APs recording and the possibility to perform whole network recordings from a cardiac beating monolayer.

High quality attenuated intracellular on fractal-like disordered gold nanostructures from Axion Biosystems have also been reported. The application of optoporation on multiwall MEA configuration paves the way for application in pharma industry, where high parallelization is required. It is also reasonable that laser optoporation is an extremely effective technique that can have an immediate impact on pharmaceutical research, enabling the reliable assessment of cardiotoxicities at the industrial scale.

High quality attenuated intracellular action potentials on nanoporous TiN electrodes from MCS-MEA have been demonstrated in the end. In conjunction with the porosity at the nanoscale, which enhances both cell adhesion and optical coupling, the low work function of TiN compensates for the poor plasmonic enhancement of the material. Since MCS-MEAs are the most exploited devices in biological research applications, they have also been exploited for a detailed investigation of the response of optoporation and meta-electrodes in different experimental situations. Different conditions have been simulated to prove the capability of attenuate APs recording for prolonged time (useful for example in the assessment of chronic drug-induced response in research field) or from multiple electrodes from the same culture in order to investigate cell’s maturation and differentiation.

The developed experimental paradigm of intracellular recording, constituted by functionally connected hiPSCs-CM cultured on meta-electrodes can constitute a next generation tool that address the CiPA requirements and therefore is an appealing candidate for cardiotoxic evaluation of chemical compound in drug development screening.